%% file: main.tex
\documentclass{hpg_sty/egpubl}
\usepackage{hpg_sty/hpg25}
 
\SpecialIssueSubmission    

\CGFStandardLicense

\usepackage[T1]{fontenc}
\usepackage{hpg_sty/dfadobe}  

\biberVersion
\BibtexOrBiblatex
\usepackage[style=alphabetic]{biblatex} 
\electronicVersion
\PrintedOrElectronic

\ifpdf \usepackage[pdftex]{graphicx} \pdfcompresslevel=9
\else \usepackage[dvips]{graphicx} \fi
\usepackage{multirow}

\usepackage{hpg_sty/egweblnk} 

\def\metricname{CGVQM}

\newcommand{\Rev}[1]{{{#1}}}

\usepackage{subcaption}

\usepackage{amsmath}
\usepackage{amssymb}
\usepackage{pifont}
\newcommand{\cmark}{\ding{51}}%
\newcommand{\xmark}{\ding{55}}%
\usepackage[table,xcdraw]{xcolor}
\usepackage{placeins}
\usepackage{xspace}
\usepackage{lscape}
\usepackage{longtable}
\title[\metricname+D]%
      {\metricname+D: Computer Graphics Video Quality Metric and Dataset}

\author[Jindal et al.]
{\parbox{\textwidth}{\centering A. Jindal, N. Sadaka, M.\,M. Thomas, A. Sochenov and A. Kaplanyan
        }
        \\
{\parbox{\textwidth}{\centering Intel Corporation, USA
       }
}
}

\begin{document}

\teaser{
  \includegraphics[width=\textwidth]{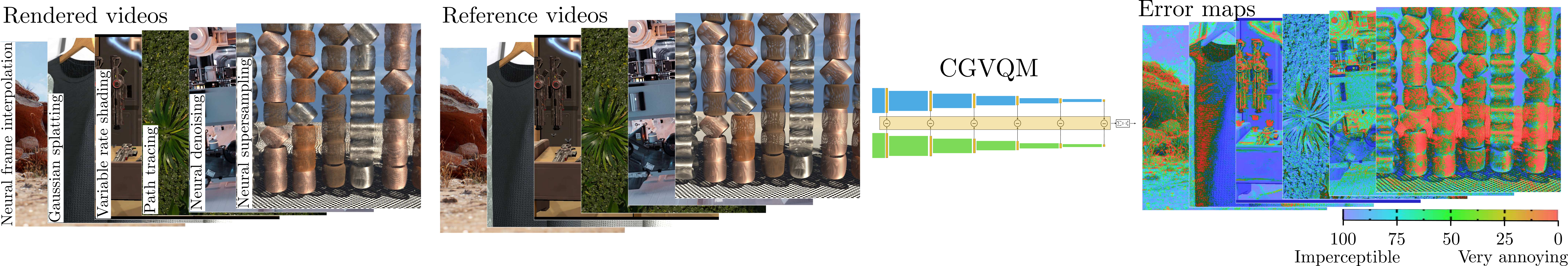}
  \centering
  \caption{We collected a video quality dataset containing distortions generated by modern rendering algorithms. Using this dataset, we demonstrate that the feature space of pre-trained 3D-CNNs exhibits a high correlation with human ratings and propose a new video quality metric, \metricname, which outperforms existing quality metrics.}
  \label{fig:teaser}
}

\maketitle
\begin{abstract}
   While existing video and image quality datasets have extensively studied natural videos and traditional distortions, the perception of synthetic content and modern rendering artifacts remains underexplored. We present a novel video quality dataset focused on distortions introduced by advanced rendering techniques, including neural supersampling, novel-view synthesis, path tracing, neural denoising, frame interpolation, and variable rate shading. Our evaluations show that existing full-reference quality metrics perform sub-optimally on these distortions, with a maximum Pearson correlation of 0.78. Additionally, we find that the feature space of pre-trained 3D CNNs aligns strongly with human perception of visual quality. We propose \metricname, a full-reference video quality metric that significantly outperforms existing metrics while generating both per-pixel error maps and global quality scores. Our dataset and metric implementation is available at \url{https://github.com/IntelLabs/CGVQM}.


\end{abstract} 

\input{secs/introduction}
\input{secs/related_work}
\input{secs/dataset}
\input{secs/3dcnn}
\input{secs/conclusion}

\section*{Acknowledgments}
We thank the anonymous reviewers for their insightful comments and helpful suggestions, which helped improve the quality of this work. We also thank Intel Labs for supporting this research, and to all the volunteers who contributed their time to participate in the user study.


\printbibliography 
\input{figs/distortions_all}

\setcounter{section}{1}
\renewcommand{\thesection}{A\arabic{section}}
\input{secs/appendix}

\end{document}

%% file: secs/introduction.tex
\section{Introduction}
\label{sec:intro}
The rapid growth of video content and streaming services has led to increasing demand for accurate video quality assessment (VQA) methods. Video quality metrics are essential for optimizing compression algorithms \cite{10.1145/3571727}, streaming protocols \cite{7524566}, and rendering techniques \cite{singh2023power}, ensuring that viewers experience the highest possible quality under varying network conditions and device constraints. Traditional approaches to assessing video quality rely on subjective human evaluations, which, although accurate, are time-consuming, costly, and difficult to scale. As a result, there has been substantial interest in developing objective video quality metrics that can automatically and efficiently predict perceived video quality.

Full-reference objective quality metrics quantify the difference between distorted and reference videos, making them valuable for evaluating compression techniques, transmission protocols, and rendering systems. However, the performance of these metrics depends heavily on the types of distortions they are designed to address. Traditional metrics like PSNR, SSIM \cite{wang2004image}, and VMAF \cite{rassool2017vmaf} perform well for conventional distortions, such as compression artifacts \cite{yeo2013rate} and transmission errors \cite{katsenou2021vmaf}, but often fail with more complex distortions introduced by modern rendering methods \cite{liang2024perceptual, Jindal2022RenderingFH}.

The progress of advanced rendering methods, such as neural supersampling, path tracing, novel-view synthesis and variable rate shading, has introduced new types of artifacts, exhibiting complex spatio-temporal patterns, that are not well addressed by SOTA video quality metrics. On top of that, real-time graphics content often presents unique visual characteristics that differ significantly from natural videos, further complicating quality assessment.

In this paper, we address this problem by introducing a new Computer Graphics Video Quality Dataset (CG-VQD), specifically focusing on distortions from real-time graphics (Section \ref{sec:dataset}). The dataset includes perceptual ratings for 80 video sequences featuring distortions from popular rendering methods such as neural supersampling, novel-view synthesis, path tracing, neural denoising, frame interpolation, and variable rate shading. Examples of these distortions include spatio-temporal aliasing, flicker, ghosting, moire, fireflies, noise, blur, tiling, and hallucinations (neural reconstruction errors).
We also investigate how well the internal activations of 3D convolutional neural networks trained for high-level classification tasks align with human perceptual judgments (Section \ref{sec:experiments}). Our experiments show that 3D CNN activations correlate far better with human ratings than existing full-reference quality metrics. However, their performance is heavily influenced by the choice of network architecture, pre-training task, dataset, and calibration with human ratings. In Section \ref{sec:metric}, we propose a new quality metric based on 3D ResNet, which outperforms current metrics and generates both per-pixel error maps and a global quality score, making it particularly suitable for computer graphics applications.

By focusing on real-time graphics and rendering artifacts, this work offers a new perspective on video quality assessment, extending its applicability to emerging areas such as gaming, streaming, and neural rendering. The dataset and metric we propose aim to bridge the gap between traditional video quality assessment methods and the needs of modern rendering technologies.

%% file: secs/related_work.tex
\section{Related work}
\label{sec:related-work}
\textbf{Video quality datasets.}
Developing reliable video quality datasets is essential for creating and evaluating video quality metrics, especially given the rapid evolution of rendering technologies, display systems, and rising user expectations. In this work, we aim to build a full-reference metric and thus only focus  on datasets that assess perceived quality differences between distorted videos and their corresponding high-quality reference versions. Several such datasets have been collected to address various types of video distortions.  MSU \cite{antsiferova2022video}, LIVE-VIDEO \cite{seshadrinathan2010study}, LIVE-Meta \cite{saha2023study}, MCML 4K \cite{live4k}, IVP \cite{zhang2011ivp}, and AVT \cite{rao2019avt} focus on video compression artifacts. LIVE-LIVESTREAM \cite{shang2021study} and EPFL-PoliMI \cite{de2009epfl}  examine network-induced distortions. Other distortion types that have been studied include flicker (LIVE-Flicker \cite{choi2015motion}), frame-rate variations (\cite{madhusudana2021subjective}, \cite{mackin2018study}), spatio-temporal subsampling (\cite{lee2021subjective}), fast motion (\cite{ebenezer2022subjective}), and mobile (\cite{livemobile}) distortions. Most of these datasets use natural videos, which can behave differently from synthetic content, such as video game footage. Addressing this gap,  Barman et al. \cite{barman2018gamingvideoset, barman2021user} investigated the impact of compression on video game content. 

While these datasets provide a broad range of natural and synthetic video distortions, they do not account for artifacts introduced by recent advancements in real-time graphics.  In Section \ref{sec:dataset}, we introduce a new video quality dataset that includes previously unstudied distortions arising from neural supersampling, path tracing, neural denoising, Gaussian splatting, frame generation, and variable rate shading, offering a more comprehensive evaluation framework for modern rendering technologies.

\textbf{Objective quality metrics.}
Video quality metrics are designed to automatically predict the perceptual quality of videos, offering a scalable alternative to subjective human evaluations. Note that, although image quality metrics are not designed for video quality assessment, they can still perform effectively by averaging predictions across individual video frames. Full-reference quality metrics map distorted and reference videos to a perceptually uniform space, where differences between them correlate with perceived quality differences. These mapping functions can be broadly categorized into four main approaches.

\textit{Psychophysical metrics} explicitly model aspects of low-level human vision, such as contrast sensitivity and masking \cite{mantiuk2021fovvideovdp, mantiuk2024colorvideovdp}. These metrics excel in generalizing to unseen display conditions and distortions, but they struggle to model higher-order visual aspects like motion and saliency. 

Another approach is to \textit{hand-craft visual features} such as the 
difference of contrast used in SSIM \cite{wang2004image} or phase congruency and gradient magnitude in FSIM \cite{zhang2011ivp}. These features can also be combined using support vector regression. Examples of such \textit{ensemble metrics} include VMAF \cite{rassool2017vmaf}, FUNQUE \cite{venkataramanan2022funque}, and 3C-FUNQUE \cite{venkataramanan2023one}. However, designing such features is time consuming and rely on carefully-constructed heuristics, making it a challenging task.

In our work, we take the \textit{data-driven approach} that offers an alternative to other classes by learning features directly from data. Data-driven methods either take pre-trained CNNs to calculate quality difference \cite{zhang2018unreasonable} or train CNNs from scratch on image/video quality datasets \cite{prashnani2018pieapp}. While these ideas have been extensively explored for image quality \cite{wang2021survey}, they remain underexplored for videos. Some methods attempt to incorporate temporal modeling by adding recurrent structures on top of spatial 2D-CNN features \cite{kim2018deep}, however, this has been shown to be ineffective \cite{fang2023study}. C3DVQA \cite{xu2020c3dvqa} uses 3D-CNNs for quality estimation, where convolutional kernel weights are learned based on video quality datasets. However, such approaches often risk over-fitting due to the limited availability of human-annotated quality ratings \cite{mikhailiuk2021consolidated}. To address this,  DeepVQUE \cite{dendi2019full} instead uses a pre-trained C3D network \cite{tran2015learning} to calculate perceived distance thus reducing the number of trainable parameters. Our approach follows a similar strategy, but we observe that the choice of 3D-CNN architecture has a significant impact on performance. Specifically, we find that 3D-ResNet outperforms the C3D network, as detailed in Section \ref{sec:metric}.

Recently, Transformer-based models have shown promising results in no-reference VQA (NR-VQA) \cite{wu2022discovqatemporaldistortioncontenttransformers}. These models often outperform CNN-based methods qualitatively, leveraging self-attention mechanisms that capture pairwise correlations across spatial and temporal dimensions. However, this advantage comes at the cost of quadratic computational complexity, increased inference time, and larger memory footprint. We leave the exploration of Transformer-based models for full-reference VQA (FR-VQA) to future work.

%% file: secs/dataset.tex
\section{Computer graphics video quality dataset (CG-VQD)}
\label{sec:dataset}

The subjective datasets used to calibrate video quality metrics must accurately reflect the types of artifacts that the metrics are intended to evaluate. This presents a challenge when little prior work exists, as is the case with advanced rendering methods. While traditional artifacts, such as those introduced by video compression, have been extensively studied, distortions from modern rendering techniques remain largely unexplored. To address this gap, we curated a dataset that captures the most relevant spatio-temporal distortions and conducted a subjective study, gathering difference mean opinion scores (DMOS\footnote{We use the ITU-T Rec. P.910 definition DMOS($v$) = MOS($v$) - MOS($v_{\textrm{ref}}$) + 100}) to create the Computer Graphics Video Quality Dataset (CG-VQD). This dataset is not intended for training deep models from scratch, but rather to provide a foundation for evaluating and calibrating video quality metrics tailored to modern rendering techniques and game engines. Additionally, it aims to help identify which rendering methods warrant deeper investigation in quality assessments.

\subsection{Source sequences}

We selected 15 open-source 3D scenes that are representative of video games and challenging for modern rendering methods (Figure \ref{fig:exp-scenes}). This dataset includes popular industry samples and demos from Amazon (\textit{Bistro} \cite{ORCAAmazonBistro}), Intel (\textit{Jungle1/2} \cite{siqueira_jungle_ruins_2025}), NVIDIA (\textit{ZeroDay} \cite{ZeroDay}), Unity (\textit{Market} \cite{unity_adventure_sample_game_2024}), and Epic (Unreal Engine demos: \textit{Park} \cite{silverTm_city_park_2024}, \textit{Dock} \cite{polypixel_medieval_docks_2020}, \textit{Clothes} \cite{zhdanov_clothing_shoe_stores_2024}, \textit{Infiltrator} \cite{epic_infiltrator_demo_2024}, \textit{Street} \cite{epic_matrix_awakens_2021}, \textit{Meerkat} \cite{epic_meerkat_demo_2021}, and \textit{Mushroom} \cite{sychov_deep_elder_caves_2024}). Additionally, we created custom scenes (\textit{Coils}, \textit{Bridge}, and \textit{House}) featuring thin geometry, bright lights, and complex textures to induce artifacts like moiré, ghosting, and flicker (more details in Appendix A1.1). The camera paths for each scene were designed manually to generate a large diversity of content.

\subsection{Distortions}

Each scene was rendered into a 3 seconds long video of resolution 1024$\times$1024 pixels at 30 fps (60 fps for \textit{Meerkat} and \textit{Infiltrator}) using one of the following six rendering techniques. We selected widely adopted rendering methods with relatively less explored quality-performance trade-offs. For each technique, parameters were adjusted to produce distortions characteristic of that method across 4–5 levels of severity. Examples of these distortions can be found in Figure \ref{fig:distortions_all}.

\textbf{Path tracing.} Path tracing algorithms are used for photorealistic visuals in graphics applications and popular game engines. To generate path tracing artifacts (noise, fireflies and shimmering), we use Real-time Path Tracing Research Framework \cite{RTPT2023} on \textit{Jungle1} and \textit{Jungle2} scenes at 4 different sampling rates (4, 16, 64, and 256 samples-per-pixel (\textit{spp})).

\textbf{Neural denoising.} Practical use of real-time path tracing methods is only possible due to good denoising methods. We use a recently proposed method called neural partitioning pyramids (NPPD) \cite{balint2023neural} to generate denoising artifacts for different input spp counts (2,4, and 8 spp) on \textit{ZeroDay} and \textit{Bistro} scenes.

\textbf{Neural supersampling.} Real-time upscaling of rendered frames is a popular technique that has several proprietary (NVIDIA DLSS \cite{DLSSNVIDIA2024}, AMD FSR \cite{FSRAMD2024}, Intel XeSS \cite{XeSSIntel2013}) and open source (Real-ESRGAN \cite{wang2021realesrgan}, FSRCNN-T \cite{FSRCNNT}) implementations. To study the artifacts caused by this, we use Intel XeSS SDK v1.3 \cite{XeSSIntel2013}. We simulate authentic distortions by first rendering the scenes at 4 reduced resolutions ($1/3$, $1/2.3$, $1/2$, and $1/1.5$) and then upscaling them to the native resolution using XeSS. Additionally, to study higher levels of distortions, we add 3 conditions where we artificially boost artifacts, by applying thresholding and scaling operations on the output of the warp module \cite{intel_xess2_whitepaper} that leads to ghosting and flicker artifacts. \textit{Coils}, \textit{Mushroom}, \textit{Street}, and \textit{Park} scenes were used for scaling distortions, and \textit{Bridge}, \textit{Park}, and \textit{Street} scenes were used for synthetic ghosting and flicker distortions.


\textbf{Gaussian splatting.} Scene representation using 3D Gaussians \cite{kerbl20233d} is a rapidly evolving area in image-based rendering. Owing to its high representational fidelity and real-time rendering capabilities, this approach is increasingly replacing neural implicit representations and gaining traction in traditional graphics pipelines. To investigate artifacts characteristic of this method, we first rendered the \textit{Dock} and \textit{Clothes} scenes as videos, then used these videos to generate 3D Gaussian representations with varying numbers of Gaussians: 750K, 375K, 175K, and 87K for \textit{Clothes}; and 1000K, 500K, 250K, and 125K for \textit{Dock}. We use the publicly available implementation for this process \cite{gaussian_splatting_repo}.


\textbf{Frame interpolation.} Frame interpolation methods are used to improve the frame rate of a given video and generate smoother-looking animations. We generate 60 fps videos of \textit{Meerkat} and \textit{Infiltrator} scenes from their 12, 15, 20, and 30 fps videos using NVIDIA Frame Rate Up Conversion (NVFRUC) method \cite{NVFRUC}.

\textbf{Adaptive variable rate shading.} Variable-rate shading is a modern graphics feature that enables fine control of the visual quality, in which each 16×16 image tile can be rendered with a different shading rate. We use a recent Adaptive Local Shading and Refresh Rate method (\textit{ALSaRR}) \cite{jindal2021perceptual} that adaptively distributes a given shading budget to maximize video quality. \textit{House} and \textit{Market} scenes were used for this method and shading budgets of 50\%, 25\%, 12.5\%, and 6.25\% pixels were chosen.

Following prior practice \cite{vcadik2012new, wolski19emap}, we used different scenes for different distortions to give us a total of 80 videos (including references). A full-factorial experimental design—pairing all scenes with all rendering methods—was not feasible, as it would require each participant to provide eight times more ratings. The choice of scene-distortion pairs was made to ensure a variety of distortions with sub-threshold, near-threshold, and suprathreshold magnitudes (as seen in Figure \ref{fig:exp_results}). The reference videos were generated using 16K spp for path tracing methods and 16$\times$ supersampling anti-aliasing (SSAA) for all other methods.

\subsection{Participants and Procedure}

20 participants aged 25-57, 18 male and 2 female, with normal or corrected-to-normal vision participated in the experiment. The experiment was authorized by an internal institutional review board and participants received token compensation. It was conducted remotely on the participants' device. Each participant reviewed a briefing and signed a written consent form before starting the experiment. All participants reported prior experience in video quality evaluation but were naïve to the content of this user study.

\begin{figure}[] 
\centering
\includegraphics[width=0.95\linewidth]{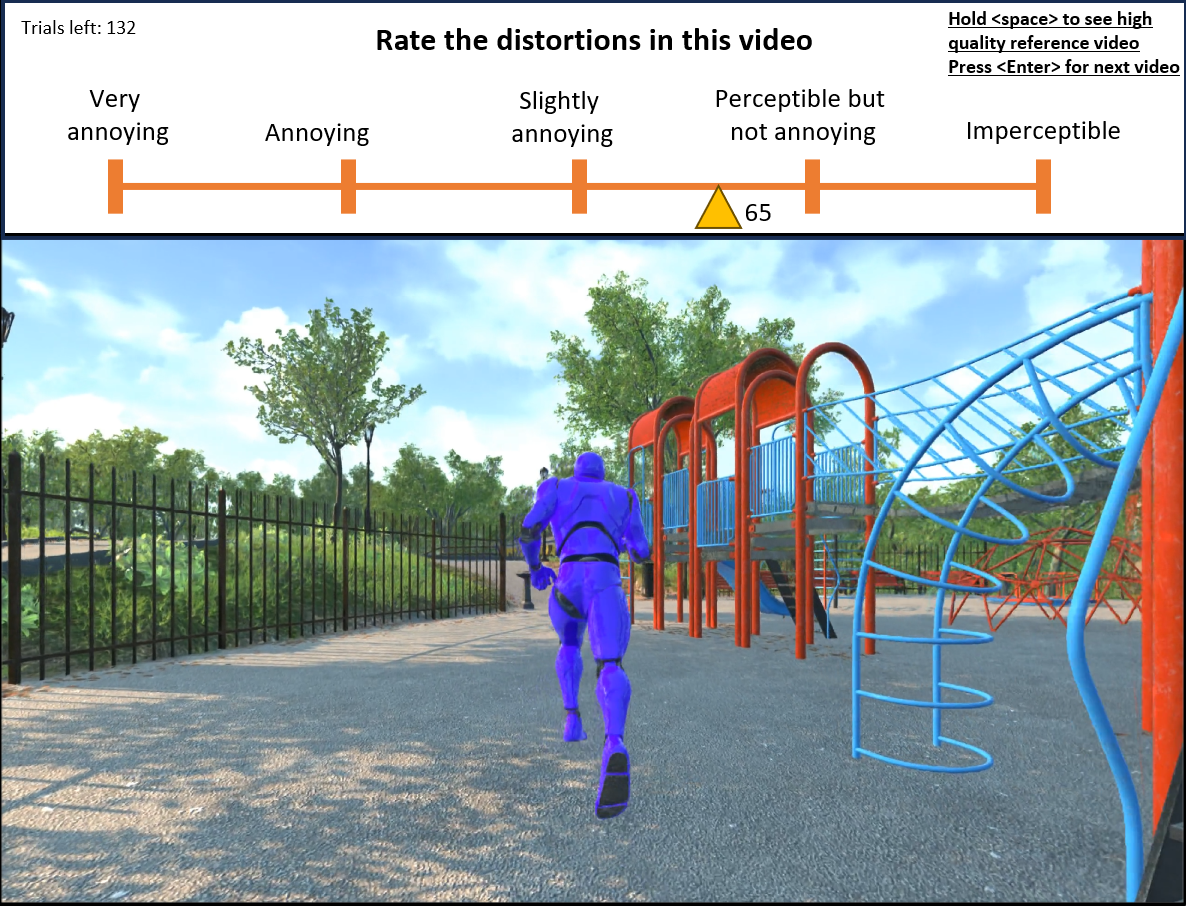}
\caption{User interface used in the experiment.}
\label{fig:userstudy_UI}   
\end{figure}

We developed the user study interface using Pygame and GLSL, which provided precise control over frame rates and resolution for flipping between reference and distorted videos. A screenshot of the software is shown in Figure \ref{fig:userstudy_UI}.
For each trial, participants were asked to ``\textit{rate the quality of distortions in the shown video w.r.t. its corresponding reference video}" on a continuous rating scale of 0 -- 100. The distortions were defined as anything that differs from reference. The scale was marked with labels ``Very annoying'', ``Annoying'', ``Slightly annoying'',  ``Perceptible but not annoying'', and ``Imperceptible'' to facilitate the subjects in making decisions. The participants could press the space bar to switch between reference and distorted videos. A short blank of 0.5s was shown before switching so that the participants could not use temporal flicker to detect the presence of artifacts. All videos were played in a loop.

Each participant performed 240 ratings (80 videos repeated 3 times each, including reference) giving us 4,800 ratings across all participants. The order in which the videos were presented was randomized. To minimize display-related artifacts, all participants completed the experiment on  desktops with screens of at least Full HD resolution and 60 Hz refresh rate in a typical office environment. Participants were also instructed to disable any color correction or resolution scaling software. They were free to take breaks as needed and the experiment took approximately 60-90 minutes to complete. Due to the remote nature of the study, we did not have control over the viewing conditions of the participants and could not screen them for vision deficiencies. A comprehensive reliability analysis is presented in Appendix A1.1 to verify the consistency and robustness of participant ratings, ensuring their suitability for subsequent quality analysis and model development.

\subsection{Results}

The user ratings from the experiments were converted to difference mean opinion scores for each distorted video using maximum likelihood estimation \cite{li2020simple} as implemented in SUREAL toolbox \cite{surreal}. The results for each video are shown in Figure \ref{fig:exp_results}. Additional details on subject bias and inconsistency, outlier rejection, and content ambiguity are provided in Appendix A1.1.

As seen in Figure \ref{fig:exp_results}, perceived quality exhibits a non-linear relationship with the distortion factor. Our dataset captures a broad range of quality levels, with some distorted videos appearing nearly indistinguishable from their reference counterparts. These conditions are particularly useful for assessing whether a metric can correctly disregard imperceptible distortions. By grading all artifacts on a unified, linear perceptual DMOS scale, we enable direct comparison across different distortions and rendering methods. This reveals interesting interactions, such as the varying DMOS ranges observed across different rendering techniques, and highlights the content-dependency of rendering quality. Notably, the results for neural supersampling and Gaussian splatting (Coils, Park, and Dock) are not always strictly monotonic with increasing distortion levels, suggesting that these methods may introduce distortions in complex, non-linear ways.

\input{figs/experiment}


%% file: figs/experiment.tex
\begin{figure*}
    \centering
    \begin{subfigure}{\textwidth}
        \centering
        \includegraphics[width=0.85\textwidth]{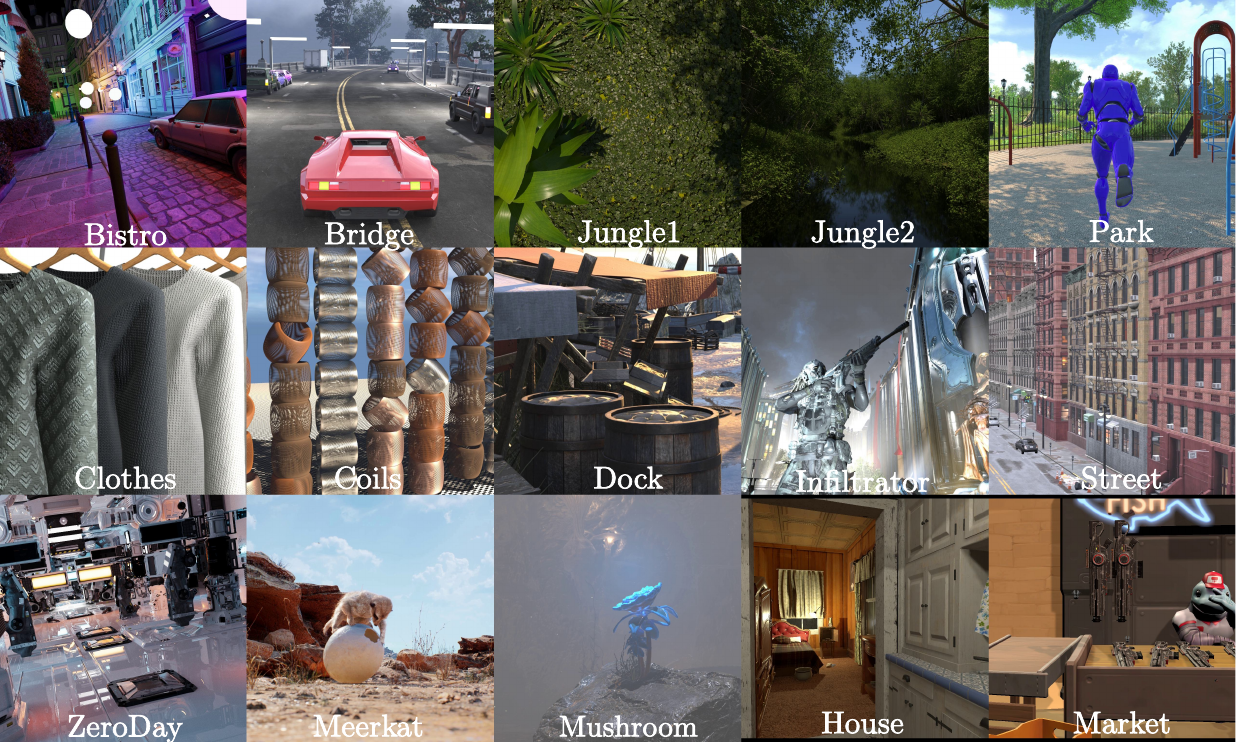}
        \caption{Screenshots of the scenes used in the experiment.}
        \label{fig:exp-scenes}
    \end{subfigure}
        \begin{subfigure}{\textwidth}
        \centering
        \includegraphics[width=\textwidth]{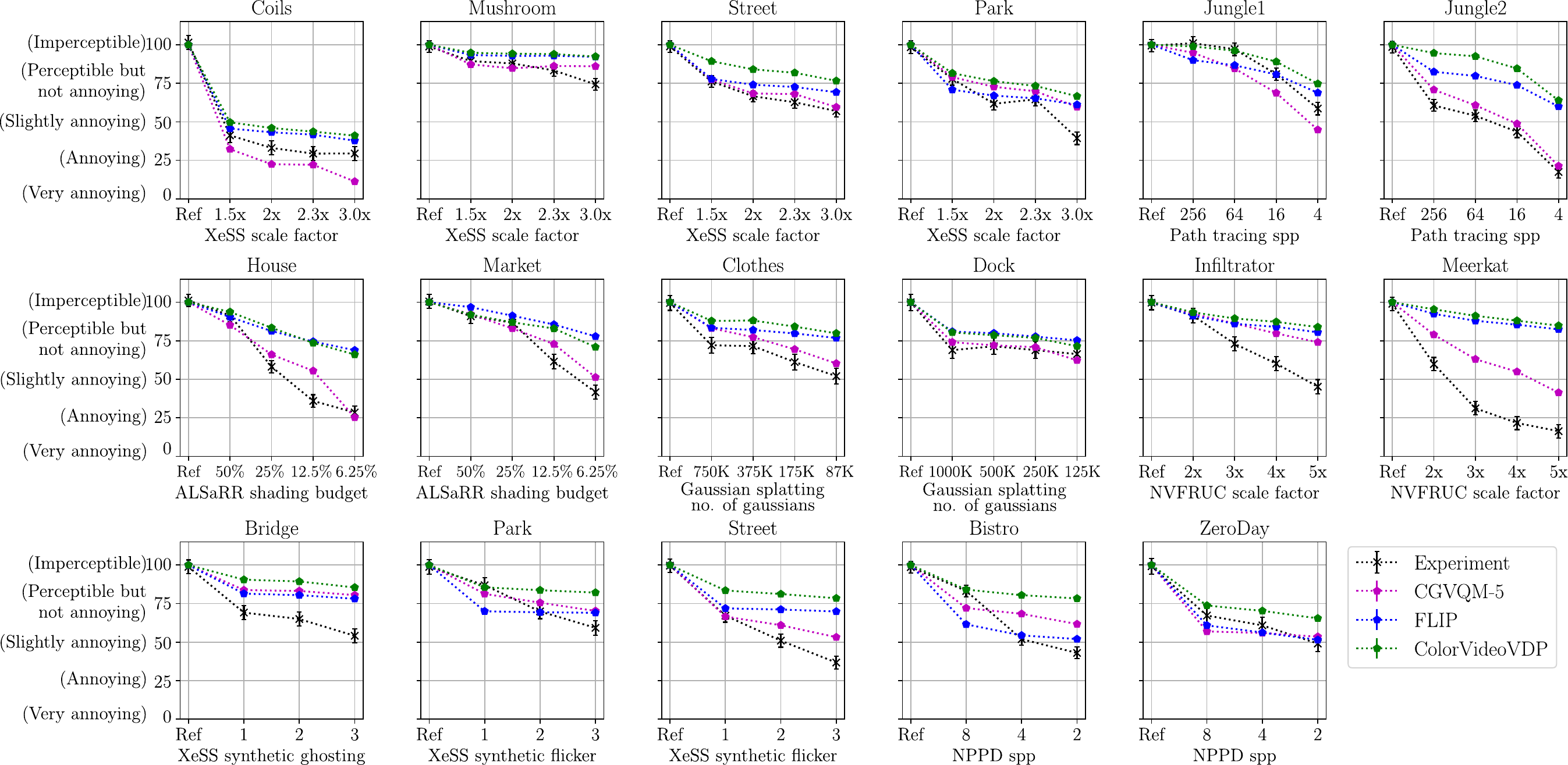}
        \caption{Experiment results and metric predictions. Each sub-plot denotes a different scene and x-axis denotes the rendering method. Error bars denote 95\% confidence interval. Perception of artifacts varies with rendering technique and content.}
        \label{fig:exp_results}
    \end{subfigure}
\caption{Experiment stimuli and results from our computer graphics video quality dataset. Examples of distortions can be found in Figure \ref{fig:distortions_all}.}
\label{fig:experiment}    
\end{figure*}

%% file: secs/3dcnn.tex
\section{3DCNN Feature spaces}

\Rev{Our goal is to develop a video quality metric that closely aligns with human perception of graphics distortions. Perceptual distances are known to be non-uniform in pixel space \cite{wang2004image}, rendering metrics like mean squared error (MSE) inadequate for assessing video quality. Recent work such as \cite{zhang2018unreasonable} has shown that deep neural networks, particularly convolutional neural networks (CNNs) trained for image classification, can extract features that better represent how people judge image quality. Differences measured in the feature space of these networks tend to be more perceptually uniform—that is, they align more closely with how humans perceive visual differences—compared to some traditional image quality metrics.}

\Rev{Inspired by this idea, we ask: can this approach be extended from images to videos? Specifically, can the internal features of 3D CNNs—networks that capture both spatial and temporal patterns—help us assess the visual quality of rendered video content more accurately? In this section, we explore this question by testing several pre-trained 3D CNN architectures and show how their learned feature spaces can be used to build a more perceptually aligned video quality metric.}

\label{sec:metric}
\subsection{Network architectures}

\Rev{To find out if network design and training influences perceptual uniformity of feature space, we tested two widely used 3D CNN architectures:
\begin{itemize}
    \item \textit{3D ResNet-18} \cite{tran2018closer}: A ResNet-based network pretrained on the Kinetics-400 dataset \cite{carreira2017quo}, which includes a wide variety of real-world human actions in videos.
    \item \textit{C3D} \cite{tran2015learning}: A 3D CNN pretrained on Sports-1M \cite{karpathy2014large}, a large video dataset focused on sports classification.
\end{itemize}
We also compared different styles of 3D convolutions \cite{tran2018closer} within the ResNet-18 architecture:
\begin{itemize}
    \item \textit{R3D}: Standard 3D convolutions applied across space and time.
    \item \textit{MC3}: A hybrid design using both 2D (spatial) and 3D (spatiotemporal) filters.
    \item \textit{R(2+1)D}: A factorized approach that applies a 2D spatial convolution followed by a 1D temporal convolution.
\end{itemize}
}

\begin{figure*}[] 
    \centering
    \begin{subfigure}[]{\textwidth}
        \centering
        \includegraphics[width=0.85\textwidth]{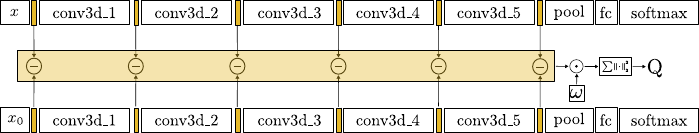}
        \caption{Computing quality from 3D Resnet-18. To compute the quality of distorted video $x_0$, w.r.t. reference video $x$, we first feed $x$ and $x_0$ to a pre-trained 3D Resnet-18 network to compute deep features, extract the output features from each block conv3d\_i, normalize them in the channel dimension, scale each channel by vector $\omega$, take the $l_2$ distance, and average across space and time to get the video quality score. $\omega$ is a free parameter calibrated over video quality datasets. Architectural details of each conv3d block can be found in \cite{tran2018closer}.}
        \label{fig:metric-arch}
    \end{subfigure}\vspace{2mm}
        \begin{subfigure}[]{\textwidth}
        \centering
        \includegraphics[width=\textwidth]{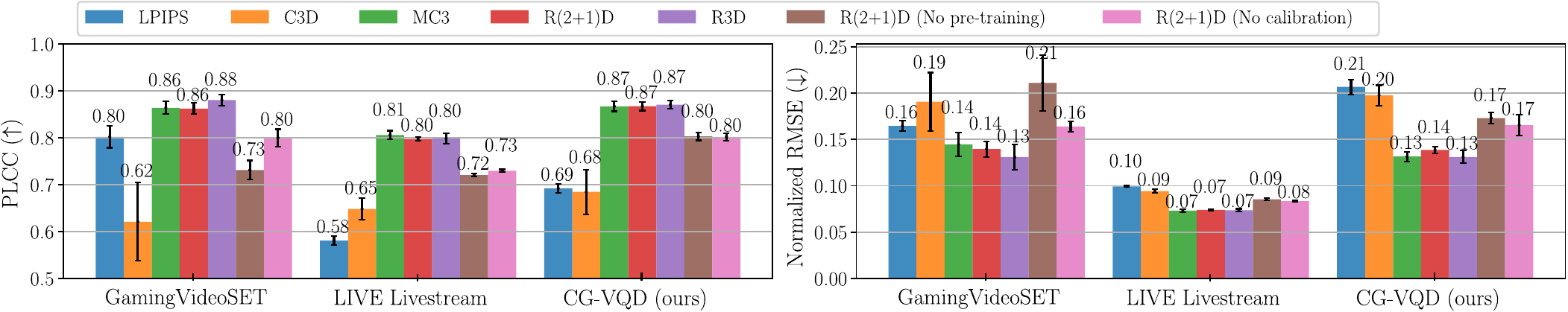}
        \caption{We compare performance of different CNNs on 3 different video quality datasets. 3D CNNs significantly outperform 2D-CNN based metric (LPIPS). Both the 3D-CNN network architecture and its pre-trained weights play an important role in our metric performance.}
        \label{fig:metric-experiments}
    \end{subfigure}\vspace{2mm}
    \begin{subfigure}[]{\textwidth}
        \centering
        \includegraphics[width=\textwidth]{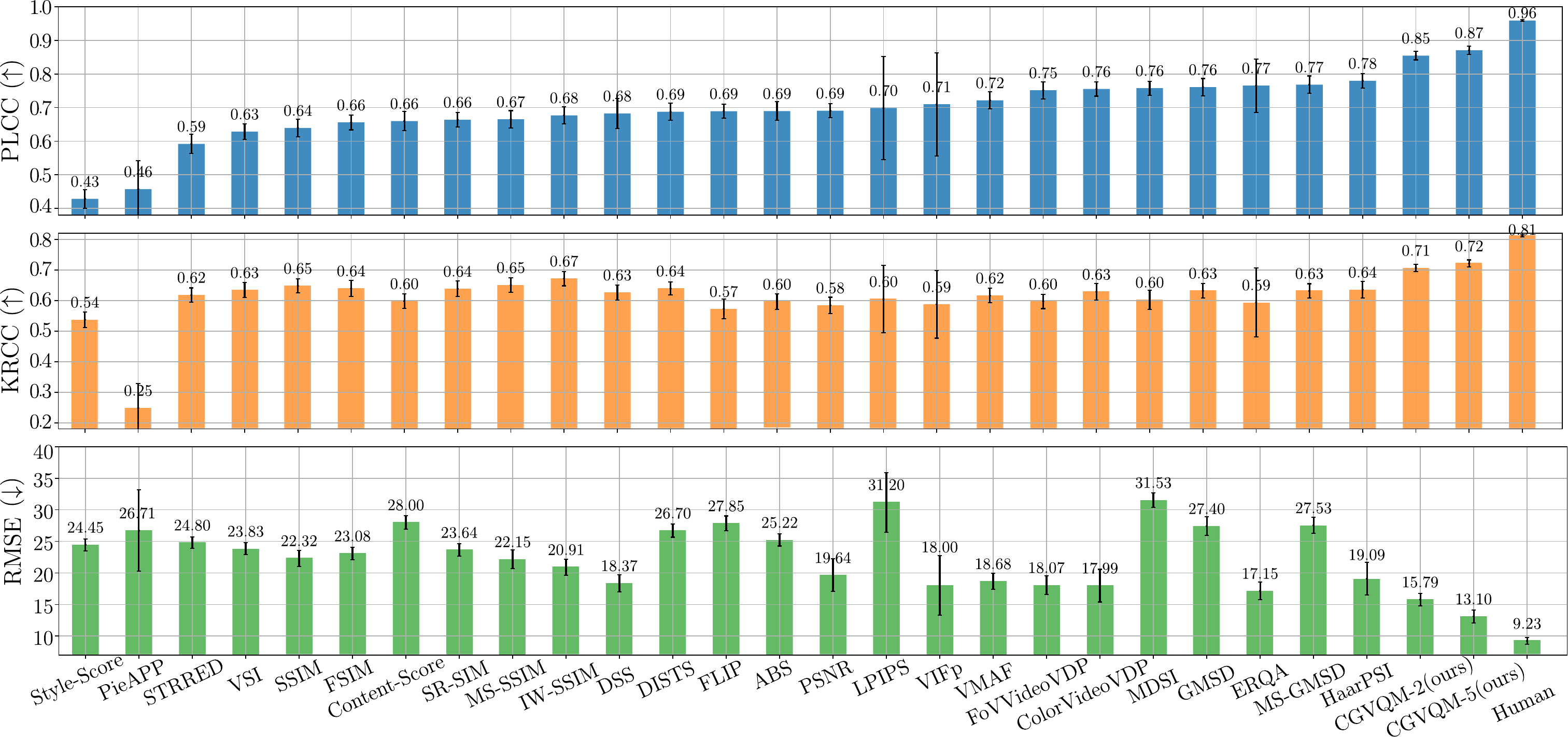}
        \caption{Comparison of quality metrics in terms of Pearson correlation (PLCC), Kendall rank correlation (KRCC), and root mean square error (RMSE) on our CG-VQD dataset. Metrics are sorted based on PLCC value. Detailed reports on other datasets can be found in the Appendix \ref{sec:benchmark}.}
        \label{fig:metric-benchmark}
    \end{subfigure}
    \caption{Details on proposed metric architecture and quantitative comparison with other quality metrics. Error bars were generated via bootstrapping and denote 95\% confidence interval.
}
\label{fig:metric}    
\end{figure*}

\Rev{\textbf{Feature distance to video quality}. Figure \ref{fig:metric-arch} and Equation \ref{eq:quality} illustrate how we obtain the perceived difference between the distorted video $x_0$ and its high-quality reference video $x$ using a 3D-CNN $\mathcal{N}$ to extract features from both videos. Let the spatiotemporal resolution of the videos be $F\times H \times W$, where $F$ is the number of frames, and $H$ and $W$ are the height and width of each frame. We pass both $x$ and $x_0$ through $\mathcal{N}$, and collect intermediate features from $L$ different layers. These features are high-dimensional tensors that capture spatial and temporal patterns at multiple scales. We then unit-normalize each feature map along the channel dimension to reduce scale differences across channels. Let's denote the normalized feature maps for layer $l$ as as $\hat{x}^l$ and $\hat{x}_0^l$, each with dimension $F_l \times H_l \times W_l \times C_l$, where $C_l$ is the number of channels in that layer. Since different channels capture different types of patterns (such as edges, motion, or textures), and some are more perceptually important than others, we apply a channel-wise weight $\omega_l \in \mathbb{R}^{C_l}$ to emphasize or de-emphasize specific channels. We finally compute the squared $l_2$ distance between $\hat{x}^l , \hat{x}_0^l$ and pool it over all dimensions and all layers to get the video quality:
}


\vspace*{-6pt}
\begin{equation}
    q(x,x_0) = \alpha - \sum_{l}\!\frac{1}{F_lH_lW_l}\sum_{F_lH_lW_l}\!
    \left\| \omega_l \odot\left( \hat{x}^l_{F_lH_lW_l} -  \hat{x}^l_{0F_lH_lW_l}\right) \right\|^2_2 ,
    \label{eq:quality}
\end{equation}
\Rev{where $\alpha$ denotes the maximum allowed quality on a video quality scale (i.e. best possible rating), $\odot$ denotes element-wise multiplication and $\omega$ are free parameters tuned on video quality datasets. In addition to predicting a single global quality score for a video, we can also generate error maps that highlight \textit{where} the perceptual differences occur between the reference and distorted videos. This is done by skipping the spatiotemporal summation and interpolating the feature stack at each layer to the resolution of input video}:
\vspace*{-1pt}
\begin{equation}
    e(x,x_0) = \sum_{l}^{}\uparrow \left( \left\| \omega_l \odot\left( \hat{x}^l_{F_lH_lW_l} - \hat{x}^l_{0F_lH_lW_l}\right) \right\| \right) ,
    \label{eq:error-map}
\end{equation}
where, $\uparrow(\cdots)_{FHW}$ denotes trilinear interpolation to $F\times H \times W$ resolution. An example error map is shown in Figure \ref{fig:error-maps}.

For our experiments, we chose the last convolutional layer from each of the 5 blocks of the 3D-ResNet-18 model and the first 5 convolutional layers from the C3D network. The selected layers give a good distribution of features at multiple scales while keeping the total number of features small. This gives us a set of five 4D features ($L$=5) for both networks. We also append the input video to this feature set to ensure an injective feature transformation, a useful mathematical property for perceptual optimizations \cite{ding2020image}. The input video and the 5 features make a set with the following dimensionality: 


\begin{center}
\renewcommand{\arraystretch}{1.5}
\resizebox{0.7\columnwidth}{!}{%
\begin{tabular}{l|l|l}
\hline
      & \multicolumn{1}{c|}{$\hat{x}_{\textrm{ResNet}}$}              & \multicolumn{1}{c}{$\hat{x}_{\textrm{C3D}}$}                   \\ \hline
      & $F\times H\times W\times 3$                                   & $F\times H\times W\times 3$                                    \\
$l_1$ & $F\times \frac{H}{2}\times \frac{W}{2}\times 64$              & $F\times \frac{H}{2}\times \frac{W}{2}\times 64$               \\
$l_2$ & $F\times \frac{H}{2}\times \frac{W}{2}\times 64$              & $F\times \frac{H}{4}\times \frac{W}{4}\times 128$              \\
$l_3$ & $\frac{F}{2}\times \frac{H}{4}\times \frac{W}{4}\times 128$   & $\frac{F}{4}\times \frac{H}{8}\times \frac{W}{8}\times 256$    \\
$l_4$ & $\frac{F}{4}\times \frac{H}{8}\times \frac{W}{8}\times 256$   & $\frac{F}{8}\times \frac{H}{16}\times \frac{W}{16}\times 512$  \\
$l_5$ & $\frac{F}{8}\times \frac{H}{16}\times \frac{W}{16}\times 512$ & $\frac{F}{16}\times \frac{H}{32}\times \frac{W}{32}\times 512$ \\ \hline
\end{tabular}
}
\end{center}
This gives us 1027 free parameters ($\omega\in \mathbb{R}^{1027}$) for ResNet and 1475 free parameters for C3D ($\omega\in \mathbb{R}^{1475}$) , which are learned by maximizing the correlation between human ratings, $h^D_{x_0}$, on a video quality assessment dataset $D$, and metric predictions, $q^D_{x_0}$:
\vspace*{0pt}
\begin{equation}
    \underbrace{\textrm{min}}_\omega \sum_{D}1-\textrm{PLCC}\left( h^D_{x_0}, q^D_{x_0} \right) ,
    \label{eq:data-ftting}
\end{equation}
where PLCC denotes Pearson correlation. Using scale-invariant PLCC as a loss allows us to simultaneously calibrate on multiple datasets with different perceptual scales. We found PLCC loss slightly outperforms the more commonly used procedure of scale normalization followed by MSE loss in our experiments. We use ``calibrate'' to denote the process of optimizing for feature weights $\omega$ and ``training'' to denote learning convolutional kernel weights. Note that the convolutional kernel weights are frozen after pre-training on classification datasets, and only $\omega$ is optimized. This approach reduces the number of free parameters, thereby mitigating the risk of overfitting on small quality datasets. The architecture for the ResNet-18 metric is illustrated in Figure \ref{fig:metric-arch}. We selected three datasets for calibration: GamingVideoSet \cite{barman2018gamingvideoset}, LIVE Livestream \cite{shang2021study}, and CG-VQD (ours). GamingVideoSet contains quality ratings for 90 video game videos with applied H.264/MPEG-AVC distortions. LIVE Livestream consists of 315 natural videos with H.264 compression, aliasing, judder, flicker, frame drops, and interlacing. Together, the three datasets span a wide range of natural and video game content and various neural and traditional spatiotemporal distortions. 
\Rev{We randomly selected 30, 63, and 15 videos for training and 60, 252, and 65 videos for testing from GamingVideoSet, Livestream, and CG-VQD datasets, respectively. To avoid overfitting, there was no overlap in scenes between the training and test sets. This approximate 20:80 split that is stricter-than-conventional 80:20 split was adopted to allocate more data to the test set, thereby improving the stability and reliability of correlation measures, which are known to be sensitive to small sample sizes.}
To manage the high resolution of some videos, which were too large to process with 3D CNNs on a GPU, we divided the videos into smaller patches ($x^p$) of resolution of $30\times512\times512$ pixels. The overall video quality was then defined as the minimum quality score among all patches $\left( \textrm{min}\left\{ q\left( x^p, x_0^p \right) \right\}_p \right)$. 

\Rev{\textit{Feature calibration}. We used the open-source PyTorch implementations of 3D ResNet-18 \cite{r3dPytorch} and C3D \cite{c3d} networks. The network weights were frozen, and all videos were decomposed into pooled CNN features once. We then optimized the feature weight $\omega$ using the Adam optimizer with default parameters and a learning rate of 1e\textsuperscript{-6}. The system was trained for 100{,}000 epochs and the $\omega$ corresponding to the lowest test loss was selected. Both training and test loss curves were observed to plateau asymptotically. No separate validation set was used due to the limited size of the available data. To estimate variability, the entire process—including random train/test splitting and feature calibration—was repeated 10 times, and the resulting error bars are reported in subsequent experiments.}

\subsection{Experiments}
\label{sec:experiments}
Results on our test sets are shown in Figure \ref{fig:metric-experiments}. We
first evaluate how well our metrics correlate with human ratings. We measure the performance of the metric using 4 standard statistical measures (Pearson linear correlation coefficient (PLCC), Spearman's rank correlation coefficient (SRCC), Kendall rank correlation coefficient (KRCC), and root mean squared error (RMSE)). 

\textbf{Effect of network architecture on performance.}
As seen in Fig. \ref{fig:metric-experiments}, 3D-CNNs exhibit strong correlation with human ratings achieving maximum PLCC values of \Rev{0.88, 0.81, and 0.87} on the GamingVideoSET, Livestream, and CG-VQD datasets, respectively. 3D ResNet consistently outperform 2D-CNN metric LPIPS with R3D-ResNet18 achieving \Rev{23\%} higher PLCC than LPIPS on average. This highlights the importance of temporal modeling for video quality assessment. \Rev{Note that the VGG network was used as the LPIPS backbone and its predictions were scaled using logistic regression (Equation~\ref{eq:metric-fit}) prior to correlation computation. LPIPS was not retrained on video quality datasets, as it lacks the capacity to model temporal distortions.}
    
\textbf{Network training.} Network architecture and pre-trained weights seems to play an important role in performance. The ResNet-18 architecture significantly outperforms C3D architecture (C3D performs worse than LPIPS on GamingVideoSET). The type of 3D convolution does not have a significant effect on performance, with MC3, R(2+1)D and R3D convolutions performing similarly. To test the impact of network pre-training, we randomized the R(2+1)D ResNet-18 network weights before calibrating the feature weights (eq. \ref{eq:data-ftting}). The resulting metric shows \Rev{12.4\%} lower PLCC on average than its pre-trained counterpart. Calibrating feature weights $\omega$ is also important and improves PLCC by \Rev{9.4\%} on average compared to uniformly setting $\omega = 1$. Significance testing results can be found in Appendix A1.3. 


\textbf{Comparison with existing metrics.}\label{sec:benchmark}
We compare the performance of our metric built using R3D-ResNet-18 with several state-of-the-art full-reference quality metrics, listed in Figure \ref{fig:metric-benchmark}. We name our metric \textbf{\metricname-5}: \textit{Computer Graphics Video Quality Metric}, where 5 indicates the use of five layers from R3D-ResNet-18. We do not compare with existing 3D-CNN full-reference metrics C3DVQA \cite{xu2020c3dvqa} and DeepVQUE \cite{dendi2019full} as their code is not publicly available. We used PLCC, SRCC, KRCC, and RMSE as evaluation criteria. Before computing
these measures on each dataset, we followed the standard protocol \cite{zheng2024video} of fitting a five-parameter function to allow and
compensate for a smooth nonlinear relationship:
\vspace*{0pt}
\begin{equation}
    \hat{q} = \eta_1 \cdot \left (  0.5 - \frac{1}{1 + e^{\eta_2 \cdot (q - \eta_3)}} \right ) + \eta_4\cdot q + \eta_5 ,
\label{eq:metric-fit}
\end{equation}
where ${\eta_i}^5_{i=1}$ are free parameters and $q$ is quality metric prediction. \Rev{This procedure was applied to all tested metrics. None of the existing metrics were re-trained on tested datasets. All psychophysical metrics were configured according to the viewing conditions of the corresponding experiment, when available; otherwise, the 'standard-4K' configuration used by ColorVideoVDP was applied. For image quality metrics, we used the implementations available in the PIQ library \cite{kastryulin2022piq,piq}.}

Given the inherently noisy nature of quality assessment, we also evaluate inter-participant agreement by randomly splitting participants into two equal groups, computing DMOS values for each group, and measuring the correlation between them. Error bars are estimated using bootstrapping. The resulting ``Human" performance serves as an upper bound for the performance of objective quality metrics.

The results on CG-VQD dataset, shown in Figure \ref{fig:metric-benchmark}, indicate that \metricname-5 is closest to human performance displaying a substantial improvement over the second-best metric, HaarPSI \cite{reisenhofer2018haar}. 2D-CNN-based metrics such as LPIPS (VGG backbone)\cite{zhang2018unreasonable}, DISTS \cite{ding2020image}, and PieAPP \cite{prashnani2018pieapp} perform significantly worse than our 3D-CNN-based \metricname-5. Additionally, \metricname-5 outperforms other video quality metrics — ColorVideoVDP \Rev{v0.4.2} \cite{mantiuk2024colorvideovdp}, FoVVideoVDP \Rev{v1.2.1} \cite{mantiuk2021fovvideovdp}, VMAF \Rev{v2.3.0} \cite{rassool2017vmaf}, ST-RRED \cite{soundararajan2012video}, and ERQA \cite{kirillova2021erqa}. Among hand-crafted feature metrics, gradient-magnitude based GMSD \cite{xue2013gradient} and MS-GMSD \cite{zhang2017gradient} deliver the best results, outperforming SSIM \cite{wang2004image}, FSIM \cite{zhang2011fsim}, MS-SSIM \cite{wang2003multiscale}, IW-SSIM \cite{wang2010information}, and VSI \cite{zhang2014vsi}. We also plot \metricname-5, ColorVideoVDP, and FLIP predictions alongside experiment results in Figure \ref{fig:exp_results}. \metricname-5 closely aligns with human ratings, while ColorVideoVDP and FLIP tend to overestimate quality in high-distortion scenarios. Additional performance indices for all three datasets are provided in Appendix Figure A6.

\Rev{\textit{Generalization ability.} To evaluate how well \metricname-5 generalizes to unseen data and distortion types, we tested it on six additional video quality datasets:
\begin{itemize}
    \item LIVE Meta \cite{saha2023study}: A dataset containing 600 video sequences of mobile games, derived from 30 reference videos compressed using various methods.
    \item CGVDS \cite{zadtootaghaj2020quality}: Includes 360 cloud gaming videos compressed using the H.264 codec. As this dataset does not include original uncompressed videos, we used the highest bitrate video for each scene as a reference and its MOS value to compute DMOS scores.
    \item LIVE Flicker \cite{choi2015motion}: Contains 72 test videos created from 6 HD reference sequences, with flickering artifacts simulated by varying the quantization level on a frame-by-frame basis.
    \item NVS \cite{nvs}: Consists of 88 videos generated using 7 different NeRF-based view synthesis methods across 16 unique scenes.
    \item AVT-VQDB-UHD-1 \cite{avtvqdb}: Includes 120 videos encoded using H.264, HEVC, and VP9 codecs, with resolutions ranging from 360p to 2160p and frame rates between 15 fps and 60 fps.
    \item BVI-HD \cite{zhang2018bvi}: Comprises 384 distorted videos derived from 32 reference sequences, featuring 12 types of compression distortions from both standard HEVC and HEVC with synthesis mode (HEVC-SYNTH).
\end{itemize}
A summary of results is presented in Table \ref{tab:cross-val}, with full details available in Appendix Table A1. While some existing metrics perform well on specific datasets, such as absolute error (ABS) on GamingVideoDataset or VMAF on video compression datasets, \metricname-5  consistently achieves high correlation with human subjective ratings across all datasets. Since none of these datasets overlap with the training data used for \metricname-5, its strong performance demonstrates its ability to generalize to a wide range of content and distortion types. Additionally, we observe that video quality metrics generally outperform image-based quality metrics on these benchmarks, underscoring the importance of modeling temporal artifacts in video quality assessment.
}
\input{tables/benchmark_half}

\begin{figure}[]
    \centering
    \includegraphics[width=0.9\linewidth]{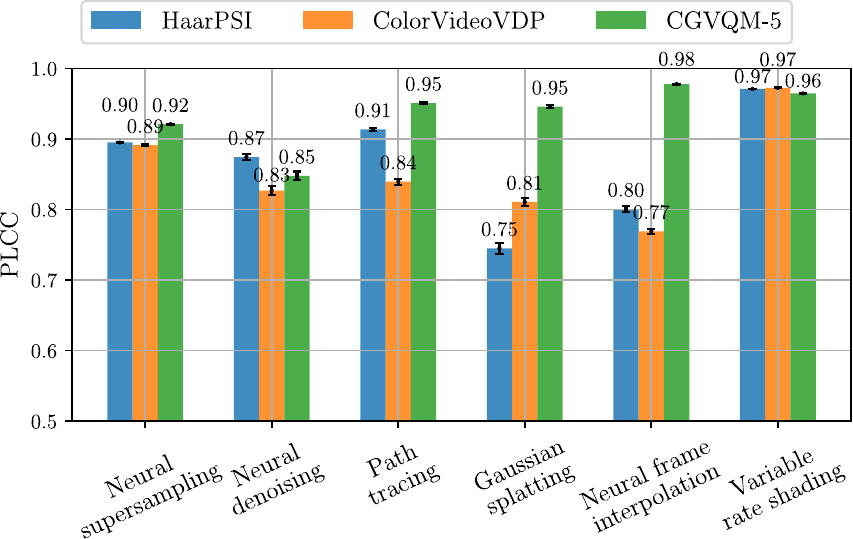}
    \caption{Metric performance for different rendering methods.}
    \label{fig:plcc-per-dist}
\end{figure}

\textbf{Performance by distortion type.} We evaluate the effectiveness of objective quality metrics across different distortion types introduced by various rendering techniques. Figure \ref{fig:plcc-per-dist} shows the PLCC values for a subset of metrics evaluated on each of the six rendering methods used in our study. Traditional metrics such as HaarPSI and ColorVideoVDP perform best on variable rate shading artifacts but exhibit significantly lower performance on Gaussian splatting and neural frame interpolation—highlighting the need for further investigation in these areas. In contrast, \metricname-5 outperforms existing metrics across all distortion types except for neural denoising. Detailed scatter plots for each metric are provided in Appendix A1.2.

\begin{figure}[]
    \centering
    \includegraphics[width=0.665\linewidth]{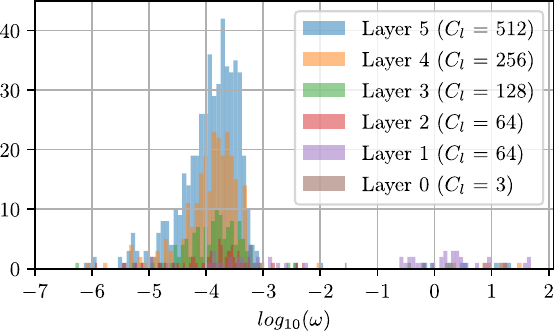}
    \includegraphics[width=0.75\linewidth]{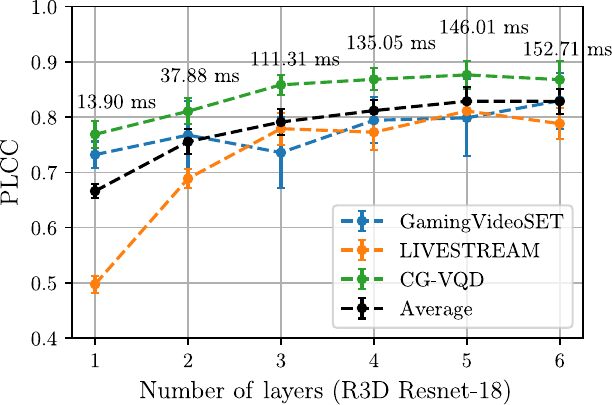}
    \caption{Feature weight ($\omega$) distribution of R3D ResNet-18 quality metric (top) and metric performance vs. number of layers ($L$). Computational performance was calculated on 32$\times$512$\times$512 patches on NVIDIA A100 GPU.}
    \label{fig:metric-weights}
\end{figure}

\textbf{Ablation studies.} We test the contribution of features from each selected layer towards metric's performance. Figure \ref{fig:metric-weights} plots the distribution of feature weights ($\omega$) across layers and how the R3D-ResNet-18 metric's performance changes with increasing number of layers. As seen in Figure \ref{fig:metric-weights}, the weights are similarly distributed across layers, however, 5\% of the total features contribute most to the final quality value. Increasing the number of layers also shows diminishing returns on average PLCC.
This presents an opportunity to reduce the feature set and optimize the computational performance of our metric. 

We define a lighter version of \metricname-5, called \textbf{\metricname-2}, that only uses features from the output layer of first 2 blocks of R3D-ResNet-18 (conv3d\_1 and conv3d\_2 in Figure \ref{fig:metric-arch}), reducing the feature set by 87\% and making it 27\% faster than \metricname-5. Using only early layers also better localizes the error as the spatio-temporal receptive fields are smaller. Fast performance, accurate error maps, and high correlation with human ratings on CG-VQD  makes it especially suitable for computer graphics applications.

\Rev{\textbf{Error localization.} In some applications, it is important not only to obtain a single overall distortion score but also to understand how distortions are spatially distributed across an image or video. The visibility of a distortion depends on both its strength (contrast detection) and the surrounding signal (contrast masking). Metrics such as SSIM, HaarPSI, and visible difference predictors (VDPs), which explicitly model aspects of low-level human vision, are well-suited for predicting the visibility of distortions.}

\Rev{In contrast, neural networks do not provide a direct mechanism to model human perception. However, recent studies suggest that when trained on natural images, these networks tend to mimic certain low-level characteristics of the human visual system \cite{tariq2020deep,hammou2025image,cai2025computer}. In this experiment, we evaluate the error localization capability of \metricname.}

\Rev{Unlike metrics that define quality as difference of global statistics, such as DISTS \cite{ding2020image} or VMAF \cite{rassool2017vmaf}), and are therefore unable to localize errors, we define video quality as the mean of feature differences (Equation \ref{eq:error-map}). This formulation enables us to capture each pixel’s relative contribution to overall quality in the form of an error map.}

\begin{figure*}[]
    \centering    
    \includegraphics[width=\linewidth]{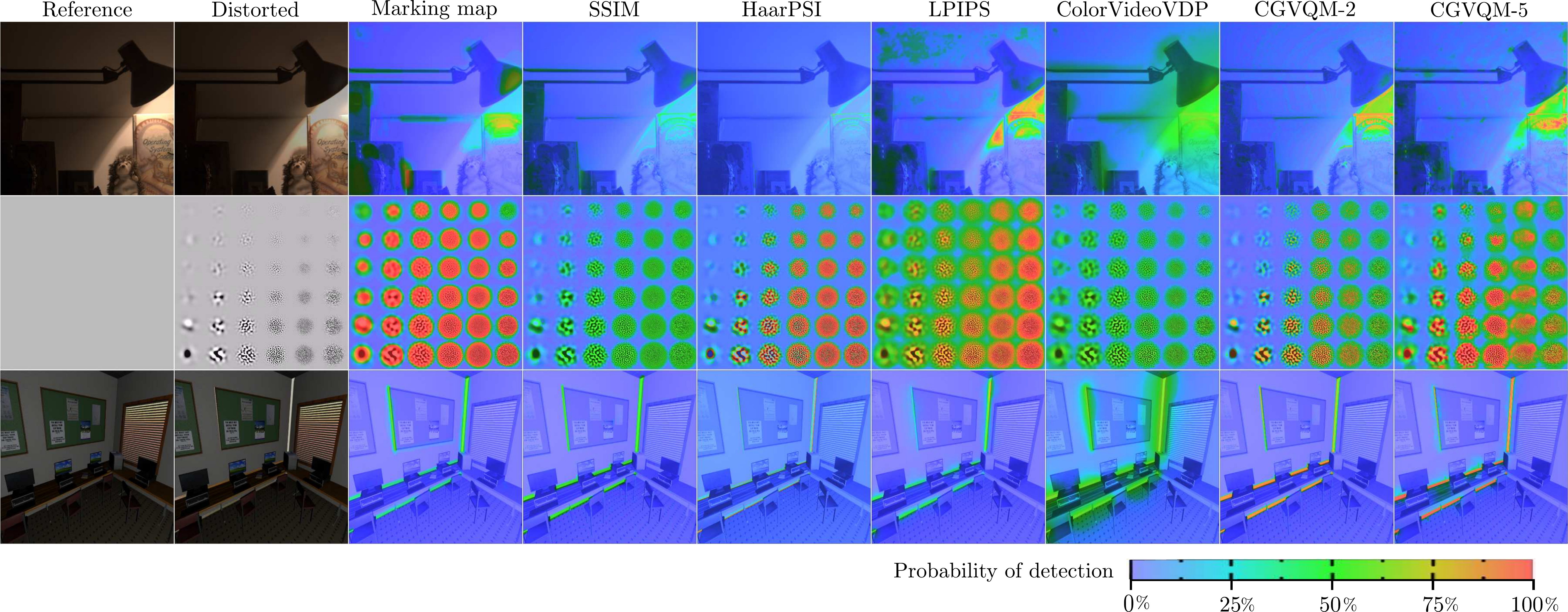}
    \vspace{-12pt}
    \setlength{\belowcaptionskip}{-13pt}
    \caption{\Rev{Reference/distorted images, observers’ markings and metric predictions for selected examples from the LocVis dataset. Metric predictions must be viewed in color.}}
    \label{fig:locvis-error-maps}
\end{figure*}

\Rev{We assess  the accuracy of error maps using LocVis dataset \cite{wolski19emap}, which contains 296 images featuring a variety of natural and synthetic scenes with localized computer graphics artifacts. Each image was annotated by 16 participants to generate per-pixel probability-of-detection maps. To compare the error maps generated by various quality metrics against these human annotations, we follow the evaluation protocol proposed by \cite{vcadik2012new}, which benchmarks each metric's ability
to identify erroneousness pixels as a binary classification task. Two statistical measures are used to assess performance: the area under the receiver operating characteristic curve (AUC-ROC) and the Matthews correlation coefficient (MCC). Prior to evaluation, each metric's predictions were fitted to the full dataset using a simple gain-gamma model. Quantitative results are presented in Table \ref{tab:locvis-emaps}, and qualitative results in Figure \ref{fig:locvis-error-maps}. The results indicate that CNN-based metrics such as LPIPS and \metricname \xspace can accurately localize and scale visual artifacts, performing on par with or better than perceptually-motivated metrics.}

\begin{table}[hbt!]
\centering
\caption{\Rev{Comparison of error maps generated by quality metrics on the LocVis dataset \cite{wolski19emap}. All results assume that regions marked by 75\% or more observers are classified as visible errors, while all other pixels are considered error-free. We report the maximum MCC value for each metric obtained by sweeping classification thresholds in the [0, 1] range. VDP metrics were configured to the viewing conditions of the original experiment.}}
\label{tab:locvis-emaps}
\resizebox{0.56\columnwidth}{!}{%
\begin{tabular}{lcc}
\hline
Metric & AUC-ROC ($\uparrow$)       & MCC ($\uparrow$)           \\ \hline
ABS                        & 0.86          & 0.27          \\
SSIM                       & 0.93          & 0.39          \\
HaarPSI                    & 0.92          & 0.38          \\
FoVVideoVDP                & 0.89          & 0.26          \\
ColorVideoVDP              & 0.92          & 0.24          \\
LPIPS                      & 0.95          & 0.48          \\
CGVQM-2                    & 0.93          & 0.4           \\
CGVQM-5                    & \textbf{0.96} & \textbf{0.49} \\ \hline
\end{tabular}%
}
\end{table}

\begin{figure}[]
    \centering
    \includegraphics[width=\linewidth]{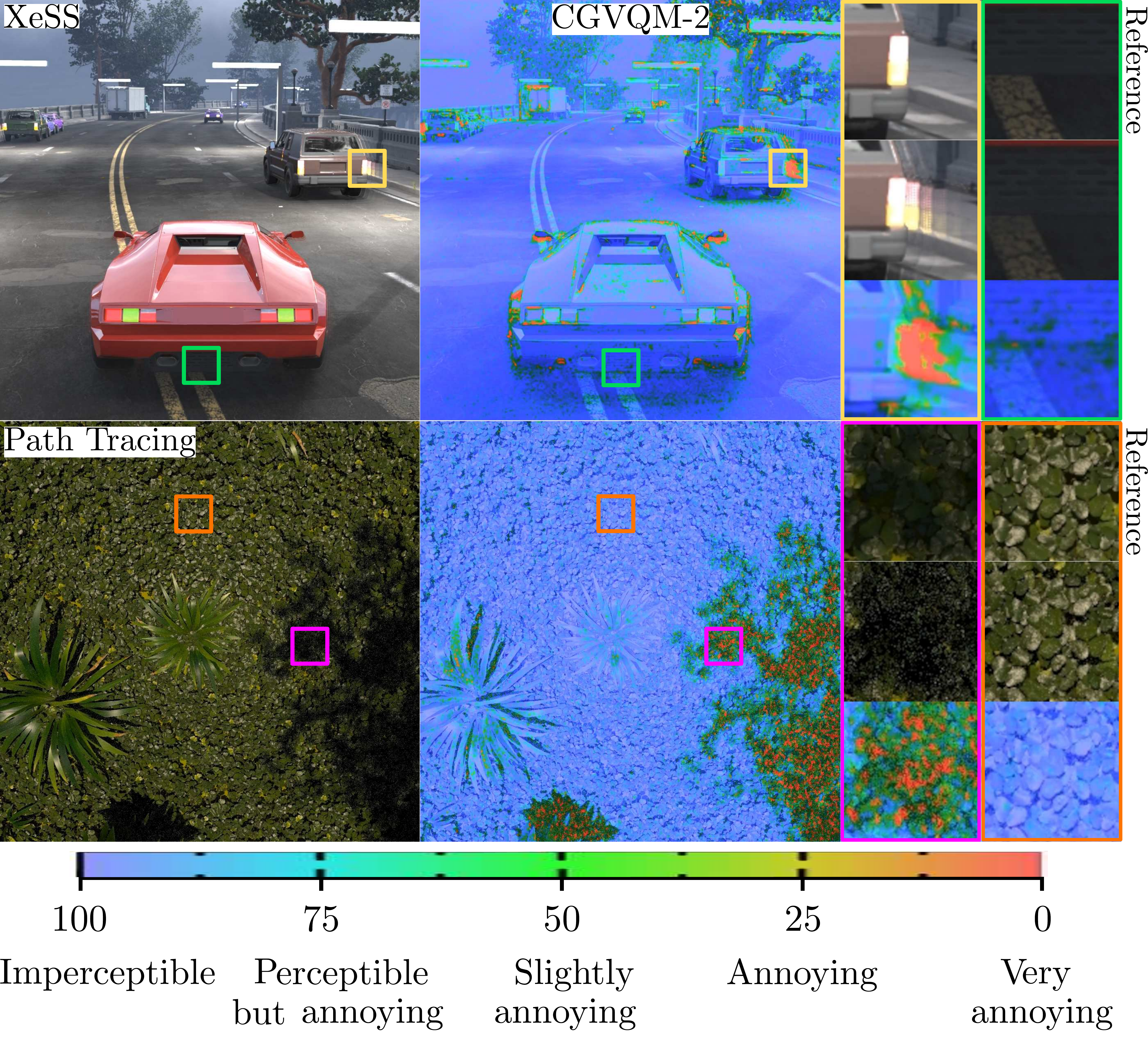}
    \vspace{-12pt}
    \setlength{\belowcaptionskip}{-10pt}
    \caption{Example error maps generated by \metricname-2 for the Bridge (3×) and Jungle2 (4spp) scenes. \metricname-2  effectively highlights ghosting and noise artifacts as perceptually significant in bright  and diffuse regions (car light and tree shadows), respectively, while appropriately down-weighting similar artifacts in visually masked areas (car shadow and leaf texture) where they are less noticeable.}
    \label{fig:error-maps}
\end{figure}

\Rev{We also include examples of error maps generated by \metricname \xspace on our CG-VQD dataset in Figure \ref{fig:error-maps}. Feature weights ($\omega$) for \metricname-2 were linearly scaled to align metric predictions with the perceptual scale used in our experiments, which ranges from ``Very Annoying" to ``Imperceptible". Additional examples are provided in Appendix Figure A4.}

\section{Limitations and future work}

\begin{figure}[]
    \centering
    \includegraphics[width=0.48\linewidth]{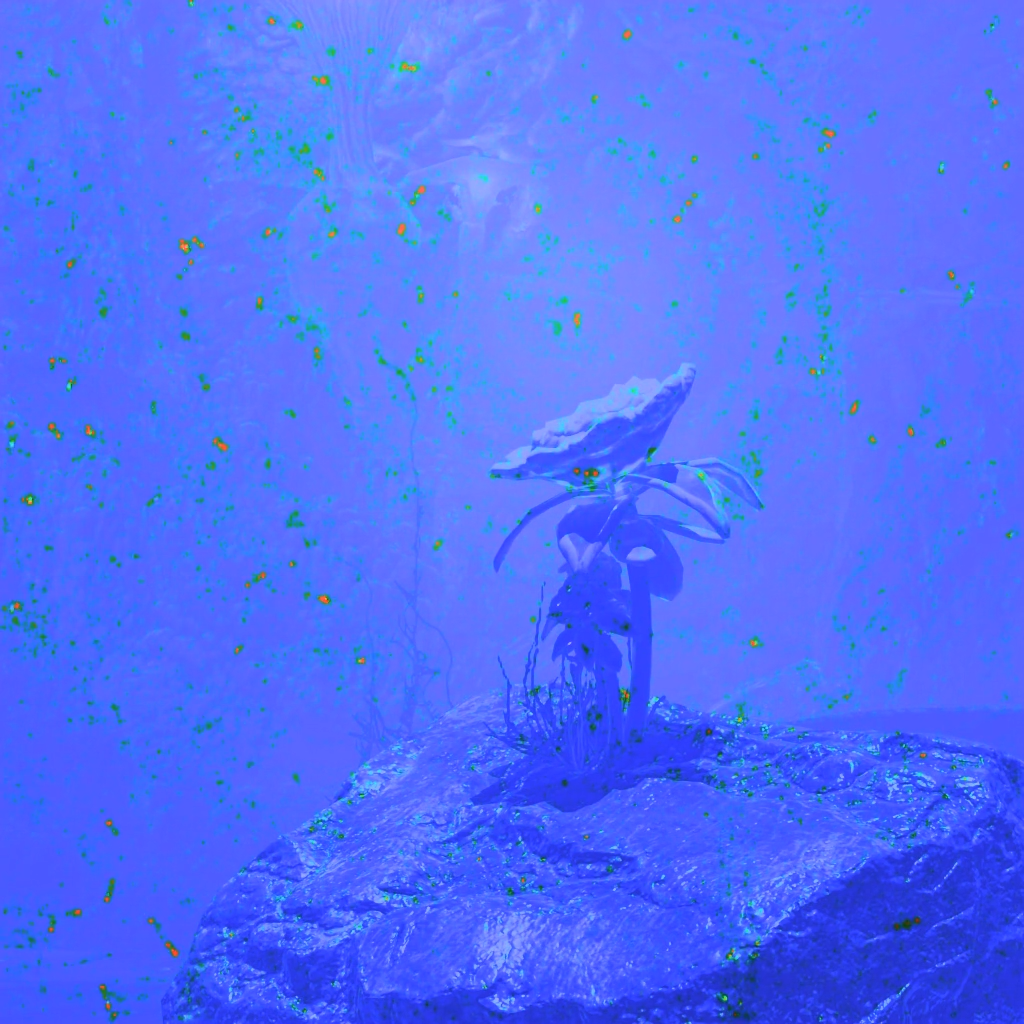}
    \includegraphics[width=0.48\linewidth]{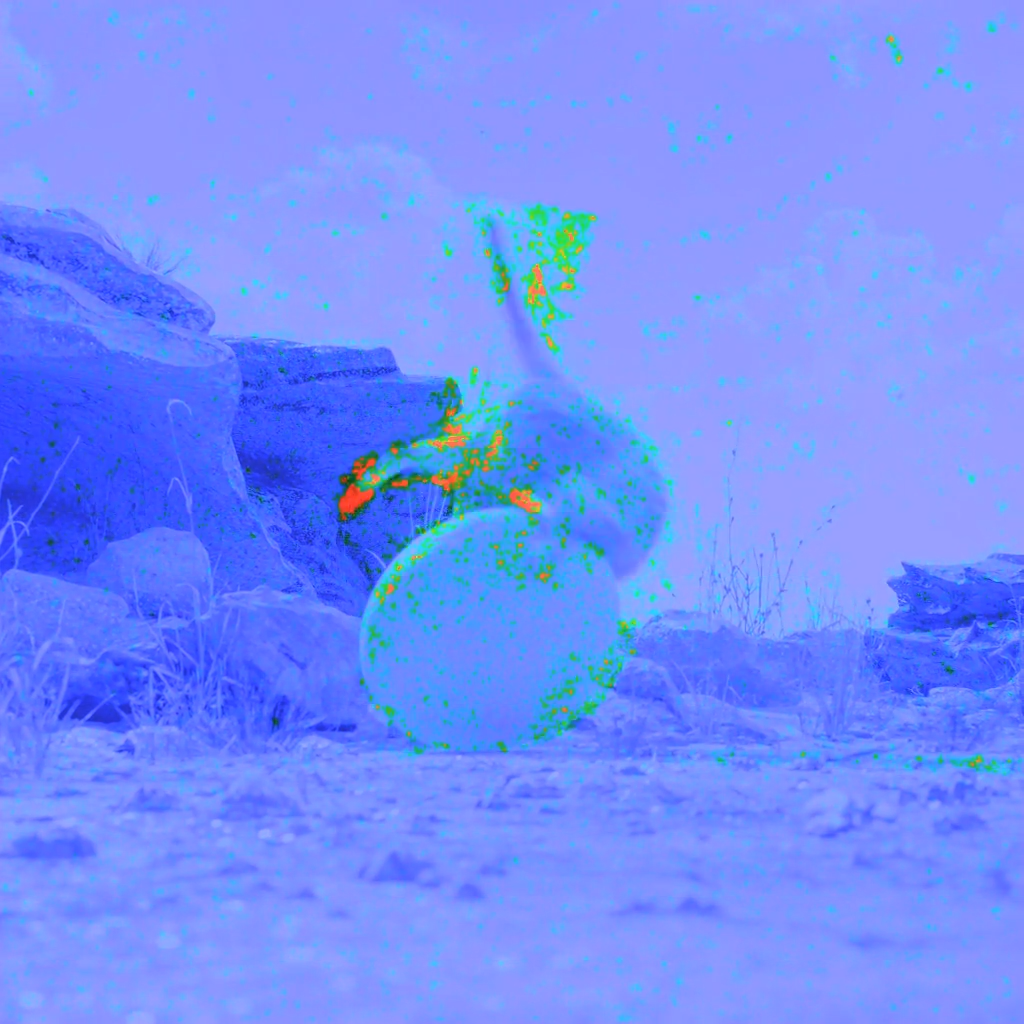}
    \caption{Forward lag in \metricname-2 error maps and exaggeration of small distortions in videos with localized distortions due to mean pooling. }
    \label{fig:limitations}
\end{figure}

In our current implementation, we use mean pooling over space and time to compute the overall quality of a video. However, when errors are localized in a small region—such as ghosting around particles in the Mushroom scene or interpolation errors in the Meerkat sequence—observers tend to rate the entire video based on these local distortions. This non-uniform perception of errors is not captured by mean pooling, which can sometimes overemphasize minor errors during data fitting, as illustrated in Figure \ref{fig:limitations}. A more perceptually accurate pooling strategy or feature distance metric could help address this issue. Note that \metricname \xspace was not trained to predict accurate error maps \cite{wolski19emap} because there is no such dataset for videos. \Rev{Unlike VDPs, which operate in calibrated photometric units to produce error maps in an interpretable probability-of-detection scale, \metricname’s error maps reflect relative contribution of local differences to overall quality and may vary in scale depending on the dataset.}


Our current model does not account for viewing conditions and display parameters, such as brightness and viewing distance, which could be addressed with display-adaptive input normalization and resampling \cite{ye2019predicting}. \Rev{Our formulation assumes that the reference and distorted videos have the same frame rate, and therefore cannot capture differences in motion quality arising from frame rate variations \cite{denes2020}}. Additionally, the 3D convolutions in our ResNet-18 implementation rely on future frames, potentially causing forward lag in the error maps. Downsampling is currently performed using convolutional striding, which may introduce aliasing in the feature space. Implementing weighted $l_2$ pooling, as proposed in \cite{ding2020image}, could further improve \metricname's performance.

Our experiments also show that pre-training 3D CNNs on auxiliary tasks significantly enhances metric performance. However, the optimal choice of pre-training task and dataset for video quality assessment remains an open question and could benefit from ongoing advances in self-supervised learning \cite{madhusudana2023conviqt}. It is worth noting that 3D CNNs can be computationally and memory-intensive, especially when processing high-resolution video inputs. Nonetheless, their spatio-temporal locality enables highly parallelizable implementations. Finally, incorporating higher-order visual features, such as optical flow or depth maps, may further improve the accuracy of the metric.


%% file: tables/benchmark_half.tex
\begin{table*}[]
\centering
\caption{\Rev{Cross-dataset validation. *Results are reported on the test splits of the GamingVideoSET, LIVE Livestream, and CG-VQD datasets, as well as on additional datasets used exclusively for testing. Background colors represent performance quantiles across all tested metrics: green (top 25\%), yellow (25–75\%), and red (bottom 25\%). CGVQM-5 demonstrates strong generalization across a wide range of content types and distortion patterns.}}
\label{tab:cross-val}
\resizebox{\textwidth}{!}{%
\begin{tabular}{l|ll|ll|ll|ll|ll|ll|ll|ll|ll}
\hline
\multicolumn{1}{c|}{}                         & \multicolumn{2}{c|}{\begin{tabular}[c]{@{}c@{}}GamingVideo-\\ Dataset*\end{tabular}}                        & \multicolumn{2}{c|}{\begin{tabular}[c]{@{}c@{}}LIVE \\ LIVESTREAM*\end{tabular}}                            & \multicolumn{2}{c|}{\begin{tabular}[c]{@{}c@{}}CG-VQD\\ (Ours)*\end{tabular}}                               & \multicolumn{2}{c|}{\begin{tabular}[c]{@{}c@{}}LIVE\\ Meta\end{tabular}}                                    & \multicolumn{2}{c|}{CGVDS}                                                                                  & \multicolumn{2}{c|}{\begin{tabular}[c]{@{}c@{}}LIVE\\ Flicker\end{tabular}}                                 & \multicolumn{2}{c|}{NVS}                                                                                    & \multicolumn{2}{c|}{\begin{tabular}[c]{@{}c@{}}AVT-VQDB-\\ UHD-1\end{tabular}}                              & \multicolumn{2}{c}{BVI-HD}                                                                                  \\ \cline{2-19} 
\multicolumn{1}{c|}{\multirow{-2}{*}{Metric}} & \multicolumn{1}{c}{PLCC}                             & \multicolumn{1}{c|}{SRCC}                            & \multicolumn{1}{c}{PLCC}                             & \multicolumn{1}{c|}{SRCC}                            & \multicolumn{1}{c}{PLCC}                             & \multicolumn{1}{c|}{SRCC}                            & \multicolumn{1}{c}{PLCC}                             & \multicolumn{1}{c|}{SRCC}                            & \multicolumn{1}{c}{PLCC}                             & \multicolumn{1}{c|}{SRCC}                            & \multicolumn{1}{c}{PLCC}                             & \multicolumn{1}{c|}{SRCC}                            & \multicolumn{1}{c}{PLCC}                             & \multicolumn{1}{c|}{SRCC}                            & \multicolumn{1}{c}{PLCC}                             & \multicolumn{1}{c|}{SRCC}                            & \multicolumn{1}{c}{PLCC}                             & \multicolumn{1}{c}{SRCC}                             \\ \hline
ABS                                           & \cellcolor[HTML]{C6EFCE}{\color[HTML]{006100} 0.853} & \cellcolor[HTML]{C6EFCE}{\color[HTML]{006100} 0.911} & \cellcolor[HTML]{FFEB9C}{\color[HTML]{9C5700} 0.649} & \cellcolor[HTML]{FFEB9C}{\color[HTML]{9C5700} 0.67}  & \cellcolor[HTML]{FFEB9C}{\color[HTML]{9C5700} 0.691} & \cellcolor[HTML]{FFEB9C}{\color[HTML]{9C5700} 0.744} & \cellcolor[HTML]{FFEB9C}{\color[HTML]{9C5700} 0.896} & \cellcolor[HTML]{FFC7CE}{\color[HTML]{9C0006} 0.926} & \cellcolor[HTML]{FFC7CE}{\color[HTML]{9C0006} 0.452} & \cellcolor[HTML]{FFC7CE}{\color[HTML]{9C0006} 0.51}  & \cellcolor[HTML]{FFEB9C}{\color[HTML]{9C5700} 0.888} & \cellcolor[HTML]{FFEB9C}{\color[HTML]{9C5700} 0.869} & \cellcolor[HTML]{FFC7CE}{\color[HTML]{9C0006} 0.492} & \cellcolor[HTML]{FFEB9C}{\color[HTML]{9C5700} 0.633} & \cellcolor[HTML]{FFEB9C}{\color[HTML]{9C5700} 0.768} & \cellcolor[HTML]{FFEB9C}{\color[HTML]{9C5700} 0.798} & \cellcolor[HTML]{FFC7CE}{\color[HTML]{9C0006} 0.416} & \cellcolor[HTML]{FFC7CE}{\color[HTML]{9C0006} 0.422} \\
PSNR                                          & \cellcolor[HTML]{FFC7CE}{\color[HTML]{9C0006} 0.537} & \cellcolor[HTML]{FFC7CE}{\color[HTML]{9C0006} 0.775} & \cellcolor[HTML]{FFC7CE}{\color[HTML]{9C0006} 0.577} & \cellcolor[HTML]{FFC7CE}{\color[HTML]{9C0006} 0.627} & \cellcolor[HTML]{FFEB9C}{\color[HTML]{9C5700} 0.697} & \cellcolor[HTML]{FFC7CE}{\color[HTML]{9C0006} 0.723} & \cellcolor[HTML]{FFC7CE}{\color[HTML]{9C0006} 0.671} & \cellcolor[HTML]{FFC7CE}{\color[HTML]{9C0006} 0.928} & \cellcolor[HTML]{FFEB9C}{\color[HTML]{9C5700} 0.514} & \cellcolor[HTML]{FFC7CE}{\color[HTML]{9C0006} 0.519} & \cellcolor[HTML]{FFC7CE}{\color[HTML]{9C0006} 0.662} & \cellcolor[HTML]{FFC7CE}{\color[HTML]{9C0006} 0.65}  & \cellcolor[HTML]{FFEB9C}{\color[HTML]{9C5700} 0.64}  & \cellcolor[HTML]{FFEB9C}{\color[HTML]{9C5700} 0.675} & \cellcolor[HTML]{FFEB9C}{\color[HTML]{9C5700} 0.791} & \cellcolor[HTML]{FFEB9C}{\color[HTML]{9C5700} 0.822} & \cellcolor[HTML]{FFEB9C}{\color[HTML]{9C5700} 0.471} & \cellcolor[HTML]{FFC7CE}{\color[HTML]{9C0006} 0.465} \\
SSIM \cite{wang2004image}                     & \cellcolor[HTML]{FFEB9C}{\color[HTML]{9C5700} 0.766} & \cellcolor[HTML]{FFEB9C}{\color[HTML]{9C5700} 0.787} & \cellcolor[HTML]{FFEB9C}{\color[HTML]{9C5700} 0.709} & \cellcolor[HTML]{C6EFCE}{\color[HTML]{006100} 0.834} & \cellcolor[HTML]{FFC7CE}{\color[HTML]{9C0006} 0.639} & \cellcolor[HTML]{C6EFCE}{\color[HTML]{006100} 0.805} & \cellcolor[HTML]{FFEB9C}{\color[HTML]{9C5700} 0.838} & \cellcolor[HTML]{FFC7CE}{\color[HTML]{9C0006} 0.909} & \cellcolor[HTML]{FFC7CE}{\color[HTML]{9C0006} 0.472} & \cellcolor[HTML]{FFEB9C}{\color[HTML]{9C5700} 0.53}  & \cellcolor[HTML]{FFEB9C}{\color[HTML]{9C5700} 0.85}  & \cellcolor[HTML]{FFEB9C}{\color[HTML]{9C5700} 0.867} & \cellcolor[HTML]{FFEB9C}{\color[HTML]{9C5700} 0.611} & \cellcolor[HTML]{FFEB9C}{\color[HTML]{9C5700} 0.635} & \cellcolor[HTML]{FFEB9C}{\color[HTML]{9C5700} 0.739} & \cellcolor[HTML]{FFEB9C}{\color[HTML]{9C5700} 0.786} & \cellcolor[HTML]{FFEB9C}{\color[HTML]{9C5700} 0.524} & \cellcolor[HTML]{FFEB9C}{\color[HTML]{9C5700} 0.621} \\
MS-SSIM \cite{wang2003multiscale}             & \cellcolor[HTML]{FFEB9C}{\color[HTML]{9C5700} 0.787} & \cellcolor[HTML]{FFEB9C}{\color[HTML]{9C5700} 0.826} & \cellcolor[HTML]{FFEB9C}{\color[HTML]{9C5700} 0.712} & \cellcolor[HTML]{FFEB9C}{\color[HTML]{9C5700} 0.748} & \cellcolor[HTML]{FFEB9C}{\color[HTML]{9C5700} 0.668} & \cellcolor[HTML]{C6EFCE}{\color[HTML]{006100} 0.804} & \cellcolor[HTML]{FFEB9C}{\color[HTML]{9C5700} 0.872} & \cellcolor[HTML]{FFEB9C}{\color[HTML]{9C5700} 0.946} & \cellcolor[HTML]{FFC7CE}{\color[HTML]{9C0006} 0.462} & \cellcolor[HTML]{FFEB9C}{\color[HTML]{9C5700} 0.524} & \cellcolor[HTML]{FFC7CE}{\color[HTML]{9C0006} 0.832} & \cellcolor[HTML]{FFEB9C}{\color[HTML]{9C5700} 0.864} & \cellcolor[HTML]{FFEB9C}{\color[HTML]{9C5700} 0.619} & \cellcolor[HTML]{FFEB9C}{\color[HTML]{9C5700} 0.655} & \cellcolor[HTML]{FFEB9C}{\color[HTML]{9C5700} 0.767} & \cellcolor[HTML]{FFEB9C}{\color[HTML]{9C5700} 0.814} & \cellcolor[HTML]{FFEB9C}{\color[HTML]{9C5700} 0.513} & \cellcolor[HTML]{FFEB9C}{\color[HTML]{9C5700} 0.602} \\
IW-SSIM \cite{wang2010information}            & \cellcolor[HTML]{FFEB9C}{\color[HTML]{9C5700} 0.779} & \cellcolor[HTML]{FFC7CE}{\color[HTML]{9C0006} 0.779} & \cellcolor[HTML]{C6EFCE}{\color[HTML]{006100} 0.772} & \cellcolor[HTML]{C6EFCE}{\color[HTML]{006100} 0.805} & \cellcolor[HTML]{FFEB9C}{\color[HTML]{9C5700} 0.688} & \cellcolor[HTML]{C6EFCE}{\color[HTML]{006100} 0.837} & \cellcolor[HTML]{FFEB9C}{\color[HTML]{9C5700} 0.86}  & \cellcolor[HTML]{FFEB9C}{\color[HTML]{9C5700} 0.955} & \cellcolor[HTML]{FFEB9C}{\color[HTML]{9C5700} 0.492} & \cellcolor[HTML]{FFEB9C}{\color[HTML]{9C5700} 0.54}  & \cellcolor[HTML]{FFC7CE}{\color[HTML]{9C0006} 0.845} & \cellcolor[HTML]{FFEB9C}{\color[HTML]{9C5700} 0.874} & \cellcolor[HTML]{FFEB9C}{\color[HTML]{9C5700} 0.671} & \cellcolor[HTML]{FFEB9C}{\color[HTML]{9C5700} 0.668} & \cellcolor[HTML]{FFEB9C}{\color[HTML]{9C5700} 0.823} & \cellcolor[HTML]{FFEB9C}{\color[HTML]{9C5700} 0.855} & \cellcolor[HTML]{FFEB9C}{\color[HTML]{9C5700} 0.565} & \cellcolor[HTML]{FFEB9C}{\color[HTML]{9C5700} 0.64}  \\
VIFp \cite{sheikh2006image}                   & \cellcolor[HTML]{FFC7CE}{\color[HTML]{9C0006} 0.721} & \cellcolor[HTML]{FFEB9C}{\color[HTML]{9C5700} 0.78}  & \cellcolor[HTML]{FFEB9C}{\color[HTML]{9C5700} 0.592} & \cellcolor[HTML]{FFC7CE}{\color[HTML]{9C0006} 0.627} & \cellcolor[HTML]{C6EFCE}{\color[HTML]{006100} 0.776} & \cellcolor[HTML]{FFEB9C}{\color[HTML]{9C5700} 0.775} & \cellcolor[HTML]{FFEB9C}{\color[HTML]{9C5700} 0.909} & \cellcolor[HTML]{FFEB9C}{\color[HTML]{9C5700} 0.941} & \cellcolor[HTML]{C6EFCE}{\color[HTML]{006100} 0.673} & \cellcolor[HTML]{C6EFCE}{\color[HTML]{006100} 0.686} & \cellcolor[HTML]{FFEB9C}{\color[HTML]{9C5700} 0.857} & \cellcolor[HTML]{FFEB9C}{\color[HTML]{9C5700} 0.859} & \cellcolor[HTML]{FFEB9C}{\color[HTML]{9C5700} 0.582} & \cellcolor[HTML]{FFEB9C}{\color[HTML]{9C5700} 0.639} & \cellcolor[HTML]{FFEB9C}{\color[HTML]{9C5700} 0.83}  & \cellcolor[HTML]{FFEB9C}{\color[HTML]{9C5700} 0.827} & \cellcolor[HTML]{C6EFCE}{\color[HTML]{006100} 0.633} & \cellcolor[HTML]{FFEB9C}{\color[HTML]{9C5700} 0.622} \\
FSIM \cite{zhang2011fsim}                     & \cellcolor[HTML]{FFEB9C}{\color[HTML]{9C5700} 0.773} & \cellcolor[HTML]{FFC7CE}{\color[HTML]{9C0006} 0.767} & \cellcolor[HTML]{FFEB9C}{\color[HTML]{9C5700} 0.655} & \cellcolor[HTML]{FFEB9C}{\color[HTML]{9C5700} 0.804} & \cellcolor[HTML]{FFEB9C}{\color[HTML]{9C5700} 0.653} & \cellcolor[HTML]{FFEB9C}{\color[HTML]{9C5700} 0.789} & \cellcolor[HTML]{FFEB9C}{\color[HTML]{9C5700} 0.856} & \cellcolor[HTML]{FFEB9C}{\color[HTML]{9C5700} 0.936} & \cellcolor[HTML]{FFEB9C}{\color[HTML]{9C5700} 0.504} & \cellcolor[HTML]{FFEB9C}{\color[HTML]{9C5700} 0.545} & \cellcolor[HTML]{FFEB9C}{\color[HTML]{9C5700} 0.889} & \cellcolor[HTML]{FFEB9C}{\color[HTML]{9C5700} 0.881} & \cellcolor[HTML]{FFC7CE}{\color[HTML]{9C0006} 0.499} & \cellcolor[HTML]{FFC7CE}{\color[HTML]{9C0006} 0.625} & \cellcolor[HTML]{FFEB9C}{\color[HTML]{9C5700} 0.724} & \cellcolor[HTML]{FFEB9C}{\color[HTML]{9C5700} 0.841} & \cellcolor[HTML]{FFEB9C}{\color[HTML]{9C5700} 0.521} & \cellcolor[HTML]{FFEB9C}{\color[HTML]{9C5700} 0.638} \\
SR-SIM\cite{zhang2012sr}                      & \cellcolor[HTML]{FFEB9C}{\color[HTML]{9C5700} 0.789} & \cellcolor[HTML]{FFEB9C}{\color[HTML]{9C5700} 0.8}   & \cellcolor[HTML]{FFC7CE}{\color[HTML]{9C0006} 0.585} & \cellcolor[HTML]{FFEB9C}{\color[HTML]{9C5700} 0.793} & \cellcolor[HTML]{FFC7CE}{\color[HTML]{9C0006} 0.651} & \cellcolor[HTML]{FFEB9C}{\color[HTML]{9C5700} 0.783} & \cellcolor[HTML]{FFC7CE}{\color[HTML]{9C0006} 0.837} & \cellcolor[HTML]{FFEB9C}{\color[HTML]{9C5700} 0.937} & \cellcolor[HTML]{FFEB9C}{\color[HTML]{9C5700} 0.485} & \cellcolor[HTML]{FFEB9C}{\color[HTML]{9C5700} 0.538} & \cellcolor[HTML]{FFEB9C}{\color[HTML]{9C5700} 0.877} & \cellcolor[HTML]{FFEB9C}{\color[HTML]{9C5700} 0.862} & \cellcolor[HTML]{FFEB9C}{\color[HTML]{9C5700} 0.631} & \cellcolor[HTML]{FFEB9C}{\color[HTML]{9C5700} 0.659} & \cellcolor[HTML]{FFC7CE}{\color[HTML]{9C0006} 0.68}  & \cellcolor[HTML]{FFEB9C}{\color[HTML]{9C5700} 0.805} & \cellcolor[HTML]{FFEB9C}{\color[HTML]{9C5700} 0.551} & \cellcolor[HTML]{FFEB9C}{\color[HTML]{9C5700} 0.642} \\
GMSD \cite{xue2013gradient}                   & \cellcolor[HTML]{C6EFCE}{\color[HTML]{006100} 0.853} & \cellcolor[HTML]{FFEB9C}{\color[HTML]{9C5700} 0.86}  & \cellcolor[HTML]{FFEB9C}{\color[HTML]{9C5700} 0.68}  & \cellcolor[HTML]{FFEB9C}{\color[HTML]{9C5700} 0.726} & \cellcolor[HTML]{FFEB9C}{\color[HTML]{9C5700} 0.746} & \cellcolor[HTML]{FFEB9C}{\color[HTML]{9C5700} 0.767} & \cellcolor[HTML]{FFEB9C}{\color[HTML]{9C5700} 0.917} & \cellcolor[HTML]{FFEB9C}{\color[HTML]{9C5700} 0.955} & \cellcolor[HTML]{FFEB9C}{\color[HTML]{9C5700} 0.584} & \cellcolor[HTML]{FFEB9C}{\color[HTML]{9C5700} 0.581} & \cellcolor[HTML]{C6EFCE}{\color[HTML]{006100} 0.948} & \cellcolor[HTML]{FFEB9C}{\color[HTML]{9C5700} 0.867} & \cellcolor[HTML]{FFEB9C}{\color[HTML]{9C5700} 0.672} & \cellcolor[HTML]{FFEB9C}{\color[HTML]{9C5700} 0.656} & \cellcolor[HTML]{C6EFCE}{\color[HTML]{006100} 0.852} & \cellcolor[HTML]{C6EFCE}{\color[HTML]{006100} 0.864} & \cellcolor[HTML]{FFEB9C}{\color[HTML]{9C5700} 0.608} & \cellcolor[HTML]{FFEB9C}{\color[HTML]{9C5700} 0.635} \\
MS-GMSD \cite{zhang2017gradient}              & \cellcolor[HTML]{C6EFCE}{\color[HTML]{006100} 0.857} & \cellcolor[HTML]{C6EFCE}{\color[HTML]{006100} 0.866} & \cellcolor[HTML]{FFEB9C}{\color[HTML]{9C5700} 0.686} & \cellcolor[HTML]{FFEB9C}{\color[HTML]{9C5700} 0.726} & \cellcolor[HTML]{FFEB9C}{\color[HTML]{9C5700} 0.754} & \cellcolor[HTML]{FFEB9C}{\color[HTML]{9C5700} 0.767} & \cellcolor[HTML]{FFEB9C}{\color[HTML]{9C5700} 0.918} & \cellcolor[HTML]{FFEB9C}{\color[HTML]{9C5700} 0.953} & \cellcolor[HTML]{FFEB9C}{\color[HTML]{9C5700} 0.574} & \cellcolor[HTML]{FFEB9C}{\color[HTML]{9C5700} 0.573} & \cellcolor[HTML]{C6EFCE}{\color[HTML]{006100} 0.945} & \cellcolor[HTML]{FFEB9C}{\color[HTML]{9C5700} 0.882} & \cellcolor[HTML]{FFEB9C}{\color[HTML]{9C5700} 0.652} & \cellcolor[HTML]{FFEB9C}{\color[HTML]{9C5700} 0.676} & \cellcolor[HTML]{FFEB9C}{\color[HTML]{9C5700} 0.848} & \cellcolor[HTML]{C6EFCE}{\color[HTML]{006100} 0.863} & \cellcolor[HTML]{FFEB9C}{\color[HTML]{9C5700} 0.604} & \cellcolor[HTML]{FFEB9C}{\color[HTML]{9C5700} 0.627} \\
VSI \cite{zhang2014vsi}                       & \cellcolor[HTML]{FFEB9C}{\color[HTML]{9C5700} 0.778} & \cellcolor[HTML]{FFEB9C}{\color[HTML]{9C5700} 0.81}  & \cellcolor[HTML]{FFEB9C}{\color[HTML]{9C5700} 0.637} & \cellcolor[HTML]{C6EFCE}{\color[HTML]{006100} 0.816} & \cellcolor[HTML]{FFC7CE}{\color[HTML]{9C0006} 0.625} & \cellcolor[HTML]{FFEB9C}{\color[HTML]{9C5700} 0.789} & \cellcolor[HTML]{FFC7CE}{\color[HTML]{9C0006} 0.832} & \cellcolor[HTML]{FFEB9C}{\color[HTML]{9C5700} 0.93}  & \cellcolor[HTML]{FFC7CE}{\color[HTML]{9C0006} 0.461} & \cellcolor[HTML]{FFC7CE}{\color[HTML]{9C0006} 0.517} & \cellcolor[HTML]{FFEB9C}{\color[HTML]{9C5700} 0.902} & \cellcolor[HTML]{FFEB9C}{\color[HTML]{9C5700} 0.862} & \cellcolor[HTML]{FFC7CE}{\color[HTML]{9C0006} 0.522} & \cellcolor[HTML]{C6EFCE}{\color[HTML]{9C5700} 0.67}  & \cellcolor[HTML]{FFEB9C}{\color[HTML]{9C5700} 0.7}   & \cellcolor[HTML]{FFEB9C}{\color[HTML]{9C5700} 0.845} & \cellcolor[HTML]{FFEB9C}{\color[HTML]{9C5700} 0.536} & \cellcolor[HTML]{FFEB9C}{\color[HTML]{9C5700} 0.631} \\
DSS \cite{balanov2015image}                   & \cellcolor[HTML]{C6EFCE}{\color[HTML]{006100} 0.842} & \cellcolor[HTML]{FFEB9C}{\color[HTML]{9C5700} 0.831} & \cellcolor[HTML]{C6EFCE}{\color[HTML]{006100} 0.781} & \cellcolor[HTML]{FFEB9C}{\color[HTML]{9C5700} 0.8}   & \cellcolor[HTML]{FFEB9C}{\color[HTML]{9C5700} 0.667} & \cellcolor[HTML]{FFEB9C}{\color[HTML]{9C5700} 0.772} & \cellcolor[HTML]{FFEB9C}{\color[HTML]{9C5700} 0.907} & \cellcolor[HTML]{C6EFCE}{\color[HTML]{006100} 0.957} & \cellcolor[HTML]{FFEB9C}{\color[HTML]{9C5700} 0.635} & \cellcolor[HTML]{FFEB9C}{\color[HTML]{9C5700} 0.637} & \cellcolor[HTML]{C6EFCE}{\color[HTML]{006100} 0.943} & \cellcolor[HTML]{C6EFCE}{\color[HTML]{006100} 0.92}  & \cellcolor[HTML]{C6EFCE}{\color[HTML]{006100} 0.764} & \cellcolor[HTML]{C6EFCE}{\color[HTML]{006100} 0.758} & \cellcolor[HTML]{C6EFCE}{\color[HTML]{006100} 0.87}  & \cellcolor[HTML]{C6EFCE}{\color[HTML]{006100} 0.879} & \cellcolor[HTML]{FFEB9C}{\color[HTML]{9C5700} 0.618} & \cellcolor[HTML]{C6EFCE}{\color[HTML]{006100} 0.67}  \\
Content-Score \cite{gatys2015neural}          & \cellcolor[HTML]{FFEB9C}{\color[HTML]{9C5700} 0.828} & \cellcolor[HTML]{C6EFCE}{\color[HTML]{006100} 0.863} & \cellcolor[HTML]{FFEB9C}{\color[HTML]{9C5700} 0.61}  & \cellcolor[HTML]{FFEB9C}{\color[HTML]{9C5700} 0.649} & \cellcolor[HTML]{FFC7CE}{\color[HTML]{9C0006} 0.651} & \cellcolor[HTML]{FFC7CE}{\color[HTML]{9C0006} 0.739} & \cellcolor[HTML]{FFEB9C}{\color[HTML]{9C5700} 0.887} & \cellcolor[HTML]{FFC7CE}{\color[HTML]{9C0006} 0.928} & \cellcolor[HTML]{FFEB9C}{\color[HTML]{9C5700} 0.548} & \cellcolor[HTML]{FFEB9C}{\color[HTML]{9C5700} 0.564} & \cellcolor[HTML]{FFEB9C}{\color[HTML]{9C5700} 0.916} & \cellcolor[HTML]{FFEB9C}{\color[HTML]{9C5700} 0.86}  & \cellcolor[HTML]{FFC7CE}{\color[HTML]{9C0006} 0.393} & \cellcolor[HTML]{FFC7CE}{\color[HTML]{9C0006} 0.566} & \cellcolor[HTML]{FFC7CE}{\color[HTML]{9C0006} 0.398} & \cellcolor[HTML]{FFC7CE}{\color[HTML]{9C0006} 0.546} & \cellcolor[HTML]{FFC7CE}{\color[HTML]{9C0006} 0.43}  & \cellcolor[HTML]{FFC7CE}{\color[HTML]{9C0006} 0.456} \\
Style-Score \cite{gatys2015neural}            & \cellcolor[HTML]{FFC7CE}{\color[HTML]{9C0006} 0.425} & \cellcolor[HTML]{FFC7CE}{\color[HTML]{9C0006} 0.456} & \cellcolor[HTML]{FFC7CE}{\color[HTML]{9C0006} 0.232} & \cellcolor[HTML]{FFC7CE}{\color[HTML]{9C0006} 0.596} & \cellcolor[HTML]{FFC7CE}{\color[HTML]{9C0006} 0.431} & \cellcolor[HTML]{FFC7CE}{\color[HTML]{9C0006} 0.667} & \cellcolor[HTML]{FFC7CE}{\color[HTML]{9C0006} 0.636} & \cellcolor[HTML]{FFC7CE}{\color[HTML]{9C0006} 0.888} & \cellcolor[HTML]{FFC7CE}{\color[HTML]{9C0006} 0.311} & \cellcolor[HTML]{FFEB9C}{\color[HTML]{9C5700} 0.586} & \cellcolor[HTML]{FFC7CE}{\color[HTML]{9C0006} 0.795} & \cellcolor[HTML]{FFEB9C}{\color[HTML]{9C5700} 0.858} & \cellcolor[HTML]{FFC7CE}{\color[HTML]{9C0006} 0.425} & \cellcolor[HTML]{FFC7CE}{\color[HTML]{9C0006} 0.612} & \cellcolor[HTML]{FFC7CE}{\color[HTML]{9C0006} 0.326} & \cellcolor[HTML]{FFC7CE}{\color[HTML]{9C0006} 0.536} & \cellcolor[HTML]{FFC7CE}{\color[HTML]{9C0006} 0.273} & \cellcolor[HTML]{FFC7CE}{\color[HTML]{9C0006} 0.364} \\
HaarPSI \cite{reisenhofer2018haar}            & \cellcolor[HTML]{FFEB9C}{\color[HTML]{9C5700} 0.829} & \cellcolor[HTML]{FFEB9C}{\color[HTML]{9C5700} 0.829} & \cellcolor[HTML]{FFEB9C}{\color[HTML]{9C5700} 0.713} & \cellcolor[HTML]{FFEB9C}{\color[HTML]{9C5700} 0.73}  & \cellcolor[HTML]{C6EFCE}{\color[HTML]{006100} 0.777} & \cellcolor[HTML]{FFEB9C}{\color[HTML]{9C5700} 0.794} & \cellcolor[HTML]{C6EFCE}{\color[HTML]{006100} 0.937} & \cellcolor[HTML]{FFEB9C}{\color[HTML]{9C5700} 0.956} & \cellcolor[HTML]{FFEB9C}{\color[HTML]{9C5700} 0.618} & \cellcolor[HTML]{FFEB9C}{\color[HTML]{9C5700} 0.615} & \cellcolor[HTML]{C6EFCE}{\color[HTML]{006100} 0.941} & \cellcolor[HTML]{FFEB9C}{\color[HTML]{9C5700} 0.861} & \cellcolor[HTML]{C6EFCE}{\color[HTML]{006100} 0.719} & \cellcolor[HTML]{C6EFCE}{\color[HTML]{006100} 0.717} & \cellcolor[HTML]{FFEB9C}{\color[HTML]{9C5700} 0.83}  & \cellcolor[HTML]{FFEB9C}{\color[HTML]{9C5700} 0.84}  & \cellcolor[HTML]{FFEB9C}{\color[HTML]{9C5700} 0.604} & \cellcolor[HTML]{FFEB9C}{\color[HTML]{9C5700} 0.615} \\
MDSI \cite{nafchi2016mean}                    & \cellcolor[HTML]{FFEB9C}{\color[HTML]{9C5700} 0.77}  & \cellcolor[HTML]{FFEB9C}{\color[HTML]{9C5700} 0.803} & \cellcolor[HTML]{FFEB9C}{\color[HTML]{9C5700} 0.705} & \cellcolor[HTML]{FFEB9C}{\color[HTML]{9C5700} 0.724} & \cellcolor[HTML]{FFEB9C}{\color[HTML]{9C5700} 0.753} & \cellcolor[HTML]{FFC7CE}{\color[HTML]{9C0006} 0.73}  & \cellcolor[HTML]{FFC7CE}{\color[HTML]{9C0006} 0.83}  & \cellcolor[HTML]{FFC7CE}{\color[HTML]{9C0006} 0.925} & \cellcolor[HTML]{FFEB9C}{\color[HTML]{9C5700} 0.632} & \cellcolor[HTML]{FFEB9C}{\color[HTML]{9C5700} 0.602} & \cellcolor[HTML]{FFEB9C}{\color[HTML]{9C5700} 0.85}  & \cellcolor[HTML]{C6EFCE}{\color[HTML]{006100} 0.889} & \cellcolor[HTML]{FFEB9C}{\color[HTML]{9C5700} 0.662} & \cellcolor[HTML]{FFEB9C}{\color[HTML]{9C5700} 0.661} & \cellcolor[HTML]{FFEB9C}{\color[HTML]{9C5700} 0.84}  & \cellcolor[HTML]{FFEB9C}{\color[HTML]{9C5700} 0.838} & \cellcolor[HTML]{C6EFCE}{\color[HTML]{006100} 0.666} & \cellcolor[HTML]{FFEB9C}{\color[HTML]{9C5700} 0.659} \\
LPIPS \cite{zhang2018unreasonable}            & \cellcolor[HTML]{FFEB9C}{\color[HTML]{9C5700} 0.794} & \cellcolor[HTML]{FFEB9C}{\color[HTML]{9C5700} 0.804} & \cellcolor[HTML]{FFC7CE}{\color[HTML]{9C0006} 0.584} & \cellcolor[HTML]{FFC7CE}{\color[HTML]{9C0006} 0.622} & \cellcolor[HTML]{FFEB9C}{\color[HTML]{9C5700} 0.696} & \cellcolor[HTML]{FFEB9C}{\color[HTML]{9C5700} 0.748} & \cellcolor[HTML]{FFEB9C}{\color[HTML]{9C5700} 0.914} & \cellcolor[HTML]{FFEB9C}{\color[HTML]{9C5700} 0.938} & \cellcolor[HTML]{C6EFCE}{\color[HTML]{006100} 0.782} & \cellcolor[HTML]{C6EFCE}{\color[HTML]{006100} 0.777} & \cellcolor[HTML]{FFEB9C}{\color[HTML]{9C5700} 0.9}   & \cellcolor[HTML]{FFC7CE}{\color[HTML]{9C0006} 0.847} & \cellcolor[HTML]{FFEB9C}{\color[HTML]{9C5700} 0.593} & \cellcolor[HTML]{FFC7CE}{\color[HTML]{9C0006} 0.547} & \cellcolor[HTML]{FFC7CE}{\color[HTML]{9C0006} 0.501} & \cellcolor[HTML]{FFC7CE}{\color[HTML]{9C0006} 0.559} & \cellcolor[HTML]{FFEB9C}{\color[HTML]{9C5700} 0.55}  & \cellcolor[HTML]{FFEB9C}{\color[HTML]{9C5700} 0.551} \\
DISTS \cite{ding2020image}                    & \cellcolor[HTML]{FFEB9C}{\color[HTML]{9C5700} 0.818} & \cellcolor[HTML]{FFEB9C}{\color[HTML]{9C5700} 0.817} & \cellcolor[HTML]{FFC7CE}{\color[HTML]{9C0006} 0.556} & \cellcolor[HTML]{FFEB9C}{\color[HTML]{9C5700} 0.694} & \cellcolor[HTML]{FFEB9C}{\color[HTML]{9C5700} 0.686} & \cellcolor[HTML]{C6EFCE}{\color[HTML]{006100} 0.795} & \cellcolor[HTML]{C6EFCE}{\color[HTML]{006100} 0.929} & \cellcolor[HTML]{FFEB9C}{\color[HTML]{9C5700} 0.951} & \cellcolor[HTML]{C6EFCE}{\color[HTML]{006100} 0.85}  & \cellcolor[HTML]{C6EFCE}{\color[HTML]{006100} 0.835} & \cellcolor[HTML]{FFEB9C}{\color[HTML]{9C5700} 0.888} & \cellcolor[HTML]{FFEB9C}{\color[HTML]{9C5700} 0.873} & \cellcolor[HTML]{C6EFCE}{\color[HTML]{006100} 0.852} & \cellcolor[HTML]{C6EFCE}{\color[HTML]{006100} 0.843} & \cellcolor[HTML]{FFC7CE}{\color[HTML]{9C0006} 0.424} & \cellcolor[HTML]{FFC7CE}{\color[HTML]{9C0006} 0.683} & \cellcolor[HTML]{FFEB9C}{\color[HTML]{9C5700} 0.565} & \cellcolor[HTML]{FFEB9C}{\color[HTML]{9C5700} 0.613} \\
STRRED \cite{soundararajan2012video}          & \cellcolor[HTML]{FFC7CE}{\color[HTML]{9C0006} 0.711} & \cellcolor[HTML]{FFEB9C}{\color[HTML]{9C5700} 0.786} & \cellcolor[HTML]{FFC7CE}{\color[HTML]{9C0006} 0.329} & \cellcolor[HTML]{FFEB9C}{\color[HTML]{9C5700} 0.785} & \cellcolor[HTML]{FFC7CE}{\color[HTML]{9C0006} 0.606} & \cellcolor[HTML]{FFEB9C}{\color[HTML]{9C5700} 0.784} & \cellcolor[HTML]{FFC7CE}{\color[HTML]{9C0006} 0.833} & \cellcolor[HTML]{FFEB9C}{\color[HTML]{9C5700} 0.946} & \cellcolor[HTML]{FFC7CE}{\color[HTML]{9C0006} 0.359} & \cellcolor[HTML]{FFC7CE}{\color[HTML]{9C0006} 0.506} & \cellcolor[HTML]{FFC7CE}{\color[HTML]{9C0006} 0.655} & \cellcolor[HTML]{FFC7CE}{\color[HTML]{9C0006} 0.732} & \cellcolor[HTML]{FFEB9C}{\color[HTML]{9C5700} 0.671} & \cellcolor[HTML]{C6EFCE}{\color[HTML]{006100} 0.752} & \cellcolor[HTML]{FFC7CE}{\color[HTML]{9C0006} 0.317} & \cellcolor[HTML]{FFC7CE}{\color[HTML]{9C0006} 0.076} & \cellcolor[HTML]{FFC7CE}{\color[HTML]{9C0006} 0.198} & \cellcolor[HTML]{FFEB9C}{\color[HTML]{9C5700} 0.589} \\
PieAPP \cite{prashnani2018pieapp}             & \cellcolor[HTML]{FFC7CE}{\color[HTML]{9C0006} 0.755} & \cellcolor[HTML]{FFC7CE}{\color[HTML]{9C0006} 0.745} & \cellcolor[HTML]{FFEB9C}{\color[HTML]{9C5700} 0.66}  & \cellcolor[HTML]{FFC7CE}{\color[HTML]{9C0006} 0.639} & \cellcolor[HTML]{FFEB9C}{\color[HTML]{9C5700} 0.73}  & \cellcolor[HTML]{FFEB9C}{\color[HTML]{9C5700} 0.749} & \cellcolor[HTML]{C6EFCE}{\color[HTML]{006100} 0.931} & \cellcolor[HTML]{C6EFCE}{\color[HTML]{006100} 0.957} & \cellcolor[HTML]{C6EFCE}{\color[HTML]{006100} 0.778} & \cellcolor[HTML]{C6EFCE}{\color[HTML]{006100} 0.78}  & \cellcolor[HTML]{C6EFCE}{\color[HTML]{006100} 0.95}  & \cellcolor[HTML]{C6EFCE}{\color[HTML]{006100} 0.939} & \cellcolor[HTML]{C6EFCE}{\color[HTML]{006100} 0.788} & \cellcolor[HTML]{C6EFCE}{\color[HTML]{006100} 0.777} & \cellcolor[HTML]{FFEB9C}{\color[HTML]{9C5700} 0.836} & \cellcolor[HTML]{FFEB9C}{\color[HTML]{9C5700} 0.824} & \cellcolor[HTML]{FFEB9C}{\color[HTML]{9C5700} 0.587} & \cellcolor[HTML]{FFEB9C}{\color[HTML]{9C5700} 0.606} \\
FLIP \cite{andersson2020flip}                 & \cellcolor[HTML]{FFEB9C}{\color[HTML]{9C5700} 0.827} & \cellcolor[HTML]{C6EFCE}{\color[HTML]{006100} 0.876} & \cellcolor[HTML]{FFEB9C}{\color[HTML]{9C5700} 0.674} & \cellcolor[HTML]{FFEB9C}{\color[HTML]{9C5700} 0.672} & \cellcolor[HTML]{FFEB9C}{\color[HTML]{9C5700} 0.693} & \cellcolor[HTML]{FFC7CE}{\color[HTML]{9C0006} 0.728} & \cellcolor[HTML]{FFEB9C}{\color[HTML]{9C5700} 0.901} & \cellcolor[HTML]{FFEB9C}{\color[HTML]{9C5700} 0.931} & \cellcolor[HTML]{FFEB9C}{\color[HTML]{9C5700} 0.57}  & \cellcolor[HTML]{FFEB9C}{\color[HTML]{9C5700} 0.577} & \cellcolor[HTML]{FFEB9C}{\color[HTML]{9C5700} 0.849} & \cellcolor[HTML]{FFC7CE}{\color[HTML]{9C0006} 0.806} & \cellcolor[HTML]{FFC7CE}{\color[HTML]{9C0006} 0.488} & \cellcolor[HTML]{FFC7CE}{\color[HTML]{9C0006} 0.596} & \cellcolor[HTML]{FFEB9C}{\color[HTML]{9C5700} 0.818} & \cellcolor[HTML]{FFEB9C}{\color[HTML]{9C5700} 0.84}  & \cellcolor[HTML]{FFC7CE}{\color[HTML]{9C0006} 0.466} & \cellcolor[HTML]{FFC7CE}{\color[HTML]{9C0006} 0.501} \\
ERQA \cite{kirillova2021erqa}                 & \cellcolor[HTML]{FFC7CE}{\color[HTML]{9C0006} 0.646} & \cellcolor[HTML]{FFC7CE}{\color[HTML]{9C0006} 0.587} & \cellcolor[HTML]{FFEB9C}{\color[HTML]{9C5700} 0.618} & \cellcolor[HTML]{FFC7CE}{\color[HTML]{9C0006} 0.623} & \cellcolor[HTML]{C6EFCE}{\color[HTML]{006100} 0.797} & \cellcolor[HTML]{FFEB9C}{\color[HTML]{9C5700} 0.787} & \cellcolor[HTML]{FFEB9C}{\color[HTML]{9C5700} 0.9}   & \cellcolor[HTML]{FFEB9C}{\color[HTML]{9C5700} 0.936} & \cellcolor[HTML]{C6EFCE}{\color[HTML]{006100} 0.756} & \cellcolor[HTML]{C6EFCE}{\color[HTML]{006100} 0.737} & \cellcolor[HTML]{FFC7CE}{\color[HTML]{9C0006} 0.834} & \cellcolor[HTML]{FFC7CE}{\color[HTML]{9C0006} 0.771} & \cellcolor[HTML]{C6EFCE}{\color[HTML]{006100} 0.722} & \cellcolor[HTML]{C6EFCE}{\color[HTML]{006100} 0.717} & \cellcolor[HTML]{FFEB9C}{\color[HTML]{9C5700} 0.7}   & \cellcolor[HTML]{FFC7CE}{\color[HTML]{9C0006} 0.728} & \cellcolor[HTML]{FFC7CE}{\color[HTML]{9C0006} 0.4}   & \cellcolor[HTML]{FFC7CE}{\color[HTML]{9C0006} 0.491} \\
VMAF \cite{rassool2017vmaf}                   & \cellcolor[HTML]{FFEB9C}{\color[HTML]{9C5700} 0.826} & \cellcolor[HTML]{FFEB9C}{\color[HTML]{9C5700} 0.818} & \cellcolor[HTML]{FFEB9C}{\color[HTML]{9C5700} 0.716} & \cellcolor[HTML]{FFEB9C}{\color[HTML]{9C5700} 0.768} & \cellcolor[HTML]{FFEB9C}{\color[HTML]{9C5700} 0.715} & \cellcolor[HTML]{FFEB9C}{\color[HTML]{9C5700} 0.79}  & \cellcolor[HTML]{C6EFCE}{\color[HTML]{006100} 0.932} & \cellcolor[HTML]{C6EFCE}{\color[HTML]{006100} 0.96}  & \cellcolor[HTML]{FFEB9C}{\color[HTML]{9C5700} 0.492} & \cellcolor[HTML]{FFC7CE}{\color[HTML]{9C0006} 0.438} & \cellcolor[HTML]{FFEB9C}{\color[HTML]{9C5700} 0.919} & \cellcolor[HTML]{FFC7CE}{\color[HTML]{9C0006} 0.847} & \cellcolor[HTML]{FFEB9C}{\color[HTML]{9C5700} 0.654} & \cellcolor[HTML]{FFEB9C}{\color[HTML]{9C5700} 0.682} & \cellcolor[HTML]{C6EFCE}{\color[HTML]{006100} 0.929} & \cellcolor[HTML]{C6EFCE}{\color[HTML]{006100} 0.902} & \cellcolor[HTML]{C6EFCE}{\color[HTML]{006100} 0.772} & \cellcolor[HTML]{C6EFCE}{\color[HTML]{006100} 0.773} \\
FoVVideoVDP \cite{mantiuk2021fovvideovdp}     & \cellcolor[HTML]{FFEB9C}{\color[HTML]{9C5700} 0.801} & \cellcolor[HTML]{FFEB9C}{\color[HTML]{9C5700} 0.791} & \cellcolor[HTML]{C6EFCE}{\color[HTML]{006100} 0.84}  & \cellcolor[HTML]{C6EFCE}{\color[HTML]{006100} 0.855} & \cellcolor[HTML]{C6EFCE}{\color[HTML]{006100} 0.755} & \cellcolor[HTML]{FFC7CE}{\color[HTML]{9C0006} 0.741} & \cellcolor[HTML]{C6EFCE}{\color[HTML]{006100} 0.937} & \cellcolor[HTML]{C6EFCE}{\color[HTML]{006100} 0.963} & \cellcolor[HTML]{FFEB9C}{\color[HTML]{9C5700} 0.609} & \cellcolor[HTML]{FFEB9C}{\color[HTML]{9C5700} 0.59}  & \cellcolor[HTML]{C6EFCE}{\color[HTML]{006100} 0.942} & \cellcolor[HTML]{C6EFCE}{\color[HTML]{006100} 0.932} & \cellcolor[HTML]{FFEB9C}{\color[HTML]{9C5700} 0.633} & \cellcolor[HTML]{FFC7CE}{\color[HTML]{9C0006} 0.625} & \cellcolor[HTML]{C6EFCE}{\color[HTML]{006100} 0.859} & \cellcolor[HTML]{FFEB9C}{\color[HTML]{9C5700} 0.85}  & \cellcolor[HTML]{C6EFCE}{\color[HTML]{006100} 0.796} & \cellcolor[HTML]{C6EFCE}{\color[HTML]{006100} 0.807} \\
ColorVideoVDP \cite{mantiuk2024colorvideovdp} & \cellcolor[HTML]{C6EFCE}{\color[HTML]{006100} 0.872} & \cellcolor[HTML]{C6EFCE}{\color[HTML]{006100} 0.898} & \cellcolor[HTML]{C6EFCE}{\color[HTML]{006100} 0.775} & \cellcolor[HTML]{FFEB9C}{\color[HTML]{9C5700} 0.8}   & \cellcolor[HTML]{FFEB9C}{\color[HTML]{9C5700} 0.744} & \cellcolor[HTML]{FFEB9C}{\color[HTML]{9C5700} 0.775} & \cellcolor[HTML]{FFEB9C}{\color[HTML]{9C5700} 0.928} & \cellcolor[HTML]{C6EFCE}{\color[HTML]{006100} 0.966} & \cellcolor[HTML]{FFEB9C}{\color[HTML]{9C5700} 0.621} & \cellcolor[HTML]{FFEB9C}{\color[HTML]{9C5700} 0.598} & \cellcolor[HTML]{FFEB9C}{\color[HTML]{9C5700} 0.852} & \cellcolor[HTML]{FFC7CE}{\color[HTML]{9C0006} 0.813} & \cellcolor[HTML]{FFEB9C}{\color[HTML]{9C5700} 0.658} & \cellcolor[HTML]{FFEB9C}{\color[HTML]{9C5700} 0.651} & \cellcolor[HTML]{C6EFCE}{\color[HTML]{006100} 0.896} & \cellcolor[HTML]{C6EFCE}{\color[HTML]{006100} 0.9}   & \cellcolor[HTML]{C6EFCE}{\color[HTML]{006100} 0.769} & \cellcolor[HTML]{C6EFCE}{\color[HTML]{006100} 0.804} \\
CGVQM-2 (ours)                                & \cellcolor[HTML]{FFEB9C}{\color[HTML]{9C5700} 0.788} & \cellcolor[HTML]{FFEB9C}{\color[HTML]{9C5700} 0.791} & \cellcolor[HTML]{C6EFCE}{\color[HTML]{006100} 0.787} & \cellcolor[HTML]{C6EFCE}{\color[HTML]{006100} 0.82}  & \cellcolor[HTML]{C6EFCE}{\color[HTML]{006100} 0.855} & \cellcolor[HTML]{C6EFCE}{\color[HTML]{006100} 0.876} & \cellcolor[HTML]{FFEB9C}{\color[HTML]{9C5700} 0.911} & \cellcolor[HTML]{FFEB9C}{\color[HTML]{9C5700} 0.943} & \cellcolor[HTML]{FFEB9C}{\color[HTML]{9C5700} 0.539} & \cellcolor[HTML]{FFC7CE}{\color[HTML]{9C0006} 0.459} & \cellcolor[HTML]{FFEB9C}{\color[HTML]{9C5700} 0.922} & \cellcolor[HTML]{C6EFCE}{\color[HTML]{006100} 0.893} & \cellcolor[HTML]{FFEB9C}{\color[HTML]{9C5700} 0.62}  & \cellcolor[HTML]{FFC7CE}{\color[HTML]{9C0006} 0.623} & \cellcolor[HTML]{C6EFCE}{\color[HTML]{006100} 0.868} & \cellcolor[HTML]{C6EFCE}{\color[HTML]{006100} 0.884} & \cellcolor[HTML]{FFEB9C}{\color[HTML]{9C5700} 0.615} & \cellcolor[HTML]{C6EFCE}{\color[HTML]{006100} 0.671} \\
CGVQM-5 (ours)                                & \cellcolor[HTML]{C6EFCE}{\color[HTML]{006100} 0.88}  & \cellcolor[HTML]{C6EFCE}{\color[HTML]{006100} 0.898} & \cellcolor[HTML]{C6EFCE}{\color[HTML]{006100} 0.798} & \cellcolor[HTML]{C6EFCE}{\color[HTML]{006100} 0.824} & \cellcolor[HTML]{C6EFCE}{\color[HTML]{006100} 0.871} & \cellcolor[HTML]{C6EFCE}{\color[HTML]{006100} 0.877} & \cellcolor[HTML]{C6EFCE}{\color[HTML]{006100} 0.938} & \cellcolor[HTML]{C6EFCE}{\color[HTML]{006100} 0.957} & \cellcolor[HTML]{C6EFCE}{\color[HTML]{006100} 0.67}  & \cellcolor[HTML]{C6EFCE}{\color[HTML]{006100} 0.664} & \cellcolor[HTML]{FFEB9C}{\color[HTML]{9C5700} 0.918} & \cellcolor[HTML]{C6EFCE}{\color[HTML]{006100} 0.92}  & \cellcolor[HTML]{C6EFCE}{\color[HTML]{006100} 0.697} & \cellcolor[HTML]{FFEB9C}{\color[HTML]{9C5700} 0.698} & \cellcolor[HTML]{FFEB9C}{\color[HTML]{9C5700} 0.836} & \cellcolor[HTML]{FFEB9C}{\color[HTML]{9C5700} 0.859} & \cellcolor[HTML]{C6EFCE}{\color[HTML]{006100} 0.69}  & \cellcolor[HTML]{C6EFCE}{\color[HTML]{006100} 0.701} \\ \hline
\end{tabular}%
}
\end{table*}

%% file: secs/conclusion.tex
\section{Conclusions}
\label{sec:conclusions}

In this paper, we introduced the Computer Graphics Video Quality Dataset (CG-VQD) to address the limitations of existing video quality metrics in evaluating distortions from modern rendering techniques. Our dataset includes perceptual ratings for distortions such as neural supersampling, novel-view synthesis, path tracing, neural denoising, frame interpolation and variable rate shading, that are typical for modern graphics systems such as game engines and offline renderers. 

We demonstrated that existing full-reference metrics perform poorly on these distortions, while 3D CNN activations show better alignment with human perception and their performance depends on network architecture, pre-training, and calibration. 
We propose \metricname, a new 3D ResNet-based full-reference video quality metric, which outperforms current SOTA metrics, providing both localized error maps and global scores, making it suitable for quality evaluation in modern real-time computer graphics rendering applications.

Future work could explore enhancing the metric with additional features like optical flow and identifying optimal pre-training tasks for improved performance. Overall, our contributions lay a foundation for advancing video quality assessment in the realm of real-time computer graphics.

%% file: figs/distortions_all.tex
\begin{figure*}
    \centering
    \includegraphics[width=\linewidth]{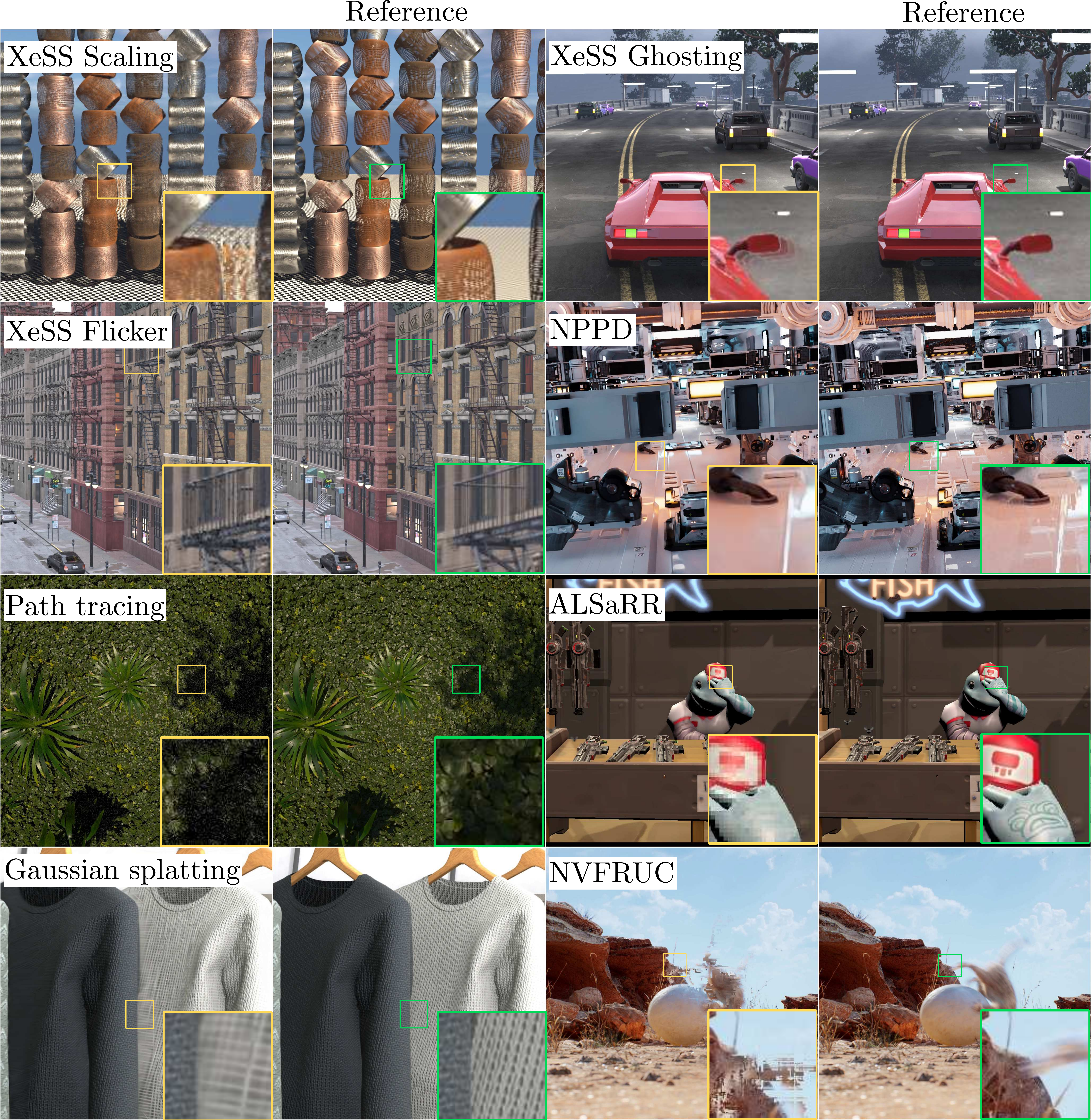}
    \caption{Examples of distortions arising from different rendering methods used in CG-VQD. Only highest levels of distortions are visualized in this figure. }
    \label{fig:distortions_all}
\end{figure*}

%% file: secs/appendix.tex
\section*{Appendix}

\renewcommand{\thefigure}{A\arabic{figure}}

\setcounter{figure}{0}

\setcounter{table}{0}
\renewcommand{\thetable}{A\arabic{table}}

\subsection{User study: Additional details}
\label{sec:userstudy-extra}



\begin{figure}[] 
    \centering
        \includegraphics[width=\linewidth]{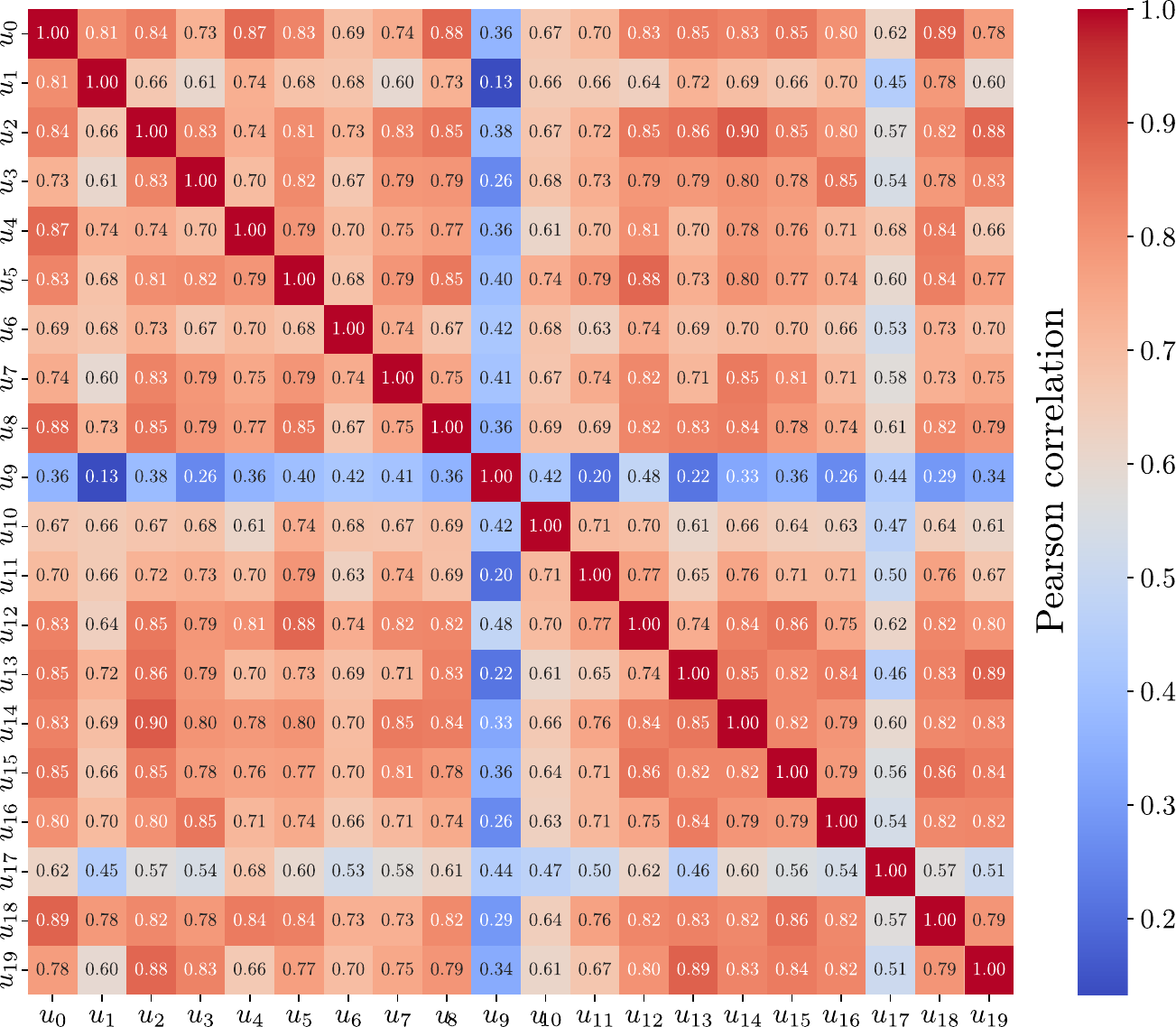}
    \caption{Inter-rater agreement analysis. We observed high correlation between ratings of each user except for $u_9$ and $u_{17}$.}
\label{fig:inter-rater-agreement}    
\end{figure}
\begin{figure}[] 
    \centering
        \includegraphics[width=\linewidth]{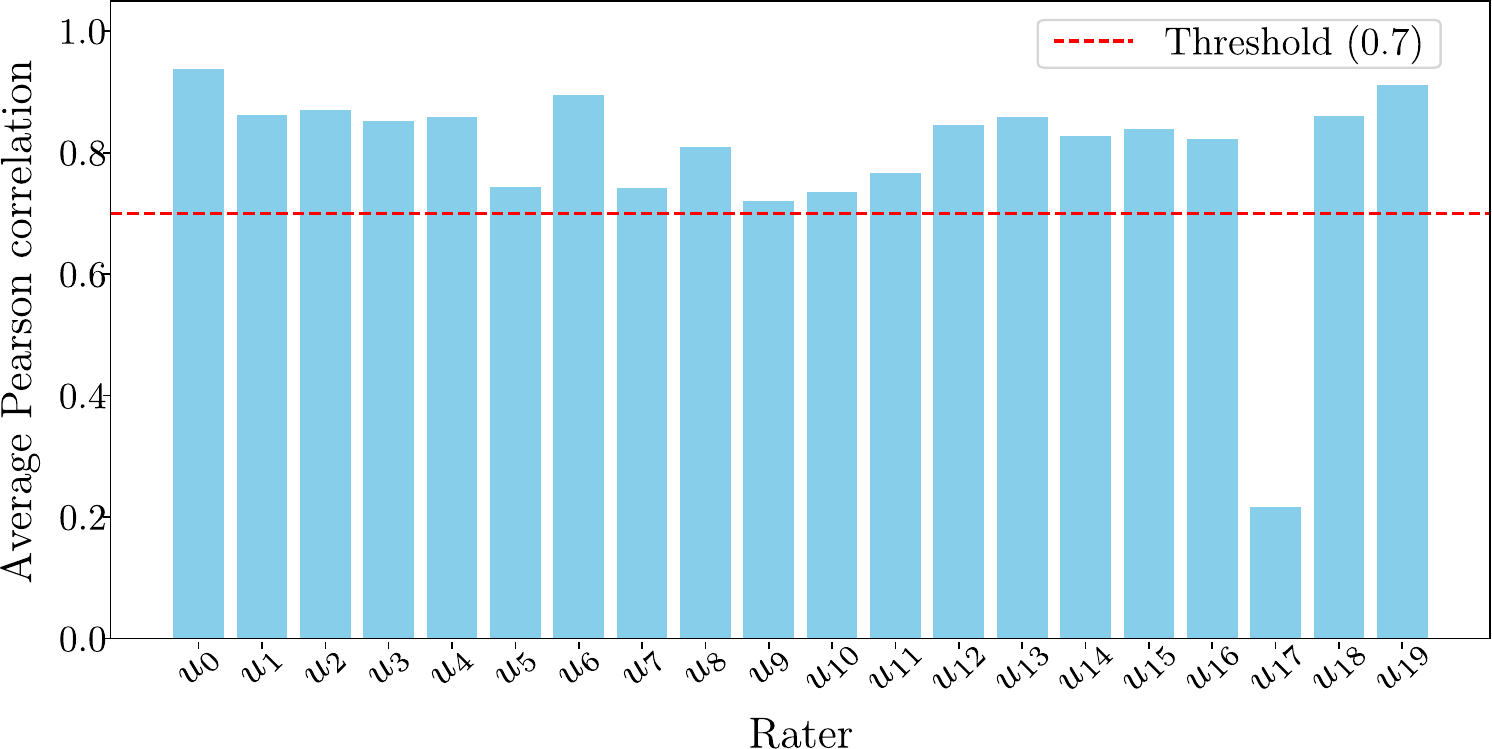}
    \caption{Intra-rater reliability analysis (average Pearson correlation across repetitions). We observed high consistency from all participants except for $u_{17}$.}
\label{fig:intra-rater-agreement}    
\end{figure}
\begin{figure}[] 
    \centering
        \includegraphics[width=\linewidth]{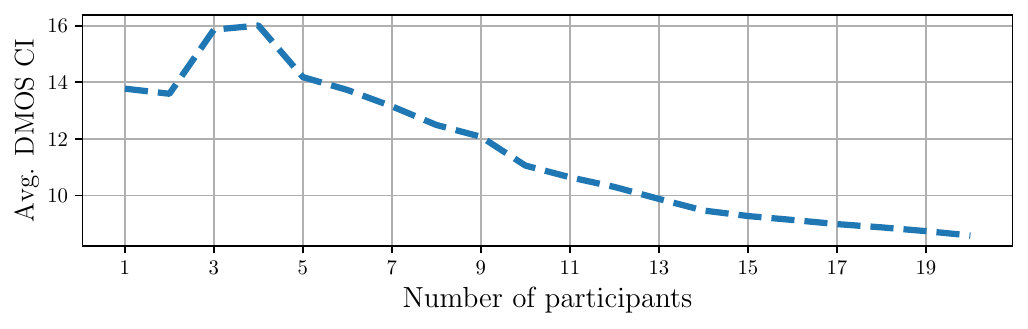}
        \includegraphics[width=\linewidth]{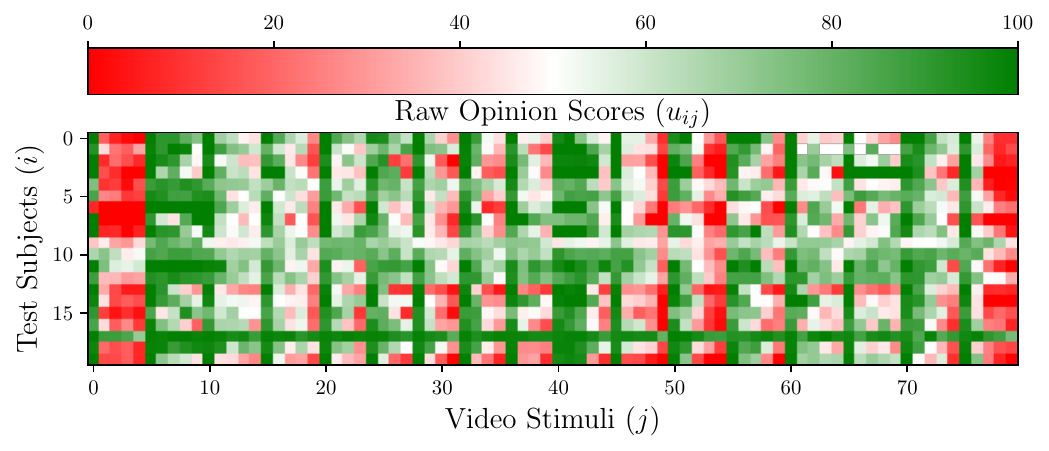}
        \includegraphics[width=\linewidth]{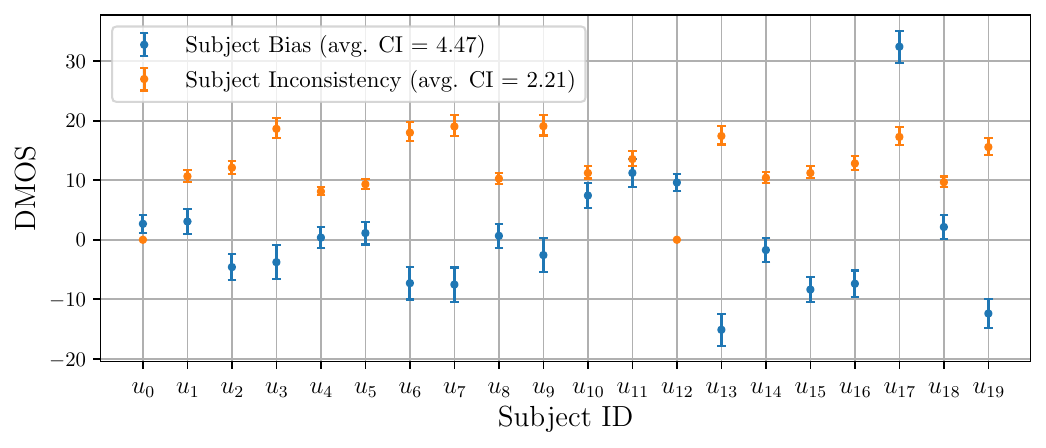}
        \includegraphics[width=\linewidth]{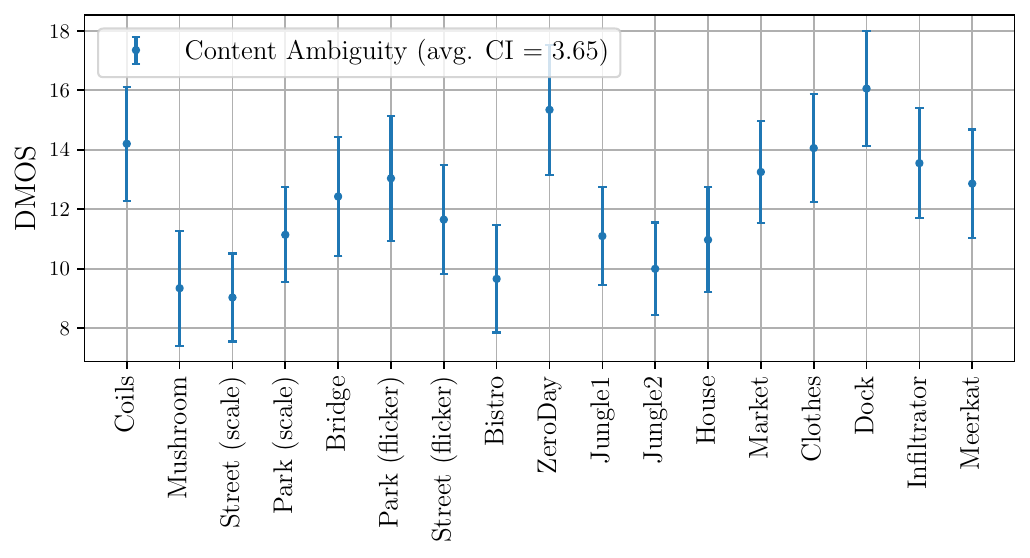}
    \caption{Additional statistical analysis on participants' data. Error bars denote 95\% confidence interval.}
\label{fig:userstudy}    
\end{figure}

\textbf{Data analysis.} To assess the quality and consistency of subjective ratings in the video quality evaluation study, we conducted several reliability analyses. We measured the intra-class correlation coefficient to be ICC (2, k) = 0.97 and ICC(3,k) = 0.98, indicating a high degree of agreement between the participants. Ratings were first averaged over three repetitions per video to stabilize individual responses. A heatmap of pairwise Pearson correlations between raters is visualized in Figure \ref{fig:inter-rater-agreement}, revealing generally high inter-rater consistency except for two participants ($u_9$ and $u_{17}$). We also evaluated intra-rater reliability by computing the average Pearson correlation between the repeated ratings of a participant for each video, capturing internal consistency over the three repetitions. Figure \ref{fig:intra-rater-agreement} visualizes these intra-rater correlations highlighting all raters (except $u_{17}$) consistently applied the same judgment. Figure \ref{fig:userstudy} shows the raw data, subject bias, inconsistency, and content ambiguity from our experiments. We iteratively apply Grubbs’ test to detect anomalous raters based on their average deviation from the group mean. One participant ($u_{17}$) was marked as outlier and excluded from our dataset. We found that 20 participants were sufficient for reliable results, as reduction in average DMOS CI per-participant decreased from 1.83 to 0.12 between 5\textsuperscript{th} and 20\textsuperscript{th} participant (Figure \ref{fig:userstudy} (top row)), with an overall fitted power law decay rate of 0.19, indicating saturation of error. These tests ensure that our raw subjective data is consistent and reliable and can be used for further quality analysis and model development.

To aggregate raw scores and determine the true quality ratings, we employed the maximum likelihood estimation method proposed by \cite{li2020simple}. This method offers an advantage over BT.500 standardized procedures by allowing ``soft" subject rejection. It jointly estimates the subjective quality of impaired videos, along with subject bias, consistency, and video content ambiguity, thereby maximizing the information extracted from the raw data.

\begin{figure*}[] 
    \centering
        \includegraphics[width=\linewidth]{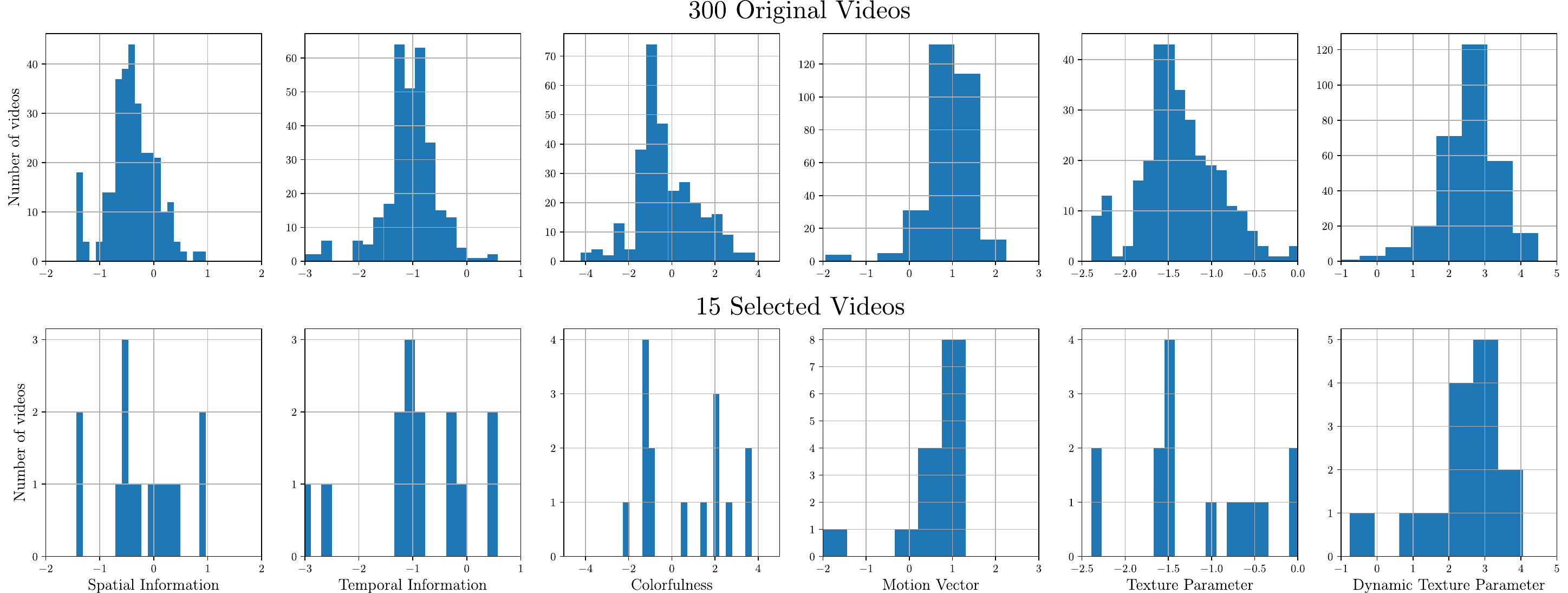}
    \caption{Visualizing data distribution of videos selected for user study and the larger dataset they were picked from. By minimizing KL-Divergence between the two distributions, we aimed for our selected videos to be representative of our larger dataset. X-axis is plotted in log scale.}
\label{fig:video_sampling}    
\end{figure*}

\textbf{Stimuli selection}
The 15 scenes used in our user study were selected from a larger in-house proprietary dataset comprising 300 scenes. Our goal was to choose a representative subset of 15 scenes to ensure that our experimental findings would generalize to the broader dataset. To achieve this, we rendered each scene into a video and parameterized each video using six features: spatial information (eq. 1 \cite{moss2015optimal}), temporal information (eq. 2 \cite{moss2015optimal}), colorfulness  (eq. 3 \cite{winkler2012analysis}), motion (eq. 4 \cite{winkler2012analysis}), texture parameter  (eq. 4 \cite{moss2015optimal}), dynamic texture parameter (eq. 5 \cite{moss2015optimal}). The distribution of these features, visualized in the first row of Figure \ref{fig:video_sampling}, generally followed a normal distribution.

To select the subset, we randomly sampled 15 videos from the dataset, treating the subsets (of size es 15 and 300) as samples from two distributions (selected and original). We then calculated the KL-divergence \cite{kullback1951information} between the 6-dimensional feature distributions of the two sets. This process was repeated 1,000,000 times, and we selected the subset with the minimum KL-divergence, which best matched the original distribution. The resulting distribution of the selected 15 videos is shown in the second row of Figure \ref{fig:video_sampling} and these videos were used in our user study.

\begin{figure}[] 
    \centering
        \includegraphics[width=\linewidth]{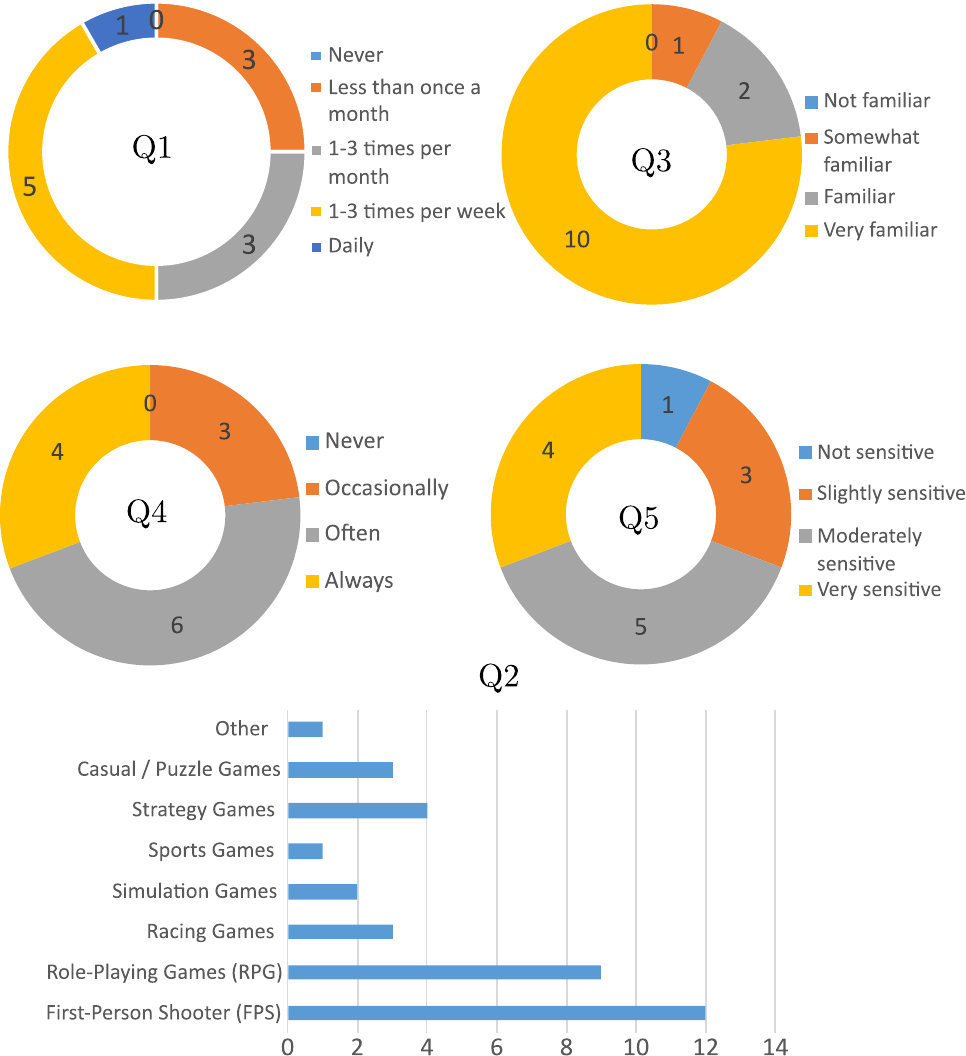}
    \caption{Post-experiment survey of participant's prior experience with graphics.}
\label{fig:survey}    
\end{figure}

\textbf{Participants experience}.
Before the study, all participants received a written briefing outlining the experimental task and the user interface. No additional training was provided. Following the user study, a voluntary survey was conducted to assess participants' prior experience with graphics content and artifacts. Thirteen out of twenty participants completed the survey. All participants reported normal or corrected-to-normal vision and no visual impairments. All but one participant indicated that they either own or regularly use gaming hardware (PC, console, or VR). Responses to the following five questions are summarized in Figure~\ref{fig:survey}:
\begin{enumerate}
\item[\textbf{Q1.}]    \textit{How often do you play video games?}
\item[\textbf{Q2.}]   \textit{What types of video games do you usually play? (Select all that apply)}
\item[\textbf{Q3.}]   \textit{How would you rate your familiarity with video game graphics quality and rendering technologies (e.g., anti-aliasing, ray tracing, upscaling)?}
\item[\textbf{Q4.}]   \textit{Have you ever noticed graphical issues or artifacts while playing games? (e.g., ghosting, aliasing, compression artifacts)}
\item[\textbf{Q5.}]   \textit{How sensitive do you think you are to visual quality issues in games?}
\end{enumerate}

\subsection{Metric Benchmark}
\begin{figure*}[] 
        \centering
        \includegraphics[width=\textwidth]{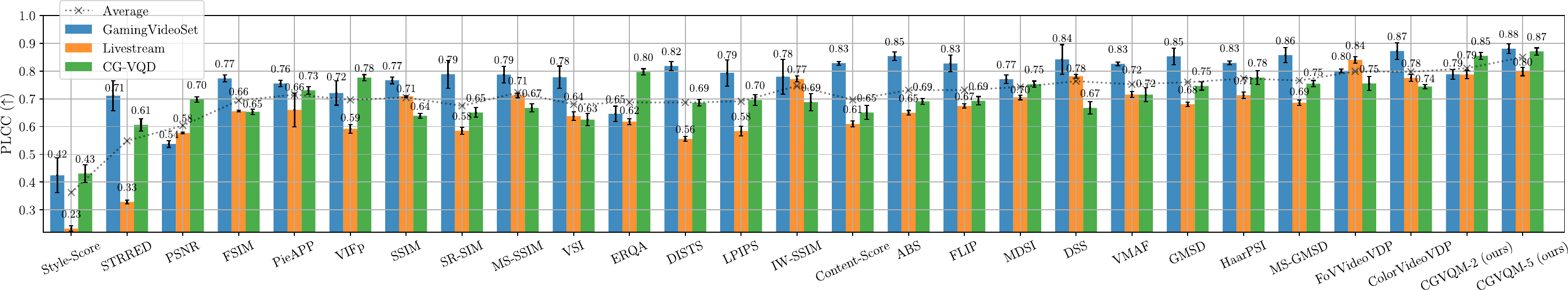}
        \includegraphics[width=\textwidth]{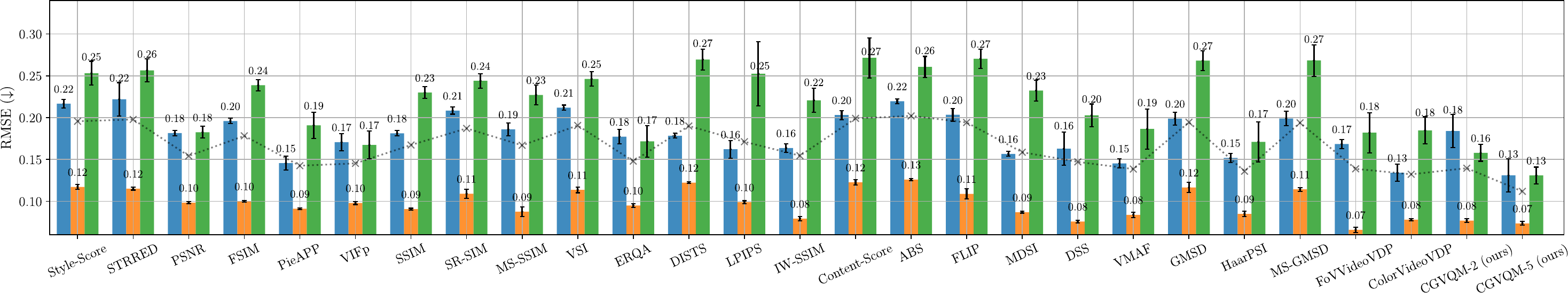}
        \caption{Comparison of quality metrics on 3 video quality datasets in terms of Pearson correlation (PLCC) and normalized Root Mean Square Error (RMSE). Error bars were generated via bootstrapping and denote 95\% confidence interval. Metrics are sorted based on PLCC value averaged across the 3 datasets.}
        \label{fig:benchmark_all}

\end{figure*}

\begin{figure*}[] 
        \centering
        \includegraphics[width=\textwidth]{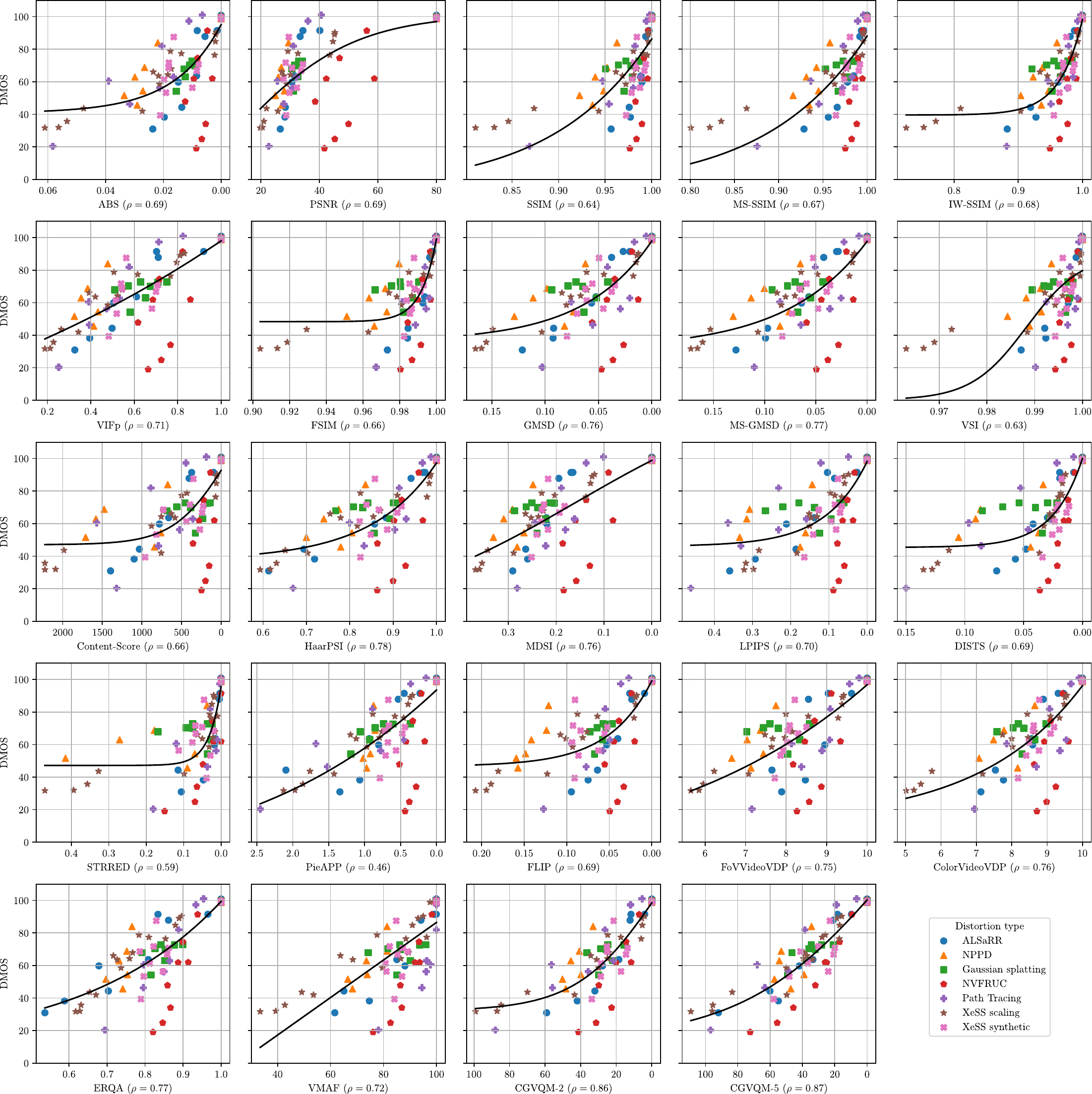}
        \caption{Comparison of human mean opinion scores against quality metric predictions on our CG-VQD dataset. $\rho$ denotes Pearson correlation.}
        \label{fig:bechmark_scatter_plot}
\end{figure*}

In Section 4.2, we compared our \metricname-5 and \metricname-2 metrics with existing full-reference quality metrics using PLCC and RMSE measures on our CG-VQD dataset. In this section, we provide additional comparisons on Livestream and GamingVideoSET datasets and significance tests on our metrics. Table \ref{tab:benchmark} summarizes the performance of all metrics on the three datasets using PLCC, SRCC, KRCC, and RMSE measures. Top-3 metrics in each column are highlighted with different shades of green. \metricname-5 consistently performs well on all 3 datasets. CGVQM-2, a lighter version of \metricname-5, performs similarly to \metricname-5 on Livestream and CG-VQD datasets but is worse on GamingVideoSET. Reducing the number of layers/features reduces \metricname's ability to generalize to different distortions. These observations are also confirmed in Table \ref{tab:benchmark-ptest} which shows that \metricname-5 is significantly better than most metrics on all datasets while \metricname-2 is only significantly better on Livestream and CG-VQD datasets. The average performance of each metric across the 3 datasets is shown in Figure \ref{fig:benchmark_all}. \metricname-5 and \metricname-2 score the highest average PLCC,  followed by ColorVideoVDP. Figure \ref{fig:bechmark_scatter_plot} shows
scatter plots of raw metric predictions  versus subjective difference mean opinion scores (DMOSs) on the CG-VQD database. From the fitted functions (Eq. (5)), one can observe that \metricname-5 is nearly linear in DMOS.

\input{tables/benchmark-all}
\input{tables/benchmark-ptest}

\subsection{3DCNN Experiments: Additional results}
\label{sec:exp-sigtest}
\begin{figure}
    \centering
    \includegraphics[width=\linewidth]{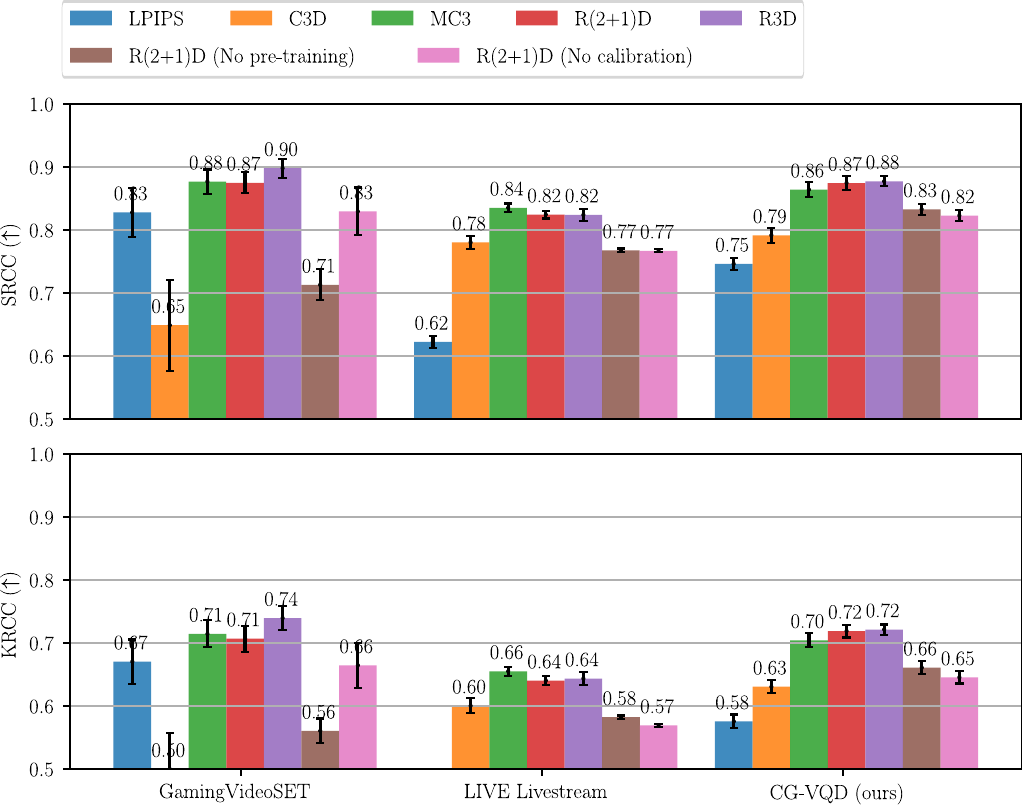}
    \caption{Comparison of different CNNs on 3 different video quality datasets using SRCC and KRCC measures.}
    \label{fig:experiments_extra}
\end{figure}

In Section 4.2, we compared the performance of different 3D-CNN architectures on video quality datasets. Here we provide additional results with SRCC and KRCC measures in Figure \ref{fig:experiments_extra} and significance tests in Table \ref{tab:arch-ptest}. 3D-ResNet performs the best on all three datasets while the type of 3D convolution does not have any significant effect on metric's performance. Significance testing was done for mean PLCC value using bootstrapping and one tailed paired t-test with significance level 0.05.

\input{tables/arch-ptest}

\subsection{CGVQD: Content and Distortions}
\label{sec:cgvqd-details}
We provide more examples of video sequences and distortions in our dataset in Figure \ref{fig:errormaps-all}. Frames from each video sequence were sampled uniformly to visualize temporal content. The videos in CG-VQD contains a wide variety of spatial and temporal content, motion, colors, and textures. The distortions include color bleeding, ghosting, moire, aliasing, noise, blur, flicker, tiling, and reconstruction errors. Such errors are often localized and adaptive to rendering parameters unlike traditional distortions.


\FloatBarrier
\begin{figure*}[htp] 
    \centering
    \begin{subfigure}[htp]{\textwidth}
        \includegraphics[width=\textwidth]{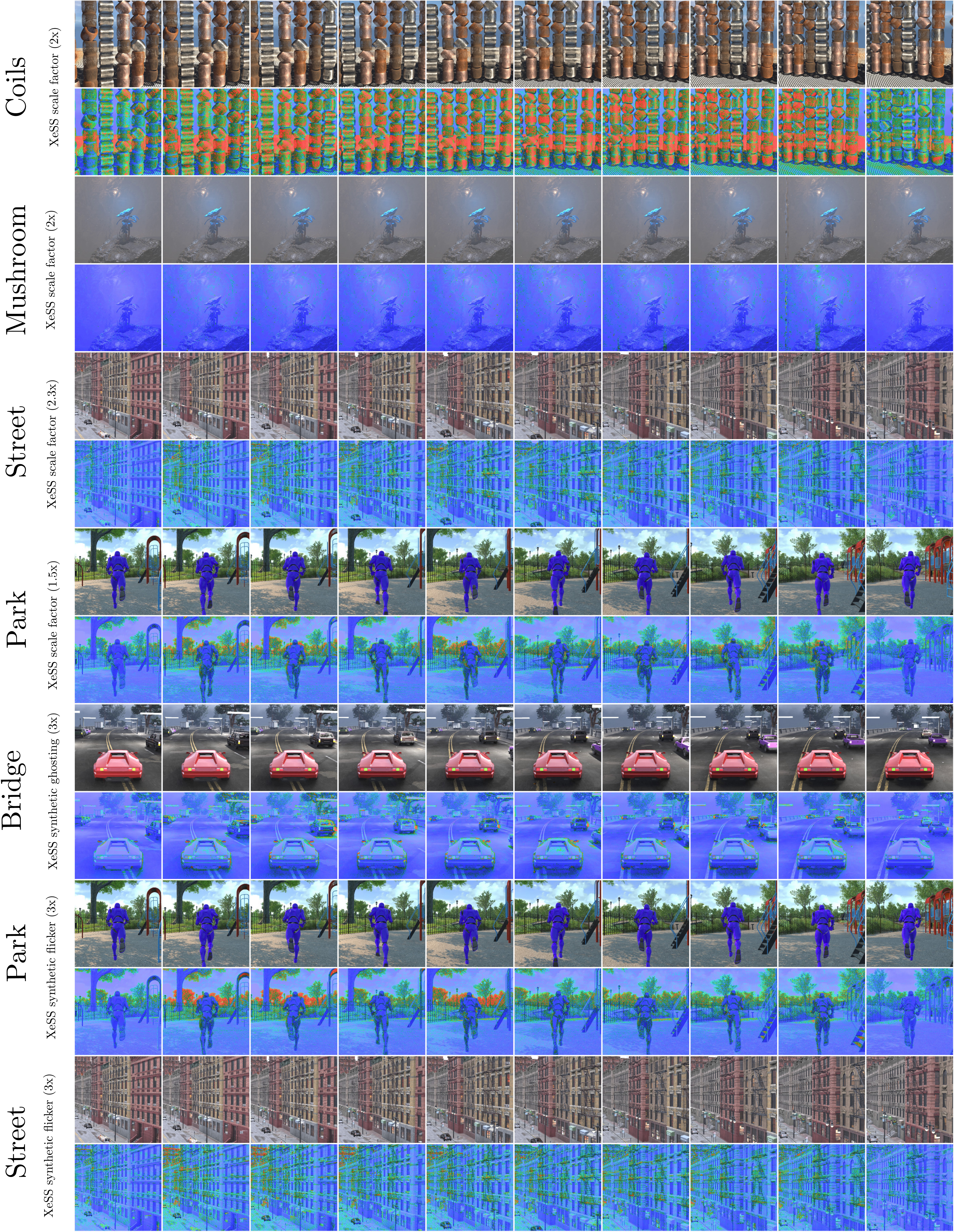}
    \end{subfigure}  
\end{figure*}
\begin{figure*}[htp]\ContinuedFloat
    \centering
    \begin{subfigure}[htp]{\textwidth}
        \includegraphics[width=\textwidth]{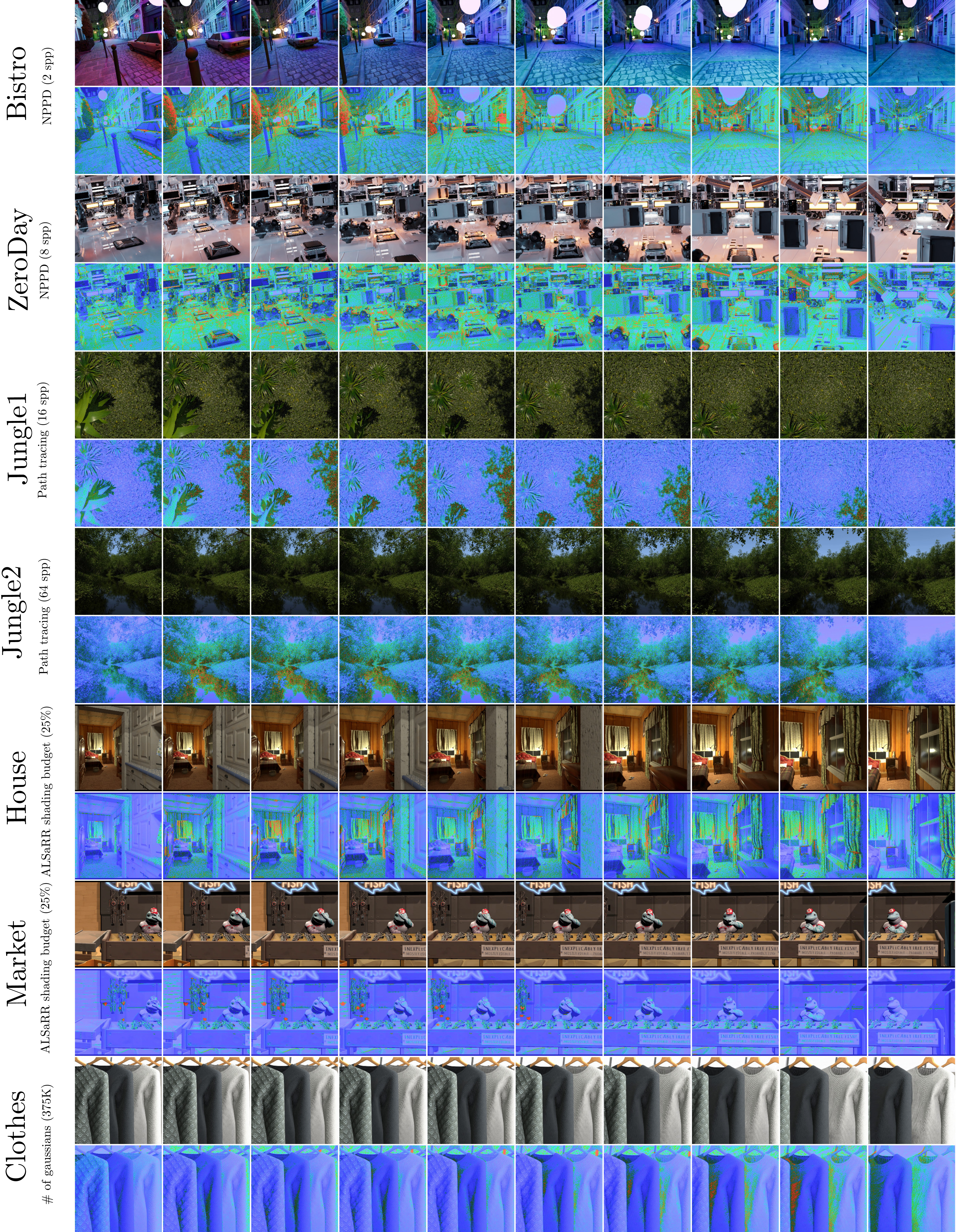}
    \end{subfigure}  
\end{figure*}
\begin{figure*}[htp]\ContinuedFloat
    \centering
    \begin{subfigure}[htp]{\textwidth}
        \includegraphics[width=\textwidth]{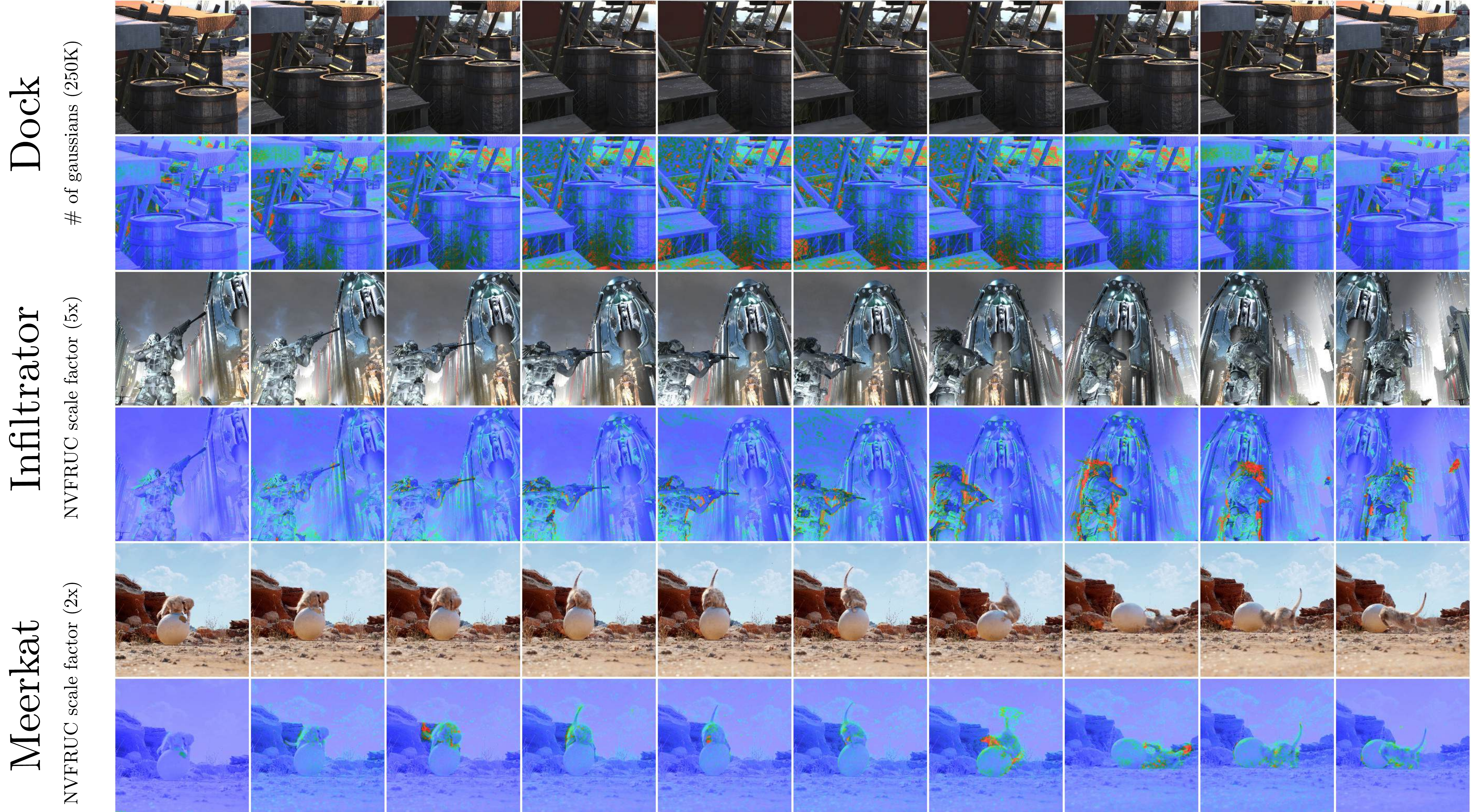}
    \end{subfigure}  
\caption{Examples of distorted videos from our dataset and their corresponding error maps generated by \metricname-2. The 10 frames were uniformly sampled from 3 second video clips.}
\label{fig:errormaps-all}
\end{figure*}

%% file: tables/benchmark-all.tex
\begin{landscape}
\begin{table*}[]
\centering
\caption{\Rev{Cross-dataset validation. *Results are reported on the test splits of the GamingVideoSET, LIVE Livestream, and CG-VQD datasets, as well as on additional datasets used exclusively for testing. Background colors represent performance quantiles across all tested metrics: green (top 25\%), yellow (25–75\%), and red (bottom 25\%). The model demonstrates strong generalization across a wide range of content types and distortion patterns.}}
\label{tab:benchmark}
\resizebox{\textwidth}{!}{%
\begin{tabular}{l|llll|llll|llll|llll|llll|llll|llll|llll|llll}
\hline
\multicolumn{1}{c|}{}                                                                       & \multicolumn{4}{c|}{GamingVideoDataset}                                                                                                                                                                                                                                                                                                  & \multicolumn{4}{c|}{LIVE LIVESTREAM}                                                                                                                                                                                                                                                                                                     & \multicolumn{4}{c|}{\textbf{CG-VQD (Ours)}}                                                                                                                                                                                                                                                                                              & \multicolumn{4}{c|}{LIVE Meta}                                                                                                                                                                                                                                                                                                           & \multicolumn{4}{c|}{CGVDS}                                                                                                                                                                                                                                                                                                               & \multicolumn{4}{c|}{LIVE Flicker}                                                                                                                                                                                                                                                                                                        & \multicolumn{4}{c|}{NVS}                                                                                                                                                                                                                                                                                                                 & \multicolumn{4}{c|}{AVT-VQDB-UHD-1}                                                                                                                                                                                                                                                                                                      & \multicolumn{4}{c}{BVI-HD}                                                                                                                                                                                                                                                                                                              \\ \cline{2-37} 
\multicolumn{1}{c|}{\multirow{-2}{*}{Metric}}                                               & \multicolumn{1}{c}{\begin{tabular}[c]{@{}c@{}}PLCC\\ ($\uparrow$)\end{tabular}} & \multicolumn{1}{c}{\begin{tabular}[c]{@{}c@{}}SRCC\\ ($\uparrow$)\end{tabular}} & \multicolumn{1}{c}{\begin{tabular}[c]{@{}c@{}}KRCC\\ ($\uparrow$)\end{tabular}} & \multicolumn{1}{c|}{\begin{tabular}[c]{@{}c@{}}RMSE\\ ($\downarrow$)\end{tabular}} & \multicolumn{1}{c}{\begin{tabular}[c]{@{}c@{}}PLCC\\ ($\uparrow$)\end{tabular}} & \multicolumn{1}{c}{\begin{tabular}[c]{@{}c@{}}SRCC\\ ($\uparrow$)\end{tabular}} & \multicolumn{1}{c}{\begin{tabular}[c]{@{}c@{}}KRCC\\ ($\uparrow$)\end{tabular}} & \multicolumn{1}{c|}{\begin{tabular}[c]{@{}c@{}}RMSE\\ ($\downarrow$)\end{tabular}} & \multicolumn{1}{c}{\begin{tabular}[c]{@{}c@{}}PLCC\\ ($\uparrow$)\end{tabular}} & \multicolumn{1}{c}{\begin{tabular}[c]{@{}c@{}}SRCC\\ ($\uparrow$)\end{tabular}} & \multicolumn{1}{c}{\begin{tabular}[c]{@{}c@{}}KRCC\\ ($\uparrow$)\end{tabular}} & \multicolumn{1}{c|}{\begin{tabular}[c]{@{}c@{}}RMSE\\ ($\downarrow$)\end{tabular}} & \multicolumn{1}{c}{\begin{tabular}[c]{@{}c@{}}PLCC\\ ($\uparrow$)\end{tabular}} & \multicolumn{1}{c}{\begin{tabular}[c]{@{}c@{}}SRCC\\ ($\uparrow$)\end{tabular}} & \multicolumn{1}{c}{\begin{tabular}[c]{@{}c@{}}KRCC\\ ($\uparrow$)\end{tabular}} & \multicolumn{1}{c|}{\begin{tabular}[c]{@{}c@{}}RMSE\\ ($\downarrow$)\end{tabular}} & \multicolumn{1}{c}{\begin{tabular}[c]{@{}c@{}}PLCC\\ ($\uparrow$)\end{tabular}} & \multicolumn{1}{c}{\begin{tabular}[c]{@{}c@{}}SRCC\\ ($\uparrow$)\end{tabular}} & \multicolumn{1}{c}{\begin{tabular}[c]{@{}c@{}}KRCC\\ ($\uparrow$)\end{tabular}} & \multicolumn{1}{c|}{\begin{tabular}[c]{@{}c@{}}RMSE\\ ($\downarrow$)\end{tabular}} & \multicolumn{1}{c}{\begin{tabular}[c]{@{}c@{}}PLCC\\ ($\uparrow$)\end{tabular}} & \multicolumn{1}{c}{\begin{tabular}[c]{@{}c@{}}SRCC\\ ($\uparrow$)\end{tabular}} & \multicolumn{1}{c}{\begin{tabular}[c]{@{}c@{}}KRCC\\ ($\uparrow$)\end{tabular}} & \multicolumn{1}{c|}{\begin{tabular}[c]{@{}c@{}}RMSE\\ ($\downarrow$)\end{tabular}} & \multicolumn{1}{c}{\begin{tabular}[c]{@{}c@{}}PLCC\\ ($\uparrow$)\end{tabular}} & \multicolumn{1}{c}{\begin{tabular}[c]{@{}c@{}}SRCC\\ ($\uparrow$)\end{tabular}} & \multicolumn{1}{c}{\begin{tabular}[c]{@{}c@{}}KRCC\\ ($\uparrow$)\end{tabular}} & \multicolumn{1}{c|}{\begin{tabular}[c]{@{}c@{}}RMSE\\ ($\downarrow$)\end{tabular}} & \multicolumn{1}{c}{\begin{tabular}[c]{@{}c@{}}PLCC\\ ($\uparrow$)\end{tabular}} & \multicolumn{1}{c}{\begin{tabular}[c]{@{}c@{}}SRCC\\ ($\uparrow$)\end{tabular}} & \multicolumn{1}{c}{\begin{tabular}[c]{@{}c@{}}KRCC\\ ($\uparrow$)\end{tabular}} & \multicolumn{1}{c|}{\begin{tabular}[c]{@{}c@{}}RMSE\\ ($\downarrow$)\end{tabular}} & \multicolumn{1}{c}{\begin{tabular}[c]{@{}c@{}}PLCC\\ ($\uparrow$)\end{tabular}} & \multicolumn{1}{c}{\begin{tabular}[c]{@{}c@{}}SRCC\\ ($\uparrow$)\end{tabular}} & \multicolumn{1}{c}{\begin{tabular}[c]{@{}c@{}}KRCC\\ ($\uparrow$)\end{tabular}} & \multicolumn{1}{c}{\begin{tabular}[c]{@{}c@{}}RMSE\\ ($\downarrow$)\end{tabular}} \\ \hline
ABS                                                                                         & \cellcolor[HTML]{C6EFCE}{\color[HTML]{006100} 0.853}                            & \cellcolor[HTML]{C6EFCE}{\color[HTML]{006100} 0.911}                            & \cellcolor[HTML]{C6EFCE}{\color[HTML]{006100} 0.753}                            & \cellcolor[HTML]{FFC7CE}{\color[HTML]{9C0006} 0.878}                               & \cellcolor[HTML]{FFEB9C}{\color[HTML]{9C5700} 0.649}                            & \cellcolor[HTML]{FFEB9C}{\color[HTML]{9C5700} 0.67}                             & \cellcolor[HTML]{FFEB9C}{\color[HTML]{9C5700} 0.498}                            & \cellcolor[HTML]{FFC7CE}{\color[HTML]{9C0006} 12.597}                              & \cellcolor[HTML]{FFEB9C}{\color[HTML]{9C5700} 0.691}                            & \cellcolor[HTML]{FFEB9C}{\color[HTML]{9C5700} 0.744}                            & \cellcolor[HTML]{FFC7CE}{\color[HTML]{9C0006} 0.576}                            & \cellcolor[HTML]{FFC7CE}{\color[HTML]{9C0006} 26.068}                              & \cellcolor[HTML]{FFEB9C}{\color[HTML]{9C5700} 0.896}                            & \cellcolor[HTML]{FFC7CE}{\color[HTML]{9C0006} 0.926}                            & \cellcolor[HTML]{FFC7CE}{\color[HTML]{9C0006} 0.78}                             & \cellcolor[HTML]{FFC7CE}{\color[HTML]{9C0006} 18.479}                              & \cellcolor[HTML]{FFC7CE}{\color[HTML]{9C0006} 0.452}                            & \cellcolor[HTML]{FFC7CE}{\color[HTML]{9C0006} 0.51}                             & \cellcolor[HTML]{FFC7CE}{\color[HTML]{9C0006} 0.375}                            & \cellcolor[HTML]{FFC7CE}{\color[HTML]{9C0006} 0.881}                               & \cellcolor[HTML]{FFEB9C}{\color[HTML]{9C5700} 0.888}                            & \cellcolor[HTML]{FFEB9C}{\color[HTML]{9C5700} 0.869}                            & \cellcolor[HTML]{FFEB9C}{\color[HTML]{9C5700} 0.694}                            & \cellcolor[HTML]{FFC7CE}{\color[HTML]{9C0006} 20.003}                              & \cellcolor[HTML]{FFC7CE}{\color[HTML]{9C0006} 0.492}                            & \cellcolor[HTML]{FFEB9C}{\color[HTML]{9C5700} 0.633}                            & \cellcolor[HTML]{FFEB9C}{\color[HTML]{9C5700} 0.464}                            & \cellcolor[HTML]{FFC7CE}{\color[HTML]{9C0006} 1.011}                               & \cellcolor[HTML]{FFEB9C}{\color[HTML]{9C5700} 0.768}                            & \cellcolor[HTML]{FFEB9C}{\color[HTML]{9C5700} 0.798}                            & \cellcolor[HTML]{FFEB9C}{\color[HTML]{9C5700} 0.631}                            & \cellcolor[HTML]{FFC7CE}{\color[HTML]{9C0006} 0.973}                               & \cellcolor[HTML]{FFC7CE}{\color[HTML]{9C0006} 0.416}                            & \cellcolor[HTML]{FFC7CE}{\color[HTML]{9C0006} 0.422}                            & \cellcolor[HTML]{FFC7CE}{\color[HTML]{9C0006} 0.278}                            & \cellcolor[HTML]{FFC7CE}{\color[HTML]{9C0006} 16.171}                             \\
PSNR                                                                                        & \cellcolor[HTML]{FFC7CE}{\color[HTML]{9C0006} 0.537}                            & \cellcolor[HTML]{FFC7CE}{\color[HTML]{9C0006} 0.775}                            & \cellcolor[HTML]{FFC7CE}{\color[HTML]{9C0006} 0.589}                            & \cellcolor[HTML]{FFEB9C}{\color[HTML]{9C5700} 0.727}                               & \cellcolor[HTML]{FFC7CE}{\color[HTML]{9C0006} 0.577}                            & \cellcolor[HTML]{FFC7CE}{\color[HTML]{9C0006} 0.627}                            & \cellcolor[HTML]{FFC7CE}{\color[HTML]{9C0006} 0.459}                            & \cellcolor[HTML]{FFEB9C}{\color[HTML]{9C5700} 9.847}                               & \cellcolor[HTML]{FFEB9C}{\color[HTML]{9C5700} 0.697}                            & \cellcolor[HTML]{FFC7CE}{\color[HTML]{9C0006} 0.723}                            & \cellcolor[HTML]{FFC7CE}{\color[HTML]{9C0006} 0.57}                             & \cellcolor[HTML]{FFEB9C}{\color[HTML]{9C5700} 18.262}                              & \cellcolor[HTML]{FFC7CE}{\color[HTML]{9C0006} 0.671}                            & \cellcolor[HTML]{FFC7CE}{\color[HTML]{9C0006} 0.928}                            & \cellcolor[HTML]{FFC7CE}{\color[HTML]{9C0006} 0.779}                            & \cellcolor[HTML]{FFC7CE}{\color[HTML]{9C0006} 15}                                  & \cellcolor[HTML]{FFEB9C}{\color[HTML]{9C5700} 0.514}                            & \cellcolor[HTML]{FFC7CE}{\color[HTML]{9C0006} 0.519}                            & \cellcolor[HTML]{FFC7CE}{\color[HTML]{9C0006} 0.378}                            & \cellcolor[HTML]{FFEB9C}{\color[HTML]{9C5700} 0.836}                               & \cellcolor[HTML]{FFC7CE}{\color[HTML]{9C0006} 0.662}                            & \cellcolor[HTML]{FFC7CE}{\color[HTML]{9C0006} 0.65}                             & \cellcolor[HTML]{FFC7CE}{\color[HTML]{9C0006} 0.457}                            & \cellcolor[HTML]{FFEB9C}{\color[HTML]{9C5700} 15.752}                              & \cellcolor[HTML]{FFEB9C}{\color[HTML]{9C5700} 0.64}                             & \cellcolor[HTML]{FFEB9C}{\color[HTML]{9C5700} 0.675}                            & \cellcolor[HTML]{FFEB9C}{\color[HTML]{9C5700} 0.503}                            & \cellcolor[HTML]{FFEB9C}{\color[HTML]{9C5700} 0.891}                               & \cellcolor[HTML]{FFEB9C}{\color[HTML]{9C5700} 0.791}                            & \cellcolor[HTML]{FFEB9C}{\color[HTML]{9C5700} 0.822}                            & \cellcolor[HTML]{FFEB9C}{\color[HTML]{9C5700} 0.653}                            & \cellcolor[HTML]{FFEB9C}{\color[HTML]{9C5700} 0.668}                               & \cellcolor[HTML]{FFEB9C}{\color[HTML]{9C5700} 0.471}                            & \cellcolor[HTML]{FFC7CE}{\color[HTML]{9C0006} 0.465}                            & \cellcolor[HTML]{FFC7CE}{\color[HTML]{9C0006} 0.315}                            & \cellcolor[HTML]{FFEB9C}{\color[HTML]{9C5700} 14.998}                             \\
\begin{tabular}[c]{@{}l@{}}SSIM\\ \cite{wang2004image}\end{tabular}                         & \cellcolor[HTML]{FFEB9C}{\color[HTML]{9C5700} 0.766}                            & \cellcolor[HTML]{FFEB9C}{\color[HTML]{9C5700} 0.787}                            & \cellcolor[HTML]{FFEB9C}{\color[HTML]{9C5700} 0.632}                            & \cellcolor[HTML]{FFEB9C}{\color[HTML]{9C5700} 0.725}                               & \cellcolor[HTML]{FFEB9C}{\color[HTML]{9C5700} 0.709}                            & \cellcolor[HTML]{C6EFCE}{\color[HTML]{006100} 0.834}                            & \cellcolor[HTML]{C6EFCE}{\color[HTML]{006100} 0.646}                            & \cellcolor[HTML]{FFEB9C}{\color[HTML]{9C5700} 9.076}                               & \cellcolor[HTML]{FFC7CE}{\color[HTML]{9C0006} 0.639}                            & \cellcolor[HTML]{C6EFCE}{\color[HTML]{006100} 0.805}                            & \cellcolor[HTML]{C6EFCE}{\color[HTML]{006100} 0.627}                            & \cellcolor[HTML]{FFEB9C}{\color[HTML]{9C5700} 22.989}                              & \cellcolor[HTML]{FFEB9C}{\color[HTML]{9C5700} 0.838}                            & \cellcolor[HTML]{FFC7CE}{\color[HTML]{9C0006} 0.909}                            & \cellcolor[HTML]{FFC7CE}{\color[HTML]{9C0006} 0.749}                            & \cellcolor[HTML]{FFEB9C}{\color[HTML]{9C5700} 11.385}                              & \cellcolor[HTML]{FFC7CE}{\color[HTML]{9C0006} 0.472}                            & \cellcolor[HTML]{FFEB9C}{\color[HTML]{9C5700} 0.53}                             & \cellcolor[HTML]{FFEB9C}{\color[HTML]{9C5700} 0.394}                            & \cellcolor[HTML]{FFC7CE}{\color[HTML]{9C0006} 0.859}                               & \cellcolor[HTML]{FFEB9C}{\color[HTML]{9C5700} 0.85}                             & \cellcolor[HTML]{FFEB9C}{\color[HTML]{9C5700} 0.867}                            & \cellcolor[HTML]{FFEB9C}{\color[HTML]{9C5700} 0.688}                            & \cellcolor[HTML]{FFEB9C}{\color[HTML]{9C5700} 13.318}                              & \cellcolor[HTML]{FFEB9C}{\color[HTML]{9C5700} 0.611}                            & \cellcolor[HTML]{FFEB9C}{\color[HTML]{9C5700} 0.635}                            & \cellcolor[HTML]{FFEB9C}{\color[HTML]{9C5700} 0.456}                            & \cellcolor[HTML]{FFEB9C}{\color[HTML]{9C5700} 0.92}                                & \cellcolor[HTML]{FFEB9C}{\color[HTML]{9C5700} 0.739}                            & \cellcolor[HTML]{FFEB9C}{\color[HTML]{9C5700} 0.786}                            & \cellcolor[HTML]{FFEB9C}{\color[HTML]{9C5700} 0.598}                            & \cellcolor[HTML]{FFEB9C}{\color[HTML]{9C5700} 0.735}                               & \cellcolor[HTML]{FFEB9C}{\color[HTML]{9C5700} 0.524}                            & \cellcolor[HTML]{FFEB9C}{\color[HTML]{9C5700} 0.621}                            & \cellcolor[HTML]{FFEB9C}{\color[HTML]{9C5700} 0.447}                            & \cellcolor[HTML]{FFEB9C}{\color[HTML]{9C5700} 14.484}                             \\
\begin{tabular}[c]{@{}l@{}}MS-SSIM\\ \cite{wang2003multiscale}\end{tabular}                 & \cellcolor[HTML]{FFEB9C}{\color[HTML]{9C5700} 0.787}                            & \cellcolor[HTML]{FFEB9C}{\color[HTML]{9C5700} 0.826}                            & \cellcolor[HTML]{FFEB9C}{\color[HTML]{9C5700} 0.67}                             & \cellcolor[HTML]{FFEB9C}{\color[HTML]{9C5700} 0.744}                               & \cellcolor[HTML]{FFEB9C}{\color[HTML]{9C5700} 0.712}                            & \cellcolor[HTML]{FFEB9C}{\color[HTML]{9C5700} 0.748}                            & \cellcolor[HTML]{FFEB9C}{\color[HTML]{9C5700} 0.561}                            & \cellcolor[HTML]{FFEB9C}{\color[HTML]{9C5700} 8.769}                               & \cellcolor[HTML]{FFEB9C}{\color[HTML]{9C5700} 0.668}                            & \cellcolor[HTML]{C6EFCE}{\color[HTML]{006100} 0.804}                            & \cellcolor[HTML]{C6EFCE}{\color[HTML]{006100} 0.627}                            & \cellcolor[HTML]{FFEB9C}{\color[HTML]{9C5700} 22.716}                              & \cellcolor[HTML]{FFEB9C}{\color[HTML]{9C5700} 0.872}                            & \cellcolor[HTML]{FFEB9C}{\color[HTML]{9C5700} 0.946}                            & \cellcolor[HTML]{FFEB9C}{\color[HTML]{9C5700} 0.813}                            & \cellcolor[HTML]{FFEB9C}{\color[HTML]{9C5700} 12.625}                              & \cellcolor[HTML]{FFC7CE}{\color[HTML]{9C0006} 0.462}                            & \cellcolor[HTML]{FFEB9C}{\color[HTML]{9C5700} 0.524}                            & \cellcolor[HTML]{FFEB9C}{\color[HTML]{9C5700} 0.389}                            & \cellcolor[HTML]{FFC7CE}{\color[HTML]{9C0006} 0.864}                               & \cellcolor[HTML]{FFC7CE}{\color[HTML]{9C0006} 0.832}                            & \cellcolor[HTML]{FFEB9C}{\color[HTML]{9C5700} 0.864}                            & \cellcolor[HTML]{FFEB9C}{\color[HTML]{9C5700} 0.678}                            & \cellcolor[HTML]{FFEB9C}{\color[HTML]{9C5700} 15.315}                              & \cellcolor[HTML]{FFEB9C}{\color[HTML]{9C5700} 0.619}                            & \cellcolor[HTML]{FFEB9C}{\color[HTML]{9C5700} 0.655}                            & \cellcolor[HTML]{FFEB9C}{\color[HTML]{9C5700} 0.475}                            & \cellcolor[HTML]{FFEB9C}{\color[HTML]{9C5700} 0.912}                               & \cellcolor[HTML]{FFEB9C}{\color[HTML]{9C5700} 0.767}                            & \cellcolor[HTML]{FFEB9C}{\color[HTML]{9C5700} 0.814}                            & \cellcolor[HTML]{FFEB9C}{\color[HTML]{9C5700} 0.643}                            & \cellcolor[HTML]{FFEB9C}{\color[HTML]{9C5700} 0.701}                               & \cellcolor[HTML]{FFEB9C}{\color[HTML]{9C5700} 0.513}                            & \cellcolor[HTML]{FFEB9C}{\color[HTML]{9C5700} 0.602}                            & \cellcolor[HTML]{FFEB9C}{\color[HTML]{9C5700} 0.427}                            & \cellcolor[HTML]{FFEB9C}{\color[HTML]{9C5700} 14.6}                               \\
\begin{tabular}[c]{@{}l@{}}IW-SSIM\\ \cite{wang2010information}\end{tabular}                & \cellcolor[HTML]{FFEB9C}{\color[HTML]{9C5700} 0.779}                            & \cellcolor[HTML]{FFC7CE}{\color[HTML]{9C0006} 0.779}                            & \cellcolor[HTML]{FFEB9C}{\color[HTML]{9C5700} 0.623}                            & \cellcolor[HTML]{FFEB9C}{\color[HTML]{9C5700} 0.655}                               & \cellcolor[HTML]{C6EFCE}{\color[HTML]{006100} 0.772}                            & \cellcolor[HTML]{C6EFCE}{\color[HTML]{006100} 0.805}                            & \cellcolor[HTML]{FFEB9C}{\color[HTML]{9C5700} 0.616}                            & \cellcolor[HTML]{C6EFCE}{\color[HTML]{006100} 7.933}                               & \cellcolor[HTML]{FFEB9C}{\color[HTML]{9C5700} 0.688}                            & \cellcolor[HTML]{C6EFCE}{\color[HTML]{006100} 0.837}                            & \cellcolor[HTML]{C6EFCE}{\color[HTML]{006100} 0.653}                            & \cellcolor[HTML]{FFEB9C}{\color[HTML]{9C5700} 22.063}                              & \cellcolor[HTML]{FFEB9C}{\color[HTML]{9C5700} 0.86}                             & \cellcolor[HTML]{FFEB9C}{\color[HTML]{9C5700} 0.955}                            & \cellcolor[HTML]{FFEB9C}{\color[HTML]{9C5700} 0.821}                            & \cellcolor[HTML]{FFEB9C}{\color[HTML]{9C5700} 11.153}                              & \cellcolor[HTML]{FFEB9C}{\color[HTML]{9C5700} 0.492}                            & \cellcolor[HTML]{FFEB9C}{\color[HTML]{9C5700} 0.54}                             & \cellcolor[HTML]{FFEB9C}{\color[HTML]{9C5700} 0.401}                            & \cellcolor[HTML]{FFEB9C}{\color[HTML]{9C5700} 0.848}                               & \cellcolor[HTML]{FFC7CE}{\color[HTML]{9C0006} 0.845}                            & \cellcolor[HTML]{FFEB9C}{\color[HTML]{9C5700} 0.874}                            & \cellcolor[HTML]{FFEB9C}{\color[HTML]{9C5700} 0.701}                            & \cellcolor[HTML]{FFEB9C}{\color[HTML]{9C5700} 13.641}                              & \cellcolor[HTML]{FFEB9C}{\color[HTML]{9C5700} 0.671}                            & \cellcolor[HTML]{FFEB9C}{\color[HTML]{9C5700} 0.668}                            & \cellcolor[HTML]{FFEB9C}{\color[HTML]{9C5700} 0.499}                            & \cellcolor[HTML]{FFEB9C}{\color[HTML]{9C5700} 0.861}                               & \cellcolor[HTML]{FFEB9C}{\color[HTML]{9C5700} 0.823}                            & \cellcolor[HTML]{FFEB9C}{\color[HTML]{9C5700} 0.855}                            & \cellcolor[HTML]{FFEB9C}{\color[HTML]{9C5700} 0.684}                            & \cellcolor[HTML]{FFEB9C}{\color[HTML]{9C5700} 0.62}                                & \cellcolor[HTML]{FFEB9C}{\color[HTML]{9C5700} 0.565}                            & \cellcolor[HTML]{FFEB9C}{\color[HTML]{9C5700} 0.64}                             & \cellcolor[HTML]{FFEB9C}{\color[HTML]{9C5700} 0.458}                            & \cellcolor[HTML]{FFEB9C}{\color[HTML]{9C5700} 14.028}                             \\
\begin{tabular}[c]{@{}l@{}}VIFp\\ \cite{sheikh2006image}\end{tabular}                       & \cellcolor[HTML]{FFC7CE}{\color[HTML]{9C0006} 0.721}                            & \cellcolor[HTML]{FFEB9C}{\color[HTML]{9C5700} 0.78}                             & \cellcolor[HTML]{FFEB9C}{\color[HTML]{9C5700} 0.624}                            & \cellcolor[HTML]{FFEB9C}{\color[HTML]{9C5700} 0.683}                               & \cellcolor[HTML]{FFEB9C}{\color[HTML]{9C5700} 0.592}                            & \cellcolor[HTML]{FFC7CE}{\color[HTML]{9C0006} 0.627}                            & \cellcolor[HTML]{FFC7CE}{\color[HTML]{9C0006} 0.46}                             & \cellcolor[HTML]{FFEB9C}{\color[HTML]{9C5700} 9.807}                               & \cellcolor[HTML]{C6EFCE}{\color[HTML]{006100} 0.776}                            & \cellcolor[HTML]{FFEB9C}{\color[HTML]{9C5700} 0.775}                            & \cellcolor[HTML]{C6EFCE}{\color[HTML]{006100} 0.617}                            & \cellcolor[HTML]{C6EFCE}{\color[HTML]{006100} 16.737}                              & \cellcolor[HTML]{FFEB9C}{\color[HTML]{9C5700} 0.909}                            & \cellcolor[HTML]{FFEB9C}{\color[HTML]{9C5700} 0.941}                            & \cellcolor[HTML]{FFEB9C}{\color[HTML]{9C5700} 0.802}                            & \cellcolor[HTML]{FFEB9C}{\color[HTML]{9C5700} 8.395}                               & \cellcolor[HTML]{C6EFCE}{\color[HTML]{006100} 0.673}                            & \cellcolor[HTML]{C6EFCE}{\color[HTML]{006100} 0.686}                            & \cellcolor[HTML]{C6EFCE}{\color[HTML]{006100} 0.515}                            & \cellcolor[HTML]{C6EFCE}{\color[HTML]{006100} 0.72}                                & \cellcolor[HTML]{FFEB9C}{\color[HTML]{9C5700} 0.857}                            & \cellcolor[HTML]{FFEB9C}{\color[HTML]{9C5700} 0.859}                            & \cellcolor[HTML]{FFEB9C}{\color[HTML]{9C5700} 0.681}                            & \cellcolor[HTML]{FFEB9C}{\color[HTML]{9C5700} 10.83}                               & \cellcolor[HTML]{FFEB9C}{\color[HTML]{9C5700} 0.582}                            & \cellcolor[HTML]{FFEB9C}{\color[HTML]{9C5700} 0.639}                            & \cellcolor[HTML]{FFEB9C}{\color[HTML]{9C5700} 0.472}                            & \cellcolor[HTML]{FFEB9C}{\color[HTML]{9C5700} 0.944}                               & \cellcolor[HTML]{FFEB9C}{\color[HTML]{9C5700} 0.83}                             & \cellcolor[HTML]{FFEB9C}{\color[HTML]{9C5700} 0.827}                            & \cellcolor[HTML]{FFEB9C}{\color[HTML]{9C5700} 0.65}                             & \cellcolor[HTML]{FFEB9C}{\color[HTML]{9C5700} 0.609}                               & \cellcolor[HTML]{C6EFCE}{\color[HTML]{006100} 0.633}                            & \cellcolor[HTML]{FFEB9C}{\color[HTML]{9C5700} 0.622}                            & \cellcolor[HTML]{FFEB9C}{\color[HTML]{9C5700} 0.444}                            & \cellcolor[HTML]{C6EFCE}{\color[HTML]{006100} 13.164}                             \\
\begin{tabular}[c]{@{}l@{}}FSIM\\ \cite{zhang2011fsim}\end{tabular}                         & \cellcolor[HTML]{FFEB9C}{\color[HTML]{9C5700} 0.773}                            & \cellcolor[HTML]{FFC7CE}{\color[HTML]{9C0006} 0.767}                            & \cellcolor[HTML]{FFC7CE}{\color[HTML]{9C0006} 0.616}                            & \cellcolor[HTML]{FFEB9C}{\color[HTML]{9C5700} 0.784}                               & \cellcolor[HTML]{FFEB9C}{\color[HTML]{9C5700} 0.655}                            & \cellcolor[HTML]{FFEB9C}{\color[HTML]{9C5700} 0.804}                            & \cellcolor[HTML]{FFEB9C}{\color[HTML]{9C5700} 0.621}                            & \cellcolor[HTML]{FFEB9C}{\color[HTML]{9C5700} 10.013}                              & \cellcolor[HTML]{FFEB9C}{\color[HTML]{9C5700} 0.653}                            & \cellcolor[HTML]{FFEB9C}{\color[HTML]{9C5700} 0.789}                            & \cellcolor[HTML]{FFEB9C}{\color[HTML]{9C5700} 0.611}                            & \cellcolor[HTML]{FFEB9C}{\color[HTML]{9C5700} 23.87}                               & \cellcolor[HTML]{FFEB9C}{\color[HTML]{9C5700} 0.856}                            & \cellcolor[HTML]{FFEB9C}{\color[HTML]{9C5700} 0.936}                            & \cellcolor[HTML]{FFEB9C}{\color[HTML]{9C5700} 0.792}                            & \cellcolor[HTML]{FFEB9C}{\color[HTML]{9C5700} 13.441}                              & \cellcolor[HTML]{FFEB9C}{\color[HTML]{9C5700} 0.504}                            & \cellcolor[HTML]{FFEB9C}{\color[HTML]{9C5700} 0.545}                            & \cellcolor[HTML]{FFEB9C}{\color[HTML]{9C5700} 0.402}                            & \cellcolor[HTML]{FFEB9C}{\color[HTML]{9C5700} 0.841}                               & \cellcolor[HTML]{FFEB9C}{\color[HTML]{9C5700} 0.889}                            & \cellcolor[HTML]{FFEB9C}{\color[HTML]{9C5700} 0.881}                            & \cellcolor[HTML]{FFEB9C}{\color[HTML]{9C5700} 0.717}                            & \cellcolor[HTML]{FFC7CE}{\color[HTML]{9C0006} 18.001}                              & \cellcolor[HTML]{FFC7CE}{\color[HTML]{9C0006} 0.499}                            & \cellcolor[HTML]{FFC7CE}{\color[HTML]{9C0006} 0.625}                            & \cellcolor[HTML]{FFC7CE}{\color[HTML]{9C0006} 0.449}                            & \cellcolor[HTML]{FFC7CE}{\color[HTML]{9C0006} 1.006}                               & \cellcolor[HTML]{FFEB9C}{\color[HTML]{9C5700} 0.724}                            & \cellcolor[HTML]{FFEB9C}{\color[HTML]{9C5700} 0.841}                            & \cellcolor[HTML]{FFEB9C}{\color[HTML]{9C5700} 0.659}                            & \cellcolor[HTML]{FFEB9C}{\color[HTML]{9C5700} 0.764}                               & \cellcolor[HTML]{FFEB9C}{\color[HTML]{9C5700} 0.521}                            & \cellcolor[HTML]{FFEB9C}{\color[HTML]{9C5700} 0.638}                            & \cellcolor[HTML]{FFEB9C}{\color[HTML]{9C5700} 0.457}                            & \cellcolor[HTML]{FFEB9C}{\color[HTML]{9C5700} 14.673}                             \\
\begin{tabular}[c]{@{}l@{}}SR-SIM\\ \cite{zhang2012sr}\end{tabular}                         & \cellcolor[HTML]{FFEB9C}{\color[HTML]{9C5700} 0.789}                            & \cellcolor[HTML]{FFEB9C}{\color[HTML]{9C5700} 0.8}                              & \cellcolor[HTML]{FFEB9C}{\color[HTML]{9C5700} 0.657}                            & \cellcolor[HTML]{FFC7CE}{\color[HTML]{9C0006} 0.834}                               & \cellcolor[HTML]{FFC7CE}{\color[HTML]{9C0006} 0.585}                            & \cellcolor[HTML]{FFEB9C}{\color[HTML]{9C5700} 0.793}                            & \cellcolor[HTML]{FFEB9C}{\color[HTML]{9C5700} 0.607}                            & \cellcolor[HTML]{FFEB9C}{\color[HTML]{9C5700} 10.911}                              & \cellcolor[HTML]{FFC7CE}{\color[HTML]{9C0006} 0.651}                            & \cellcolor[HTML]{FFEB9C}{\color[HTML]{9C5700} 0.783}                            & \cellcolor[HTML]{FFEB9C}{\color[HTML]{9C5700} 0.604}                            & \cellcolor[HTML]{FFEB9C}{\color[HTML]{9C5700} 24.398}                              & \cellcolor[HTML]{FFC7CE}{\color[HTML]{9C0006} 0.837}                            & \cellcolor[HTML]{FFEB9C}{\color[HTML]{9C5700} 0.937}                            & \cellcolor[HTML]{FFEB9C}{\color[HTML]{9C5700} 0.793}                            & \cellcolor[HTML]{FFC7CE}{\color[HTML]{9C0006} 17.275}                              & \cellcolor[HTML]{FFEB9C}{\color[HTML]{9C5700} 0.485}                            & \cellcolor[HTML]{FFEB9C}{\color[HTML]{9C5700} 0.538}                            & \cellcolor[HTML]{FFEB9C}{\color[HTML]{9C5700} 0.395}                            & \cellcolor[HTML]{FFEB9C}{\color[HTML]{9C5700} 0.852}                               & \cellcolor[HTML]{FFEB9C}{\color[HTML]{9C5700} 0.877}                            & \cellcolor[HTML]{FFEB9C}{\color[HTML]{9C5700} 0.862}                            & \cellcolor[HTML]{FFEB9C}{\color[HTML]{9C5700} 0.672}                            & \cellcolor[HTML]{FFC7CE}{\color[HTML]{9C0006} 19.893}                              & \cellcolor[HTML]{FFEB9C}{\color[HTML]{9C5700} 0.631}                            & \cellcolor[HTML]{FFEB9C}{\color[HTML]{9C5700} 0.659}                            & \cellcolor[HTML]{FFEB9C}{\color[HTML]{9C5700} 0.491}                            & \cellcolor[HTML]{FFEB9C}{\color[HTML]{9C5700} 0.9}                                 & \cellcolor[HTML]{FFC7CE}{\color[HTML]{9C0006} 0.68}                             & \cellcolor[HTML]{FFEB9C}{\color[HTML]{9C5700} 0.805}                            & \cellcolor[HTML]{FFEB9C}{\color[HTML]{9C5700} 0.622}                            & \cellcolor[HTML]{FFEB9C}{\color[HTML]{9C5700} 0.918}                               & \cellcolor[HTML]{FFEB9C}{\color[HTML]{9C5700} 0.551}                            & \cellcolor[HTML]{FFEB9C}{\color[HTML]{9C5700} 0.642}                            & \cellcolor[HTML]{FFEB9C}{\color[HTML]{9C5700} 0.462}                            & \cellcolor[HTML]{FFC7CE}{\color[HTML]{9C0006} 15.635}                             \\
\begin{tabular}[c]{@{}l@{}}GMSD\\ \cite{xue2013gradient}\end{tabular}                       & \cellcolor[HTML]{C6EFCE}{\color[HTML]{006100} 0.853}                            & \cellcolor[HTML]{FFEB9C}{\color[HTML]{9C5700} 0.86}                             & \cellcolor[HTML]{C6EFCE}{\color[HTML]{006100} 0.689}                            & \cellcolor[HTML]{FFEB9C}{\color[HTML]{9C5700} 0.795}                               & \cellcolor[HTML]{FFEB9C}{\color[HTML]{9C5700} 0.68}                             & \cellcolor[HTML]{FFEB9C}{\color[HTML]{9C5700} 0.726}                            & \cellcolor[HTML]{FFEB9C}{\color[HTML]{9C5700} 0.556}                            & \cellcolor[HTML]{FFC7CE}{\color[HTML]{9C0006} 11.657}                              & \cellcolor[HTML]{FFEB9C}{\color[HTML]{9C5700} 0.746}                            & \cellcolor[HTML]{FFEB9C}{\color[HTML]{9C5700} 0.767}                            & \cellcolor[HTML]{FFEB9C}{\color[HTML]{9C5700} 0.594}                            & \cellcolor[HTML]{FFC7CE}{\color[HTML]{9C0006} 26.808}                              & \cellcolor[HTML]{FFEB9C}{\color[HTML]{9C5700} 0.917}                            & \cellcolor[HTML]{FFEB9C}{\color[HTML]{9C5700} 0.955}                            & \cellcolor[HTML]{C6EFCE}{\color[HTML]{006100} 0.829}                            & \cellcolor[HTML]{FFEB9C}{\color[HTML]{9C5700} 8.066}                               & \cellcolor[HTML]{FFEB9C}{\color[HTML]{9C5700} 0.584}                            & \cellcolor[HTML]{FFEB9C}{\color[HTML]{9C5700} 0.581}                            & \cellcolor[HTML]{FFEB9C}{\color[HTML]{9C5700} 0.427}                            & \cellcolor[HTML]{FFEB9C}{\color[HTML]{9C5700} 0.79}                                & \cellcolor[HTML]{C6EFCE}{\color[HTML]{006100} 0.948}                            & \cellcolor[HTML]{FFEB9C}{\color[HTML]{9C5700} 0.867}                            & \cellcolor[HTML]{FFEB9C}{\color[HTML]{9C5700} 0.704}                            & \cellcolor[HTML]{C6EFCE}{\color[HTML]{006100} 6.705}                               & \cellcolor[HTML]{FFEB9C}{\color[HTML]{9C5700} 0.672}                            & \cellcolor[HTML]{FFEB9C}{\color[HTML]{9C5700} 0.656}                            & \cellcolor[HTML]{FFEB9C}{\color[HTML]{9C5700} 0.487}                            & \cellcolor[HTML]{FFEB9C}{\color[HTML]{9C5700} 0.859}                               & \cellcolor[HTML]{C6EFCE}{\color[HTML]{006100} 0.852}                            & \cellcolor[HTML]{C6EFCE}{\color[HTML]{006100} 0.864}                            & \cellcolor[HTML]{C6EFCE}{\color[HTML]{006100} 0.694}                            & \cellcolor[HTML]{C6EFCE}{\color[HTML]{006100} 0.571}                               & \cellcolor[HTML]{FFEB9C}{\color[HTML]{9C5700} 0.608}                            & \cellcolor[HTML]{FFEB9C}{\color[HTML]{9C5700} 0.635}                            & \cellcolor[HTML]{FFEB9C}{\color[HTML]{9C5700} 0.46}                             & \cellcolor[HTML]{FFEB9C}{\color[HTML]{9C5700} 13.497}                             \\
\begin{tabular}[c]{@{}l@{}}MS-GMSD\\ \cite{zhang2017gradient}\end{tabular}                  & \cellcolor[HTML]{C6EFCE}{\color[HTML]{006100} 0.857}                            & \cellcolor[HTML]{C6EFCE}{\color[HTML]{006100} 0.866}                            & \cellcolor[HTML]{C6EFCE}{\color[HTML]{006100} 0.695}                            & \cellcolor[HTML]{FFEB9C}{\color[HTML]{9C5700} 0.796}                               & \cellcolor[HTML]{FFEB9C}{\color[HTML]{9C5700} 0.686}                            & \cellcolor[HTML]{FFEB9C}{\color[HTML]{9C5700} 0.726}                            & \cellcolor[HTML]{FFEB9C}{\color[HTML]{9C5700} 0.555}                            & \cellcolor[HTML]{FFEB9C}{\color[HTML]{9C5700} 11.429}                              & \cellcolor[HTML]{FFEB9C}{\color[HTML]{9C5700} 0.754}                            & \cellcolor[HTML]{FFEB9C}{\color[HTML]{9C5700} 0.767}                            & \cellcolor[HTML]{FFEB9C}{\color[HTML]{9C5700} 0.598}                            & \cellcolor[HTML]{FFC7CE}{\color[HTML]{9C0006} 26.828}                              & \cellcolor[HTML]{FFEB9C}{\color[HTML]{9C5700} 0.918}                            & \cellcolor[HTML]{FFEB9C}{\color[HTML]{9C5700} 0.953}                            & \cellcolor[HTML]{FFEB9C}{\color[HTML]{9C5700} 0.825}                            & \cellcolor[HTML]{FFEB9C}{\color[HTML]{9C5700} 9.933}                               & \cellcolor[HTML]{FFEB9C}{\color[HTML]{9C5700} 0.574}                            & \cellcolor[HTML]{FFEB9C}{\color[HTML]{9C5700} 0.573}                            & \cellcolor[HTML]{FFEB9C}{\color[HTML]{9C5700} 0.421}                            & \cellcolor[HTML]{FFEB9C}{\color[HTML]{9C5700} 0.797}                               & \cellcolor[HTML]{C6EFCE}{\color[HTML]{006100} 0.945}                            & \cellcolor[HTML]{FFEB9C}{\color[HTML]{9C5700} 0.882}                            & \cellcolor[HTML]{FFEB9C}{\color[HTML]{9C5700} 0.72}                             & \cellcolor[HTML]{FFEB9C}{\color[HTML]{9C5700} 11.235}                              & \cellcolor[HTML]{FFEB9C}{\color[HTML]{9C5700} 0.652}                            & \cellcolor[HTML]{FFEB9C}{\color[HTML]{9C5700} 0.676}                            & \cellcolor[HTML]{FFEB9C}{\color[HTML]{9C5700} 0.51}                             & \cellcolor[HTML]{FFEB9C}{\color[HTML]{9C5700} 0.888}                               & \cellcolor[HTML]{FFEB9C}{\color[HTML]{9C5700} 0.848}                            & \cellcolor[HTML]{C6EFCE}{\color[HTML]{006100} 0.863}                            & \cellcolor[HTML]{C6EFCE}{\color[HTML]{006100} 0.695}                            & \cellcolor[HTML]{FFEB9C}{\color[HTML]{9C5700} 0.579}                               & \cellcolor[HTML]{FFEB9C}{\color[HTML]{9C5700} 0.604}                            & \cellcolor[HTML]{FFEB9C}{\color[HTML]{9C5700} 0.627}                            & \cellcolor[HTML]{FFEB9C}{\color[HTML]{9C5700} 0.453}                            & \cellcolor[HTML]{FFEB9C}{\color[HTML]{9C5700} 13.554}                             \\
\begin{tabular}[c]{@{}l@{}}VSI\\ \cite{zhang2014vsi}\end{tabular}                           & \cellcolor[HTML]{FFEB9C}{\color[HTML]{9C5700} 0.778}                            & \cellcolor[HTML]{FFEB9C}{\color[HTML]{9C5700} 0.81}                             & \cellcolor[HTML]{FFEB9C}{\color[HTML]{9C5700} 0.67}                             & \cellcolor[HTML]{FFC7CE}{\color[HTML]{9C0006} 0.848}                               & \cellcolor[HTML]{FFEB9C}{\color[HTML]{9C5700} 0.637}                            & \cellcolor[HTML]{C6EFCE}{\color[HTML]{006100} 0.816}                            & \cellcolor[HTML]{C6EFCE}{\color[HTML]{006100} 0.632}                            & \cellcolor[HTML]{FFEB9C}{\color[HTML]{9C5700} 11.369}                              & \cellcolor[HTML]{FFC7CE}{\color[HTML]{9C0006} 0.625}                            & \cellcolor[HTML]{FFEB9C}{\color[HTML]{9C5700} 0.789}                            & \cellcolor[HTML]{FFEB9C}{\color[HTML]{9C5700} 0.607}                            & \cellcolor[HTML]{FFEB9C}{\color[HTML]{9C5700} 24.63}                               & \cellcolor[HTML]{FFC7CE}{\color[HTML]{9C0006} 0.832}                            & \cellcolor[HTML]{FFEB9C}{\color[HTML]{9C5700} 0.93}                             & \cellcolor[HTML]{FFEB9C}{\color[HTML]{9C5700} 0.782}                            & \cellcolor[HTML]{FFC7CE}{\color[HTML]{9C0006} 18.647}                              & \cellcolor[HTML]{FFC7CE}{\color[HTML]{9C0006} 0.461}                            & \cellcolor[HTML]{FFC7CE}{\color[HTML]{9C0006} 0.517}                            & \cellcolor[HTML]{FFC7CE}{\color[HTML]{9C0006} 0.379}                            & \cellcolor[HTML]{FFC7CE}{\color[HTML]{9C0006} 0.875}                               & \cellcolor[HTML]{FFEB9C}{\color[HTML]{9C5700} 0.902}                            & \cellcolor[HTML]{FFEB9C}{\color[HTML]{9C5700} 0.862}                            & \cellcolor[HTML]{FFEB9C}{\color[HTML]{9C5700} 0.688}                            & \cellcolor[HTML]{FFC7CE}{\color[HTML]{9C0006} 20.413}                              & \cellcolor[HTML]{FFC7CE}{\color[HTML]{9C0006} 0.522}                            & \cellcolor[HTML]{FFEB9C}{\color[HTML]{9C5700} 0.67}                             & \cellcolor[HTML]{FFEB9C}{\color[HTML]{9C5700} 0.497}                            & \cellcolor[HTML]{FFC7CE}{\color[HTML]{9C0006} 0.99}                                & \cellcolor[HTML]{FFEB9C}{\color[HTML]{9C5700} 0.7}                              & \cellcolor[HTML]{FFEB9C}{\color[HTML]{9C5700} 0.845}                            & \cellcolor[HTML]{FFEB9C}{\color[HTML]{9C5700} 0.673}                            & \cellcolor[HTML]{FFC7CE}{\color[HTML]{9C0006} 1.045}                               & \cellcolor[HTML]{FFEB9C}{\color[HTML]{9C5700} 0.536}                            & \cellcolor[HTML]{FFEB9C}{\color[HTML]{9C5700} 0.631}                            & \cellcolor[HTML]{FFEB9C}{\color[HTML]{9C5700} 0.446}                            & \cellcolor[HTML]{FFC7CE}{\color[HTML]{9C0006} 16.335}                             \\
\begin{tabular}[c]{@{}l@{}}DSS\\ \cite{balanov2015image}\end{tabular}                       & \cellcolor[HTML]{C6EFCE}{\color[HTML]{006100} 0.842}                            & \cellcolor[HTML]{FFEB9C}{\color[HTML]{9C5700} 0.831}                            & \cellcolor[HTML]{FFEB9C}{\color[HTML]{9C5700} 0.673}                            & \cellcolor[HTML]{FFEB9C}{\color[HTML]{9C5700} 0.652}                               & \cellcolor[HTML]{C6EFCE}{\color[HTML]{006100} 0.781}                            & \cellcolor[HTML]{FFEB9C}{\color[HTML]{9C5700} 0.8}                              & \cellcolor[HTML]{FFEB9C}{\color[HTML]{9C5700} 0.614}                            & \cellcolor[HTML]{C6EFCE}{\color[HTML]{006100} 7.595}                               & \cellcolor[HTML]{FFEB9C}{\color[HTML]{9C5700} 0.667}                            & \cellcolor[HTML]{FFEB9C}{\color[HTML]{9C5700} 0.772}                            & \cellcolor[HTML]{FFEB9C}{\color[HTML]{9C5700} 0.59}                             & \cellcolor[HTML]{FFEB9C}{\color[HTML]{9C5700} 20.277}                              & \cellcolor[HTML]{FFEB9C}{\color[HTML]{9C5700} 0.907}                            & \cellcolor[HTML]{C6EFCE}{\color[HTML]{006100} 0.957}                            & \cellcolor[HTML]{FFEB9C}{\color[HTML]{9C5700} 0.823}                            & \cellcolor[HTML]{FFEB9C}{\color[HTML]{9C5700} 8.509}                               & \cellcolor[HTML]{FFEB9C}{\color[HTML]{9C5700} 0.635}                            & \cellcolor[HTML]{FFEB9C}{\color[HTML]{9C5700} 0.637}                            & \cellcolor[HTML]{FFEB9C}{\color[HTML]{9C5700} 0.471}                            & \cellcolor[HTML]{FFEB9C}{\color[HTML]{9C5700} 0.753}                               & \cellcolor[HTML]{C6EFCE}{\color[HTML]{006100} 0.943}                            & \cellcolor[HTML]{C6EFCE}{\color[HTML]{006100} 0.92}                             & \cellcolor[HTML]{C6EFCE}{\color[HTML]{006100} 0.779}                            & \cellcolor[HTML]{C6EFCE}{\color[HTML]{006100} 7.01}                                & \cellcolor[HTML]{C6EFCE}{\color[HTML]{006100} 0.764}                            & \cellcolor[HTML]{C6EFCE}{\color[HTML]{006100} 0.758}                            & \cellcolor[HTML]{C6EFCE}{\color[HTML]{006100} 0.581}                            & \cellcolor[HTML]{C6EFCE}{\color[HTML]{006100} 0.748}                               & \cellcolor[HTML]{C6EFCE}{\color[HTML]{006100} 0.87}                             & \cellcolor[HTML]{C6EFCE}{\color[HTML]{006100} 0.879}                            & \cellcolor[HTML]{C6EFCE}{\color[HTML]{006100} 0.714}                            & \cellcolor[HTML]{C6EFCE}{\color[HTML]{006100} 0.539}                               & \cellcolor[HTML]{FFEB9C}{\color[HTML]{9C5700} 0.618}                            & \cellcolor[HTML]{C6EFCE}{\color[HTML]{006100} 0.67}                             & \cellcolor[HTML]{C6EFCE}{\color[HTML]{006100} 0.489}                            & \cellcolor[HTML]{FFEB9C}{\color[HTML]{9C5700} 13.36}                              \\
\begin{tabular}[c]{@{}l@{}}Content-\\ Score\\ \cite{gatys2015neural}\end{tabular}           & \cellcolor[HTML]{FFEB9C}{\color[HTML]{9C5700} 0.828}                            & \cellcolor[HTML]{C6EFCE}{\color[HTML]{006100} 0.863}                            & \cellcolor[HTML]{FFEB9C}{\color[HTML]{9C5700} 0.678}                            & \cellcolor[HTML]{FFEB9C}{\color[HTML]{9C5700} 0.813}                               & \cellcolor[HTML]{FFEB9C}{\color[HTML]{9C5700} 0.61}                             & \cellcolor[HTML]{FFEB9C}{\color[HTML]{9C5700} 0.649}                            & \cellcolor[HTML]{FFEB9C}{\color[HTML]{9C5700} 0.478}                            & \cellcolor[HTML]{FFC7CE}{\color[HTML]{9C0006} 12.276}                              & \cellcolor[HTML]{FFC7CE}{\color[HTML]{9C0006} 0.651}                            & \cellcolor[HTML]{FFC7CE}{\color[HTML]{9C0006} 0.739}                            & \cellcolor[HTML]{FFC7CE}{\color[HTML]{9C0006} 0.576}                            & \cellcolor[HTML]{FFC7CE}{\color[HTML]{9C0006} 27.149}                              & \cellcolor[HTML]{FFEB9C}{\color[HTML]{9C5700} 0.887}                            & \cellcolor[HTML]{FFC7CE}{\color[HTML]{9C0006} 0.928}                            & \cellcolor[HTML]{FFC7CE}{\color[HTML]{9C0006} 0.778}                            & \cellcolor[HTML]{FFEB9C}{\color[HTML]{9C5700} 9.591}                               & \cellcolor[HTML]{FFEB9C}{\color[HTML]{9C5700} 0.548}                            & \cellcolor[HTML]{FFEB9C}{\color[HTML]{9C5700} 0.564}                            & \cellcolor[HTML]{FFEB9C}{\color[HTML]{9C5700} 0.41}                             & \cellcolor[HTML]{FFEB9C}{\color[HTML]{9C5700} 0.815}                               & \cellcolor[HTML]{FFEB9C}{\color[HTML]{9C5700} 0.916}                            & \cellcolor[HTML]{FFEB9C}{\color[HTML]{9C5700} 0.86}                             & \cellcolor[HTML]{FFEB9C}{\color[HTML]{9C5700} 0.701}                            & \cellcolor[HTML]{FFEB9C}{\color[HTML]{9C5700} 10.941}                              & \cellcolor[HTML]{FFC7CE}{\color[HTML]{9C0006} 0.393}                            & \cellcolor[HTML]{FFC7CE}{\color[HTML]{9C0006} 0.566}                            & \cellcolor[HTML]{FFC7CE}{\color[HTML]{9C0006} 0.408}                            & \cellcolor[HTML]{FFC7CE}{\color[HTML]{9C0006} 1.067}                               & \cellcolor[HTML]{FFC7CE}{\color[HTML]{9C0006} 0.398}                            & \cellcolor[HTML]{FFC7CE}{\color[HTML]{9C0006} 0.546}                            & \cellcolor[HTML]{FFC7CE}{\color[HTML]{9C0006} 0.415}                            & \cellcolor[HTML]{FFC7CE}{\color[HTML]{9C0006} 1.002}                               & \cellcolor[HTML]{FFC7CE}{\color[HTML]{9C0006} 0.43}                             & \cellcolor[HTML]{FFC7CE}{\color[HTML]{9C0006} 0.456}                            & \cellcolor[HTML]{FFC7CE}{\color[HTML]{9C0006} 0.305}                            & \cellcolor[HTML]{FFEB9C}{\color[HTML]{9C5700} 15.348}                             \\
\begin{tabular}[c]{@{}l@{}}Style-\\ Score\\ \cite{gatys2015neural}\end{tabular}             & \cellcolor[HTML]{FFC7CE}{\color[HTML]{9C0006} 0.425}                            & \cellcolor[HTML]{FFC7CE}{\color[HTML]{9C0006} 0.456}                            & \cellcolor[HTML]{FFC7CE}{\color[HTML]{9C0006} 0.335}                            & \cellcolor[HTML]{FFC7CE}{\color[HTML]{9C0006} 0.867}                               & \cellcolor[HTML]{FFC7CE}{\color[HTML]{9C0006} 0.232}                            & \cellcolor[HTML]{FFC7CE}{\color[HTML]{9C0006} 0.596}                            & \cellcolor[HTML]{FFC7CE}{\color[HTML]{9C0006} 0.422}                            & \cellcolor[HTML]{FFC7CE}{\color[HTML]{9C0006} 11.728}                              & \cellcolor[HTML]{FFC7CE}{\color[HTML]{9C0006} 0.431}                            & \cellcolor[HTML]{FFC7CE}{\color[HTML]{9C0006} 0.667}                            & \cellcolor[HTML]{FFC7CE}{\color[HTML]{9C0006} 0.51}                             & \cellcolor[HTML]{FFEB9C}{\color[HTML]{9C5700} 25.313}                              & \cellcolor[HTML]{FFC7CE}{\color[HTML]{9C0006} 0.636}                            & \cellcolor[HTML]{FFC7CE}{\color[HTML]{9C0006} 0.888}                            & \cellcolor[HTML]{FFC7CE}{\color[HTML]{9C0006} 0.719}                            & \cellcolor[HTML]{FFC7CE}{\color[HTML]{9C0006} 20.174}                              & \cellcolor[HTML]{FFC7CE}{\color[HTML]{9C0006} 0.311}                            & \cellcolor[HTML]{FFEB9C}{\color[HTML]{9C5700} 0.586}                            & \cellcolor[HTML]{FFEB9C}{\color[HTML]{9C5700} 0.421}                            & \cellcolor[HTML]{FFC7CE}{\color[HTML]{9C0006} 0.972}                               & \cellcolor[HTML]{FFC7CE}{\color[HTML]{9C0006} 0.795}                            & \cellcolor[HTML]{FFEB9C}{\color[HTML]{9C5700} 0.858}                            & \cellcolor[HTML]{FFEB9C}{\color[HTML]{9C5700} 0.698}                            & \cellcolor[HTML]{FFC7CE}{\color[HTML]{9C0006} 21.014}                              & \cellcolor[HTML]{FFC7CE}{\color[HTML]{9C0006} 0.425}                            & \cellcolor[HTML]{FFC7CE}{\color[HTML]{9C0006} 0.612}                            & \cellcolor[HTML]{FFC7CE}{\color[HTML]{9C0006} 0.435}                            & \cellcolor[HTML]{FFC7CE}{\color[HTML]{9C0006} 1.16}                                & \cellcolor[HTML]{FFC7CE}{\color[HTML]{9C0006} 0.326}                            & \cellcolor[HTML]{FFC7CE}{\color[HTML]{9C0006} 0.536}                            & \cellcolor[HTML]{FFC7CE}{\color[HTML]{9C0006} 0.409}                            & \cellcolor[HTML]{FFC7CE}{\color[HTML]{9C0006} 1.038}                               & \cellcolor[HTML]{FFC7CE}{\color[HTML]{9C0006} 0.273}                            & \cellcolor[HTML]{FFC7CE}{\color[HTML]{9C0006} 0.364}                            & \cellcolor[HTML]{FFC7CE}{\color[HTML]{9C0006} 0.237}                            & \cellcolor[HTML]{FFC7CE}{\color[HTML]{9C0006} 16.996}                             \\
\begin{tabular}[c]{@{}l@{}}HaarPSI\\ \cite{reisenhofer2018haar}\end{tabular}                & \cellcolor[HTML]{FFEB9C}{\color[HTML]{9C5700} 0.829}                            & \cellcolor[HTML]{FFEB9C}{\color[HTML]{9C5700} 0.829}                            & \cellcolor[HTML]{FFEB9C}{\color[HTML]{9C5700} 0.672}                            & \cellcolor[HTML]{C6EFCE}{\color[HTML]{006100} 0.608}                               & \cellcolor[HTML]{FFEB9C}{\color[HTML]{9C5700} 0.713}                            & \cellcolor[HTML]{FFEB9C}{\color[HTML]{9C5700} 0.73}                             & \cellcolor[HTML]{FFEB9C}{\color[HTML]{9C5700} 0.552}                            & \cellcolor[HTML]{FFEB9C}{\color[HTML]{9C5700} 8.503}                               & \cellcolor[HTML]{C6EFCE}{\color[HTML]{006100} 0.777}                            & \cellcolor[HTML]{FFEB9C}{\color[HTML]{9C5700} 0.794}                            & \cellcolor[HTML]{FFEB9C}{\color[HTML]{9C5700} 0.613}                            & \cellcolor[HTML]{C6EFCE}{\color[HTML]{006100} 17.106}                              & \cellcolor[HTML]{C6EFCE}{\color[HTML]{006100} 0.937}                            & \cellcolor[HTML]{FFEB9C}{\color[HTML]{9C5700} 0.956}                            & \cellcolor[HTML]{FFEB9C}{\color[HTML]{9C5700} 0.821}                            & \cellcolor[HTML]{C6EFCE}{\color[HTML]{006100} 7.043}                               & \cellcolor[HTML]{FFEB9C}{\color[HTML]{9C5700} 0.618}                            & \cellcolor[HTML]{FFEB9C}{\color[HTML]{9C5700} 0.615}                            & \cellcolor[HTML]{FFEB9C}{\color[HTML]{9C5700} 0.454}                            & \cellcolor[HTML]{FFEB9C}{\color[HTML]{9C5700} 0.766}                               & \cellcolor[HTML]{C6EFCE}{\color[HTML]{006100} 0.941}                            & \cellcolor[HTML]{FFEB9C}{\color[HTML]{9C5700} 0.861}                            & \cellcolor[HTML]{FFEB9C}{\color[HTML]{9C5700} 0.67}                             & \cellcolor[HTML]{C6EFCE}{\color[HTML]{006100} 7.149}                               & \cellcolor[HTML]{C6EFCE}{\color[HTML]{006100} 0.719}                            & \cellcolor[HTML]{C6EFCE}{\color[HTML]{006100} 0.717}                            & \cellcolor[HTML]{C6EFCE}{\color[HTML]{006100} 0.543}                            & \cellcolor[HTML]{C6EFCE}{\color[HTML]{006100} 0.807}                               & \cellcolor[HTML]{FFEB9C}{\color[HTML]{9C5700} 0.83}                             & \cellcolor[HTML]{FFEB9C}{\color[HTML]{9C5700} 0.84}                             & \cellcolor[HTML]{FFEB9C}{\color[HTML]{9C5700} 0.671}                            & \cellcolor[HTML]{FFEB9C}{\color[HTML]{9C5700} 0.609}                               & \cellcolor[HTML]{FFEB9C}{\color[HTML]{9C5700} 0.604}                            & \cellcolor[HTML]{FFEB9C}{\color[HTML]{9C5700} 0.615}                            & \cellcolor[HTML]{FFEB9C}{\color[HTML]{9C5700} 0.435}                            & \cellcolor[HTML]{FFEB9C}{\color[HTML]{9C5700} 13.558}                             \\
\begin{tabular}[c]{@{}l@{}}MDSI\\ \cite{nafchi2016mean}\end{tabular}                        & \cellcolor[HTML]{FFEB9C}{\color[HTML]{9C5700} 0.77}                             & \cellcolor[HTML]{FFEB9C}{\color[HTML]{9C5700} 0.803}                            & \cellcolor[HTML]{FFEB9C}{\color[HTML]{9C5700} 0.665}                            & \cellcolor[HTML]{C6EFCE}{\color[HTML]{006100} 0.627}                               & \cellcolor[HTML]{FFEB9C}{\color[HTML]{9C5700} 0.705}                            & \cellcolor[HTML]{FFEB9C}{\color[HTML]{9C5700} 0.724}                            & \cellcolor[HTML]{FFEB9C}{\color[HTML]{9C5700} 0.542}                            & \cellcolor[HTML]{FFEB9C}{\color[HTML]{9C5700} 8.706}                               & \cellcolor[HTML]{FFEB9C}{\color[HTML]{9C5700} 0.753}                            & \cellcolor[HTML]{FFC7CE}{\color[HTML]{9C0006} 0.73}                             & \cellcolor[HTML]{FFC7CE}{\color[HTML]{9C0006} 0.563}                            & \cellcolor[HTML]{FFEB9C}{\color[HTML]{9C5700} 23.23}                               & \cellcolor[HTML]{FFC7CE}{\color[HTML]{9C0006} 0.83}                             & \cellcolor[HTML]{FFC7CE}{\color[HTML]{9C0006} 0.925}                            & \cellcolor[HTML]{FFC7CE}{\color[HTML]{9C0006} 0.775}                            & \cellcolor[HTML]{FFEB9C}{\color[HTML]{9C5700} 11.252}                              & \cellcolor[HTML]{FFEB9C}{\color[HTML]{9C5700} 0.632}                            & \cellcolor[HTML]{FFEB9C}{\color[HTML]{9C5700} 0.602}                            & \cellcolor[HTML]{FFEB9C}{\color[HTML]{9C5700} 0.444}                            & \cellcolor[HTML]{FFEB9C}{\color[HTML]{9C5700} 0.755}                               & \cellcolor[HTML]{FFEB9C}{\color[HTML]{9C5700} 0.85}                             & \cellcolor[HTML]{C6EFCE}{\color[HTML]{006100} 0.889}                            & \cellcolor[HTML]{C6EFCE}{\color[HTML]{006100} 0.73}                             & \cellcolor[HTML]{FFEB9C}{\color[HTML]{9C5700} 11.082}                              & \cellcolor[HTML]{FFEB9C}{\color[HTML]{9C5700} 0.662}                            & \cellcolor[HTML]{FFEB9C}{\color[HTML]{9C5700} 0.661}                            & \cellcolor[HTML]{FFEB9C}{\color[HTML]{9C5700} 0.486}                            & \cellcolor[HTML]{FFEB9C}{\color[HTML]{9C5700} 0.87}                                & \cellcolor[HTML]{FFEB9C}{\color[HTML]{9C5700} 0.84}                             & \cellcolor[HTML]{FFEB9C}{\color[HTML]{9C5700} 0.838}                            & \cellcolor[HTML]{FFEB9C}{\color[HTML]{9C5700} 0.671}                            & \cellcolor[HTML]{FFEB9C}{\color[HTML]{9C5700} 0.593}                               & \cellcolor[HTML]{C6EFCE}{\color[HTML]{006100} 0.666}                            & \cellcolor[HTML]{FFEB9C}{\color[HTML]{9C5700} 0.659}                            & \cellcolor[HTML]{C6EFCE}{\color[HTML]{006100} 0.478}                            & \cellcolor[HTML]{C6EFCE}{\color[HTML]{006100} 12.686}                             \\
\begin{tabular}[c]{@{}l@{}}LPIPS\\ \cite{zhang2018unreasonable}\end{tabular}                & \cellcolor[HTML]{FFEB9C}{\color[HTML]{9C5700} 0.794}                            & \cellcolor[HTML]{FFEB9C}{\color[HTML]{9C5700} 0.804}                            & \cellcolor[HTML]{FFEB9C}{\color[HTML]{9C5700} 0.649}                            & \cellcolor[HTML]{FFEB9C}{\color[HTML]{9C5700} 0.649}                               & \cellcolor[HTML]{FFC7CE}{\color[HTML]{9C0006} 0.584}                            & \cellcolor[HTML]{FFC7CE}{\color[HTML]{9C0006} 0.622}                            & \cellcolor[HTML]{FFC7CE}{\color[HTML]{9C0006} 0.452}                            & \cellcolor[HTML]{FFEB9C}{\color[HTML]{9C5700} 9.897}                               & \cellcolor[HTML]{FFEB9C}{\color[HTML]{9C5700} 0.696}                            & \cellcolor[HTML]{FFEB9C}{\color[HTML]{9C5700} 0.748}                            & \cellcolor[HTML]{FFEB9C}{\color[HTML]{9C5700} 0.578}                            & \cellcolor[HTML]{FFEB9C}{\color[HTML]{9C5700} 25.247}                              & \cellcolor[HTML]{FFEB9C}{\color[HTML]{9C5700} 0.914}                            & \cellcolor[HTML]{FFEB9C}{\color[HTML]{9C5700} 0.938}                            & \cellcolor[HTML]{FFEB9C}{\color[HTML]{9C5700} 0.795}                            & \cellcolor[HTML]{FFEB9C}{\color[HTML]{9C5700} 8.177}                               & \cellcolor[HTML]{C6EFCE}{\color[HTML]{006100} 0.782}                            & \cellcolor[HTML]{C6EFCE}{\color[HTML]{006100} 0.777}                            & \cellcolor[HTML]{C6EFCE}{\color[HTML]{006100} 0.597}                            & \cellcolor[HTML]{C6EFCE}{\color[HTML]{006100} 0.607}                               & \cellcolor[HTML]{FFEB9C}{\color[HTML]{9C5700} 0.9}                              & \cellcolor[HTML]{FFC7CE}{\color[HTML]{9C0006} 0.847}                            & \cellcolor[HTML]{FFEB9C}{\color[HTML]{9C5700} 0.678}                            & \cellcolor[HTML]{FFEB9C}{\color[HTML]{9C5700} 9.155}                               & \cellcolor[HTML]{FFEB9C}{\color[HTML]{9C5700} 0.593}                            & \cellcolor[HTML]{FFC7CE}{\color[HTML]{9C0006} 0.547}                            & \cellcolor[HTML]{FFC7CE}{\color[HTML]{9C0006} 0.377}                            & \cellcolor[HTML]{FFEB9C}{\color[HTML]{9C5700} 0.934}                               & \cellcolor[HTML]{FFC7CE}{\color[HTML]{9C0006} 0.501}                            & \cellcolor[HTML]{FFC7CE}{\color[HTML]{9C0006} 0.559}                            & \cellcolor[HTML]{FFC7CE}{\color[HTML]{9C0006} 0.426}                            & \cellcolor[HTML]{FFEB9C}{\color[HTML]{9C5700} 0.945}                               & \cellcolor[HTML]{FFEB9C}{\color[HTML]{9C5700} 0.55}                             & \cellcolor[HTML]{FFEB9C}{\color[HTML]{9C5700} 0.551}                            & \cellcolor[HTML]{FFEB9C}{\color[HTML]{9C5700} 0.384}                            & \cellcolor[HTML]{FFEB9C}{\color[HTML]{9C5700} 14.203}                             \\
\begin{tabular}[c]{@{}l@{}}DISTS\\ \cite{ding2020image}\end{tabular}                        & \cellcolor[HTML]{FFEB9C}{\color[HTML]{9C5700} 0.818}                            & \cellcolor[HTML]{FFEB9C}{\color[HTML]{9C5700} 0.817}                            & \cellcolor[HTML]{FFEB9C}{\color[HTML]{9C5700} 0.652}                            & \cellcolor[HTML]{FFEB9C}{\color[HTML]{9C5700} 0.714}                               & \cellcolor[HTML]{FFC7CE}{\color[HTML]{9C0006} 0.556}                            & \cellcolor[HTML]{FFEB9C}{\color[HTML]{9C5700} 0.694}                            & \cellcolor[HTML]{FFEB9C}{\color[HTML]{9C5700} 0.497}                            & \cellcolor[HTML]{FFC7CE}{\color[HTML]{9C0006} 12.232}                              & \cellcolor[HTML]{FFEB9C}{\color[HTML]{9C5700} 0.686}                            & \cellcolor[HTML]{C6EFCE}{\color[HTML]{006100} 0.795}                            & \cellcolor[HTML]{FFEB9C}{\color[HTML]{9C5700} 0.608}                            & \cellcolor[HTML]{FFC7CE}{\color[HTML]{9C0006} 26.941}                              & \cellcolor[HTML]{C6EFCE}{\color[HTML]{006100} 0.929}                            & \cellcolor[HTML]{FFEB9C}{\color[HTML]{9C5700} 0.951}                            & \cellcolor[HTML]{FFEB9C}{\color[HTML]{9C5700} 0.809}                            & \cellcolor[HTML]{FFEB9C}{\color[HTML]{9C5700} 8.623}                               & \cellcolor[HTML]{C6EFCE}{\color[HTML]{006100} 0.85}                             & \cellcolor[HTML]{C6EFCE}{\color[HTML]{006100} 0.835}                            & \cellcolor[HTML]{C6EFCE}{\color[HTML]{006100} 0.658}                            & \cellcolor[HTML]{C6EFCE}{\color[HTML]{006100} 0.514}                               & \cellcolor[HTML]{FFEB9C}{\color[HTML]{9C5700} 0.888}                            & \cellcolor[HTML]{FFEB9C}{\color[HTML]{9C5700} 0.873}                            & \cellcolor[HTML]{FFEB9C}{\color[HTML]{9C5700} 0.712}                            & \cellcolor[HTML]{FFEB9C}{\color[HTML]{9C5700} 11.957}                              & \cellcolor[HTML]{C6EFCE}{\color[HTML]{006100} 0.852}                            & \cellcolor[HTML]{C6EFCE}{\color[HTML]{006100} 0.843}                            & \cellcolor[HTML]{C6EFCE}{\color[HTML]{006100} 0.665}                            & \cellcolor[HTML]{C6EFCE}{\color[HTML]{006100} 0.623}                               & \cellcolor[HTML]{FFC7CE}{\color[HTML]{9C0006} 0.424}                            & \cellcolor[HTML]{FFC7CE}{\color[HTML]{9C0006} 0.683}                            & \cellcolor[HTML]{FFC7CE}{\color[HTML]{9C0006} 0.53}                             & \cellcolor[HTML]{FFC7CE}{\color[HTML]{9C0006} 0.993}                               & \cellcolor[HTML]{FFEB9C}{\color[HTML]{9C5700} 0.565}                            & \cellcolor[HTML]{FFEB9C}{\color[HTML]{9C5700} 0.613}                            & \cellcolor[HTML]{FFEB9C}{\color[HTML]{9C5700} 0.433}                            & \cellcolor[HTML]{FFEB9C}{\color[HTML]{9C5700} 14.027}                             \\
\begin{tabular}[c]{@{}l@{}}STRRED\\ \cite{soundararajan2012video}\end{tabular}              & \cellcolor[HTML]{FFC7CE}{\color[HTML]{9C0006} 0.711}                            & \cellcolor[HTML]{FFEB9C}{\color[HTML]{9C5700} 0.786}                            & \cellcolor[HTML]{FFEB9C}{\color[HTML]{9C5700} 0.634}                            & \cellcolor[HTML]{FFC7CE}{\color[HTML]{9C0006} 0.888}                               & \cellcolor[HTML]{FFC7CE}{\color[HTML]{9C0006} 0.329}                            & \cellcolor[HTML]{FFEB9C}{\color[HTML]{9C5700} 0.785}                            & \cellcolor[HTML]{FFEB9C}{\color[HTML]{9C5700} 0.595}                            & \cellcolor[HTML]{FFC7CE}{\color[HTML]{9C0006} 11.511}                              & \cellcolor[HTML]{FFC7CE}{\color[HTML]{9C0006} 0.606}                            & \cellcolor[HTML]{FFEB9C}{\color[HTML]{9C5700} 0.784}                            & \cellcolor[HTML]{FFEB9C}{\color[HTML]{9C5700} 0.601}                            & \cellcolor[HTML]{FFEB9C}{\color[HTML]{9C5700} 25.647}                              & \cellcolor[HTML]{FFC7CE}{\color[HTML]{9C0006} 0.833}                            & \cellcolor[HTML]{FFEB9C}{\color[HTML]{9C5700} 0.946}                            & \cellcolor[HTML]{FFEB9C}{\color[HTML]{9C5700} 0.807}                            & \cellcolor[HTML]{FFC7CE}{\color[HTML]{9C0006} 17.368}                              & \cellcolor[HTML]{FFC7CE}{\color[HTML]{9C0006} 0.359}                            & \cellcolor[HTML]{FFC7CE}{\color[HTML]{9C0006} 0.506}                            & \cellcolor[HTML]{FFC7CE}{\color[HTML]{9C0006} 0.371}                            & \cellcolor[HTML]{FFC7CE}{\color[HTML]{9C0006} 0.909}                               & \cellcolor[HTML]{FFC7CE}{\color[HTML]{9C0006} 0.655}                            & \cellcolor[HTML]{FFC7CE}{\color[HTML]{9C0006} 0.732}                            & \cellcolor[HTML]{FFC7CE}{\color[HTML]{9C0006} 0.567}                            & \cellcolor[HTML]{FFC7CE}{\color[HTML]{9C0006} 20.215}                              & \cellcolor[HTML]{FFEB9C}{\color[HTML]{9C5700} 0.671}                            & \cellcolor[HTML]{C6EFCE}{\color[HTML]{006100} 0.752}                            & \cellcolor[HTML]{C6EFCE}{\color[HTML]{006100} 0.586}                            & \cellcolor[HTML]{FFEB9C}{\color[HTML]{9C5700} 0.861}                               & \cellcolor[HTML]{FFC7CE}{\color[HTML]{9C0006} 0.317}                            & \cellcolor[HTML]{FFC7CE}{\color[HTML]{9C0006} 0.076}                            & \cellcolor[HTML]{FFC7CE}{\color[HTML]{9C0006} 0.087}                            & \cellcolor[HTML]{FFC7CE}{\color[HTML]{9C0006} 1.036}                               & \cellcolor[HTML]{FFC7CE}{\color[HTML]{9C0006} 0.198}                            & \cellcolor[HTML]{FFEB9C}{\color[HTML]{9C5700} 0.589}                            & \cellcolor[HTML]{FFEB9C}{\color[HTML]{9C5700} 0.428}                            & \cellcolor[HTML]{FFC7CE}{\color[HTML]{9C0006} 16.666}                             \\
\begin{tabular}[c]{@{}l@{}}PieAPP\\ \cite{prashnani2018pieapp}\end{tabular}                 & \cellcolor[HTML]{FFC7CE}{\color[HTML]{9C0006} 0.755}                            & \cellcolor[HTML]{FFC7CE}{\color[HTML]{9C0006} 0.745}                            & \cellcolor[HTML]{FFC7CE}{\color[HTML]{9C0006} 0.564}                            & \cellcolor[HTML]{C6EFCE}{\color[HTML]{006100} 0.583}                               & \cellcolor[HTML]{FFEB9C}{\color[HTML]{9C5700} 0.66}                             & \cellcolor[HTML]{FFC7CE}{\color[HTML]{9C0006} 0.639}                            & \cellcolor[HTML]{FFC7CE}{\color[HTML]{9C0006} 0.477}                            & \cellcolor[HTML]{FFEB9C}{\color[HTML]{9C5700} 9.129}                               & \cellcolor[HTML]{FFEB9C}{\color[HTML]{9C5700} 0.73}                             & \cellcolor[HTML]{FFEB9C}{\color[HTML]{9C5700} 0.749}                            & \cellcolor[HTML]{FFEB9C}{\color[HTML]{9C5700} 0.589}                            & \cellcolor[HTML]{FFEB9C}{\color[HTML]{9C5700} 19.077}                              & \cellcolor[HTML]{C6EFCE}{\color[HTML]{006100} 0.931}                            & \cellcolor[HTML]{C6EFCE}{\color[HTML]{006100} 0.957}                            & \cellcolor[HTML]{C6EFCE}{\color[HTML]{006100} 0.829}                            & \cellcolor[HTML]{C6EFCE}{\color[HTML]{006100} 7.343}                               & \cellcolor[HTML]{C6EFCE}{\color[HTML]{006100} 0.778}                            & \cellcolor[HTML]{C6EFCE}{\color[HTML]{006100} 0.78}                             & \cellcolor[HTML]{C6EFCE}{\color[HTML]{006100} 0.588}                            & \cellcolor[HTML]{C6EFCE}{\color[HTML]{006100} 0.612}                               & \cellcolor[HTML]{C6EFCE}{\color[HTML]{006100} 0.95}                             & \cellcolor[HTML]{C6EFCE}{\color[HTML]{006100} 0.939}                            & \cellcolor[HTML]{C6EFCE}{\color[HTML]{006100} 0.798}                            & \cellcolor[HTML]{C6EFCE}{\color[HTML]{006100} 6.557}                               & \cellcolor[HTML]{C6EFCE}{\color[HTML]{006100} 0.788}                            & \cellcolor[HTML]{C6EFCE}{\color[HTML]{006100} 0.777}                            & \cellcolor[HTML]{C6EFCE}{\color[HTML]{006100} 0.594}                            & \cellcolor[HTML]{C6EFCE}{\color[HTML]{006100} 0.714}                               & \cellcolor[HTML]{FFEB9C}{\color[HTML]{9C5700} 0.836}                            & \cellcolor[HTML]{FFEB9C}{\color[HTML]{9C5700} 0.824}                            & \cellcolor[HTML]{FFEB9C}{\color[HTML]{9C5700} 0.646}                            & \cellcolor[HTML]{FFEB9C}{\color[HTML]{9C5700} 0.599}                               & \cellcolor[HTML]{FFEB9C}{\color[HTML]{9C5700} 0.587}                            & \cellcolor[HTML]{FFEB9C}{\color[HTML]{9C5700} 0.606}                            & \cellcolor[HTML]{FFEB9C}{\color[HTML]{9C5700} 0.437}                            & \cellcolor[HTML]{FFEB9C}{\color[HTML]{9C5700} 13.767}                             \\
\begin{tabular}[c]{@{}l@{}}FLIP\\ \cite{andersson2020flip}\end{tabular}                     & \cellcolor[HTML]{FFEB9C}{\color[HTML]{9C5700} 0.827}                            & \cellcolor[HTML]{C6EFCE}{\color[HTML]{006100} 0.876}                            & \cellcolor[HTML]{C6EFCE}{\color[HTML]{006100} 0.707}                            & \cellcolor[HTML]{FFC7CE}{\color[HTML]{9C0006} 0.814}                               & \cellcolor[HTML]{FFEB9C}{\color[HTML]{9C5700} 0.674}                            & \cellcolor[HTML]{FFEB9C}{\color[HTML]{9C5700} 0.672}                            & \cellcolor[HTML]{FFEB9C}{\color[HTML]{9C5700} 0.5}                              & \cellcolor[HTML]{FFEB9C}{\color[HTML]{9C5700} 10.892}                              & \cellcolor[HTML]{FFEB9C}{\color[HTML]{9C5700} 0.693}                            & \cellcolor[HTML]{FFC7CE}{\color[HTML]{9C0006} 0.728}                            & \cellcolor[HTML]{FFC7CE}{\color[HTML]{9C0006} 0.551}                            & \cellcolor[HTML]{FFC7CE}{\color[HTML]{9C0006} 27.03}                               & \cellcolor[HTML]{FFEB9C}{\color[HTML]{9C5700} 0.901}                            & \cellcolor[HTML]{FFEB9C}{\color[HTML]{9C5700} 0.931}                            & \cellcolor[HTML]{FFEB9C}{\color[HTML]{9C5700} 0.789}                            & \cellcolor[HTML]{FFEB9C}{\color[HTML]{9C5700} 9.533}                               & \cellcolor[HTML]{FFEB9C}{\color[HTML]{9C5700} 0.57}                             & \cellcolor[HTML]{FFEB9C}{\color[HTML]{9C5700} 0.577}                            & \cellcolor[HTML]{FFEB9C}{\color[HTML]{9C5700} 0.429}                            & \cellcolor[HTML]{FFEB9C}{\color[HTML]{9C5700} 0.8}                                 & \cellcolor[HTML]{FFEB9C}{\color[HTML]{9C5700} 0.849}                            & \cellcolor[HTML]{FFC7CE}{\color[HTML]{9C0006} 0.806}                            & \cellcolor[HTML]{FFC7CE}{\color[HTML]{9C0006} 0.613}                            & \cellcolor[HTML]{FFEB9C}{\color[HTML]{9C5700} 12.057}                              & \cellcolor[HTML]{FFC7CE}{\color[HTML]{9C0006} 0.488}                            & \cellcolor[HTML]{FFC7CE}{\color[HTML]{9C0006} 0.596}                            & \cellcolor[HTML]{FFC7CE}{\color[HTML]{9C0006} 0.427}                            & \cellcolor[HTML]{FFC7CE}{\color[HTML]{9C0006} 1.013}                               & \cellcolor[HTML]{FFEB9C}{\color[HTML]{9C5700} 0.818}                            & \cellcolor[HTML]{FFEB9C}{\color[HTML]{9C5700} 0.84}                             & \cellcolor[HTML]{FFEB9C}{\color[HTML]{9C5700} 0.667}                            & \cellcolor[HTML]{FFEB9C}{\color[HTML]{9C5700} 0.629}                               & \cellcolor[HTML]{FFC7CE}{\color[HTML]{9C0006} 0.466}                            & \cellcolor[HTML]{FFC7CE}{\color[HTML]{9C0006} 0.501}                            & \cellcolor[HTML]{FFC7CE}{\color[HTML]{9C0006} 0.333}                            & \cellcolor[HTML]{FFEB9C}{\color[HTML]{9C5700} 15.043}                             \\
\begin{tabular}[c]{@{}l@{}}ERQA\\ \cite{kirillova2021erqa}\end{tabular}                     & \cellcolor[HTML]{FFC7CE}{\color[HTML]{9C0006} 0.646}                            & \cellcolor[HTML]{FFC7CE}{\color[HTML]{9C0006} 0.587}                            & \cellcolor[HTML]{FFC7CE}{\color[HTML]{9C0006} 0.433}                            & \cellcolor[HTML]{FFEB9C}{\color[HTML]{9C5700} 0.709}                               & \cellcolor[HTML]{FFEB9C}{\color[HTML]{9C5700} 0.618}                            & \cellcolor[HTML]{FFC7CE}{\color[HTML]{9C0006} 0.623}                            & \cellcolor[HTML]{FFC7CE}{\color[HTML]{9C0006} 0.455}                            & \cellcolor[HTML]{FFEB9C}{\color[HTML]{9C5700} 9.512}                               & \cellcolor[HTML]{C6EFCE}{\color[HTML]{006100} 0.797}                            & \cellcolor[HTML]{FFEB9C}{\color[HTML]{9C5700} 0.787}                            & \cellcolor[HTML]{FFEB9C}{\color[HTML]{9C5700} 0.616}                            & \cellcolor[HTML]{C6EFCE}{\color[HTML]{006100} 17.162}                              & \cellcolor[HTML]{FFEB9C}{\color[HTML]{9C5700} 0.9}                              & \cellcolor[HTML]{FFEB9C}{\color[HTML]{9C5700} 0.936}                            & \cellcolor[HTML]{FFEB9C}{\color[HTML]{9C5700} 0.794}                            & \cellcolor[HTML]{FFEB9C}{\color[HTML]{9C5700} 8.81}                                & \cellcolor[HTML]{C6EFCE}{\color[HTML]{006100} 0.756}                            & \cellcolor[HTML]{C6EFCE}{\color[HTML]{006100} 0.737}                            & \cellcolor[HTML]{C6EFCE}{\color[HTML]{006100} 0.548}                            & \cellcolor[HTML]{C6EFCE}{\color[HTML]{006100} 0.638}                               & \cellcolor[HTML]{FFC7CE}{\color[HTML]{9C0006} 0.834}                            & \cellcolor[HTML]{FFC7CE}{\color[HTML]{9C0006} 0.771}                            & \cellcolor[HTML]{FFC7CE}{\color[HTML]{9C0006} 0.574}                            & \cellcolor[HTML]{FFEB9C}{\color[HTML]{9C5700} 11.605}                              & \cellcolor[HTML]{C6EFCE}{\color[HTML]{006100} 0.722}                            & \cellcolor[HTML]{C6EFCE}{\color[HTML]{006100} 0.717}                            & \cellcolor[HTML]{C6EFCE}{\color[HTML]{006100} 0.546}                            & \cellcolor[HTML]{C6EFCE}{\color[HTML]{006100} 0.803}                               & \cellcolor[HTML]{FFEB9C}{\color[HTML]{9C5700} 0.7}                              & \cellcolor[HTML]{FFC7CE}{\color[HTML]{9C0006} 0.728}                            & \cellcolor[HTML]{FFC7CE}{\color[HTML]{9C0006} 0.552}                            & \cellcolor[HTML]{FFEB9C}{\color[HTML]{9C5700} 0.78}                                & \cellcolor[HTML]{FFC7CE}{\color[HTML]{9C0006} 0.4}                              & \cellcolor[HTML]{FFC7CE}{\color[HTML]{9C0006} 0.491}                            & \cellcolor[HTML]{FFC7CE}{\color[HTML]{9C0006} 0.346}                            & \cellcolor[HTML]{FFC7CE}{\color[HTML]{9C0006} 15.584}                             \\
\begin{tabular}[c]{@{}l@{}}VMAF\\ \cite{rassool2017vmaf}\end{tabular}                       & \cellcolor[HTML]{FFEB9C}{\color[HTML]{9C5700} 0.826}                            & \cellcolor[HTML]{FFEB9C}{\color[HTML]{9C5700} 0.818}                            & \cellcolor[HTML]{FFEB9C}{\color[HTML]{9C5700} 0.642}                            & \cellcolor[HTML]{C6EFCE}{\color[HTML]{006100} 0.581}                               & \cellcolor[HTML]{FFEB9C}{\color[HTML]{9C5700} 0.716}                            & \cellcolor[HTML]{FFEB9C}{\color[HTML]{9C5700} 0.768}                            & \cellcolor[HTML]{FFEB9C}{\color[HTML]{9C5700} 0.575}                            & \cellcolor[HTML]{FFEB9C}{\color[HTML]{9C5700} 8.393}                               & \cellcolor[HTML]{FFEB9C}{\color[HTML]{9C5700} 0.715}                            & \cellcolor[HTML]{FFEB9C}{\color[HTML]{9C5700} 0.79}                             & \cellcolor[HTML]{FFEB9C}{\color[HTML]{9C5700} 0.606}                            & \cellcolor[HTML]{FFEB9C}{\color[HTML]{9C5700} 18.647}                              & \cellcolor[HTML]{C6EFCE}{\color[HTML]{006100} 0.932}                            & \cellcolor[HTML]{C6EFCE}{\color[HTML]{006100} 0.96}                             & \cellcolor[HTML]{C6EFCE}{\color[HTML]{006100} 0.829}                            & \cellcolor[HTML]{C6EFCE}{\color[HTML]{006100} 7.303}                               & \cellcolor[HTML]{FFEB9C}{\color[HTML]{9C5700} 0.492}                            & \cellcolor[HTML]{FFC7CE}{\color[HTML]{9C0006} 0.438}                            & \cellcolor[HTML]{FFC7CE}{\color[HTML]{9C0006} 0.31}                             & \cellcolor[HTML]{FFEB9C}{\color[HTML]{9C5700} 0.848}                               & \cellcolor[HTML]{FFEB9C}{\color[HTML]{9C5700} 0.919}                            & \cellcolor[HTML]{FFC7CE}{\color[HTML]{9C0006} 0.847}                            & \cellcolor[HTML]{FFC7CE}{\color[HTML]{9C0006} 0.65}                             & \cellcolor[HTML]{FFEB9C}{\color[HTML]{9C5700} 8.289}                               & \cellcolor[HTML]{FFEB9C}{\color[HTML]{9C5700} 0.654}                            & \cellcolor[HTML]{FFEB9C}{\color[HTML]{9C5700} 0.682}                            & \cellcolor[HTML]{FFEB9C}{\color[HTML]{9C5700} 0.511}                            & \cellcolor[HTML]{FFEB9C}{\color[HTML]{9C5700} 0.878}                               & \cellcolor[HTML]{C6EFCE}{\color[HTML]{006100} 0.929}                            & \cellcolor[HTML]{C6EFCE}{\color[HTML]{006100} 0.902}                            & \cellcolor[HTML]{C6EFCE}{\color[HTML]{006100} 0.741}                            & \cellcolor[HTML]{C6EFCE}{\color[HTML]{006100} 0.403}                               & \cellcolor[HTML]{C6EFCE}{\color[HTML]{006100} 0.772}                            & \cellcolor[HTML]{C6EFCE}{\color[HTML]{006100} 0.773}                            & \cellcolor[HTML]{C6EFCE}{\color[HTML]{006100} 0.556}                            & \cellcolor[HTML]{C6EFCE}{\color[HTML]{006100} 10.802}                             \\
\begin{tabular}[c]{@{}l@{}}FoV-\\ VideoVDP\\ \cite{mantiuk2021fovvideovdp}\end{tabular}     & \cellcolor[HTML]{FFEB9C}{\color[HTML]{9C5700} 0.801}                            & \cellcolor[HTML]{FFEB9C}{\color[HTML]{9C5700} 0.791}                            & \cellcolor[HTML]{FFEB9C}{\color[HTML]{9C5700} 0.661}                            & \cellcolor[HTML]{FFEB9C}{\color[HTML]{9C5700} 0.674}                               & \cellcolor[HTML]{C6EFCE}{\color[HTML]{006100} 0.84}                             & \cellcolor[HTML]{C6EFCE}{\color[HTML]{006100} 0.855}                            & \cellcolor[HTML]{C6EFCE}{\color[HTML]{006100} 0.682}                            & \cellcolor[HTML]{C6EFCE}{\color[HTML]{006100} 6.606}                               & \cellcolor[HTML]{C6EFCE}{\color[HTML]{006100} 0.755}                            & \cellcolor[HTML]{FFC7CE}{\color[HTML]{9C0006} 0.741}                            & \cellcolor[HTML]{FFC7CE}{\color[HTML]{9C0006} 0.565}                            & \cellcolor[HTML]{C6EFCE}{\color[HTML]{006100} 18.207}                              & \cellcolor[HTML]{C6EFCE}{\color[HTML]{006100} 0.937}                            & \cellcolor[HTML]{C6EFCE}{\color[HTML]{006100} 0.963}                            & \cellcolor[HTML]{C6EFCE}{\color[HTML]{006100} 0.841}                            & \cellcolor[HTML]{C6EFCE}{\color[HTML]{006100} 7.229}                               & \cellcolor[HTML]{FFEB9C}{\color[HTML]{9C5700} 0.609}                            & \cellcolor[HTML]{FFEB9C}{\color[HTML]{9C5700} 0.59}                             & \cellcolor[HTML]{FFEB9C}{\color[HTML]{9C5700} 0.436}                            & \cellcolor[HTML]{FFEB9C}{\color[HTML]{9C5700} 0.773}                               & \cellcolor[HTML]{C6EFCE}{\color[HTML]{006100} 0.942}                            & \cellcolor[HTML]{C6EFCE}{\color[HTML]{006100} 0.932}                            & \cellcolor[HTML]{C6EFCE}{\color[HTML]{006100} 0.805}                            & \cellcolor[HTML]{C6EFCE}{\color[HTML]{006100} 7.071}                               & \cellcolor[HTML]{FFEB9C}{\color[HTML]{9C5700} 0.633}                            & \cellcolor[HTML]{FFC7CE}{\color[HTML]{9C0006} 0.625}                            & \cellcolor[HTML]{FFEB9C}{\color[HTML]{9C5700} 0.454}                            & \cellcolor[HTML]{FFEB9C}{\color[HTML]{9C5700} 0.898}                               & \cellcolor[HTML]{C6EFCE}{\color[HTML]{006100} 0.859}                            & \cellcolor[HTML]{FFEB9C}{\color[HTML]{9C5700} 0.85}                             & \cellcolor[HTML]{FFEB9C}{\color[HTML]{9C5700} 0.663}                            & \cellcolor[HTML]{C6EFCE}{\color[HTML]{006100} 0.559}                               & \cellcolor[HTML]{C6EFCE}{\color[HTML]{006100} 0.796}                            & \cellcolor[HTML]{C6EFCE}{\color[HTML]{006100} 0.807}                            & \cellcolor[HTML]{C6EFCE}{\color[HTML]{006100} 0.592}                            & \cellcolor[HTML]{C6EFCE}{\color[HTML]{006100} 10.288}                             \\
\begin{tabular}[c]{@{}l@{}}Color-\\ VideoVDP\\ \cite{mantiuk2024colorvideovdp}\end{tabular} & \cellcolor[HTML]{C6EFCE}{\color[HTML]{006100} 0.872}                            & \cellcolor[HTML]{C6EFCE}{\color[HTML]{006100} 0.898}                            & \cellcolor[HTML]{C6EFCE}{\color[HTML]{006100} 0.743}                            & \cellcolor[HTML]{C6EFCE}{\color[HTML]{006100} 0.537}                               & \cellcolor[HTML]{C6EFCE}{\color[HTML]{006100} 0.775}                            & \cellcolor[HTML]{FFEB9C}{\color[HTML]{9C5700} 0.8}                              & \cellcolor[HTML]{C6EFCE}{\color[HTML]{006100} 0.623}                            & \cellcolor[HTML]{C6EFCE}{\color[HTML]{006100} 7.812}                               & \cellcolor[HTML]{FFEB9C}{\color[HTML]{9C5700} 0.744}                            & \cellcolor[HTML]{FFEB9C}{\color[HTML]{9C5700} 0.775}                            & \cellcolor[HTML]{FFEB9C}{\color[HTML]{9C5700} 0.601}                            & \cellcolor[HTML]{FFEB9C}{\color[HTML]{9C5700} 18.474}                              & \cellcolor[HTML]{FFEB9C}{\color[HTML]{9C5700} 0.928}                            & \cellcolor[HTML]{C6EFCE}{\color[HTML]{006100} 0.966}                            & \cellcolor[HTML]{C6EFCE}{\color[HTML]{006100} 0.852}                            & \cellcolor[HTML]{C6EFCE}{\color[HTML]{006100} 6.698}                               & \cellcolor[HTML]{FFEB9C}{\color[HTML]{9C5700} 0.621}                            & \cellcolor[HTML]{FFEB9C}{\color[HTML]{9C5700} 0.598}                            & \cellcolor[HTML]{FFEB9C}{\color[HTML]{9C5700} 0.443}                            & \cellcolor[HTML]{FFEB9C}{\color[HTML]{9C5700} 0.763}                               & \cellcolor[HTML]{FFEB9C}{\color[HTML]{9C5700} 0.852}                            & \cellcolor[HTML]{FFC7CE}{\color[HTML]{9C0006} 0.813}                            & \cellcolor[HTML]{FFC7CE}{\color[HTML]{9C0006} 0.61}                             & \cellcolor[HTML]{FFEB9C}{\color[HTML]{9C5700} 10.998}                              & \cellcolor[HTML]{FFEB9C}{\color[HTML]{9C5700} 0.658}                            & \cellcolor[HTML]{FFEB9C}{\color[HTML]{9C5700} 0.651}                            & \cellcolor[HTML]{FFEB9C}{\color[HTML]{9C5700} 0.476}                            & \cellcolor[HTML]{FFEB9C}{\color[HTML]{9C5700} 0.874}                               & \cellcolor[HTML]{C6EFCE}{\color[HTML]{006100} 0.896}                            & \cellcolor[HTML]{C6EFCE}{\color[HTML]{006100} 0.9}                              & \cellcolor[HTML]{C6EFCE}{\color[HTML]{006100} 0.739}                            & \cellcolor[HTML]{C6EFCE}{\color[HTML]{006100} 0.485}                               & \cellcolor[HTML]{C6EFCE}{\color[HTML]{006100} 0.769}                            & \cellcolor[HTML]{C6EFCE}{\color[HTML]{006100} 0.804}                            & \cellcolor[HTML]{C6EFCE}{\color[HTML]{006100} 0.587}                            & \cellcolor[HTML]{C6EFCE}{\color[HTML]{006100} 10.864}                             \\
\textbf{\begin{tabular}[c]{@{}l@{}}CGVQM-2\\ (ours)\end{tabular}}                           & \cellcolor[HTML]{FFEB9C}{\color[HTML]{9C5700} 0.788}                            & \cellcolor[HTML]{FFEB9C}{\color[HTML]{9C5700} 0.791}                            & \cellcolor[HTML]{FFC7CE}{\color[HTML]{9C0006} 0.614}                            & \cellcolor[HTML]{FFEB9C}{\color[HTML]{9C5700} 0.736}                               & \cellcolor[HTML]{C6EFCE}{\color[HTML]{006100} 0.787}                            & \cellcolor[HTML]{C6EFCE}{\color[HTML]{006100} 0.82}                             & \cellcolor[HTML]{C6EFCE}{\color[HTML]{006100} 0.638}                            & \cellcolor[HTML]{C6EFCE}{\color[HTML]{006100} 7.708}                               & \cellcolor[HTML]{C6EFCE}{\color[HTML]{006100} 0.855}                            & \cellcolor[HTML]{C6EFCE}{\color[HTML]{006100} 0.876}                            & \cellcolor[HTML]{C6EFCE}{\color[HTML]{006100} 0.707}                            & \cellcolor[HTML]{C6EFCE}{\color[HTML]{006100} 15.791}                              & \cellcolor[HTML]{FFEB9C}{\color[HTML]{9C5700} 0.911}                            & \cellcolor[HTML]{FFEB9C}{\color[HTML]{9C5700} 0.943}                            & \cellcolor[HTML]{FFEB9C}{\color[HTML]{9C5700} 0.802}                            & \cellcolor[HTML]{FFEB9C}{\color[HTML]{9C5700} 8.441}                               & \cellcolor[HTML]{FFEB9C}{\color[HTML]{9C5700} 0.539}                            & \cellcolor[HTML]{FFC7CE}{\color[HTML]{9C0006} 0.459}                            & \cellcolor[HTML]{FFC7CE}{\color[HTML]{9C0006} 0.342}                            & \cellcolor[HTML]{FFEB9C}{\color[HTML]{9C5700} 0.834}                               & \cellcolor[HTML]{FFEB9C}{\color[HTML]{9C5700} 0.922}                            & \cellcolor[HTML]{C6EFCE}{\color[HTML]{006100} 0.893}                            & \cellcolor[HTML]{C6EFCE}{\color[HTML]{006100} 0.756}                            & \cellcolor[HTML]{C6EFCE}{\color[HTML]{006100} 8.135}                               & \cellcolor[HTML]{FFEB9C}{\color[HTML]{9C5700} 0.62}                             & \cellcolor[HTML]{FFC7CE}{\color[HTML]{9C0006} 0.623}                            & \cellcolor[HTML]{FFC7CE}{\color[HTML]{9C0006} 0.449}                            & \cellcolor[HTML]{FFEB9C}{\color[HTML]{9C5700} 0.911}                               & \cellcolor[HTML]{C6EFCE}{\color[HTML]{006100} 0.868}                            & \cellcolor[HTML]{C6EFCE}{\color[HTML]{006100} 0.884}                            & \cellcolor[HTML]{C6EFCE}{\color[HTML]{006100} 0.705}                            & \cellcolor[HTML]{C6EFCE}{\color[HTML]{006100} 0.542}                               & \cellcolor[HTML]{FFEB9C}{\color[HTML]{9C5700} 0.615}                            & \cellcolor[HTML]{C6EFCE}{\color[HTML]{006100} 0.671}                            & \cellcolor[HTML]{FFEB9C}{\color[HTML]{9C5700} 0.468}                            & \cellcolor[HTML]{FFEB9C}{\color[HTML]{9C5700} 13.406}                             \\
\textbf{\begin{tabular}[c]{@{}l@{}}CGVQM-5\\ (ours)\end{tabular}}                           & \cellcolor[HTML]{C6EFCE}{\color[HTML]{006100} 0.88}                             & \cellcolor[HTML]{C6EFCE}{\color[HTML]{006100} 0.898}                            & \cellcolor[HTML]{C6EFCE}{\color[HTML]{006100} 0.74}                             & \cellcolor[HTML]{C6EFCE}{\color[HTML]{006100} 0.524}                               & \cellcolor[HTML]{C6EFCE}{\color[HTML]{006100} 0.798}                            & \cellcolor[HTML]{C6EFCE}{\color[HTML]{006100} 0.824}                            & \cellcolor[HTML]{C6EFCE}{\color[HTML]{006100} 0.644}                            & \cellcolor[HTML]{C6EFCE}{\color[HTML]{006100} 7.369}                               & \cellcolor[HTML]{C6EFCE}{\color[HTML]{006100} 0.871}                            & \cellcolor[HTML]{C6EFCE}{\color[HTML]{006100} 0.877}                            & \cellcolor[HTML]{C6EFCE}{\color[HTML]{006100} 0.721}                            & \cellcolor[HTML]{C6EFCE}{\color[HTML]{006100} 13.1}                                & \cellcolor[HTML]{C6EFCE}{\color[HTML]{006100} 0.938}                            & \cellcolor[HTML]{C6EFCE}{\color[HTML]{006100} 0.957}                            & \cellcolor[HTML]{C6EFCE}{\color[HTML]{006100} 0.829}                            & \cellcolor[HTML]{C6EFCE}{\color[HTML]{006100} 7.145}                               & \cellcolor[HTML]{C6EFCE}{\color[HTML]{006100} 0.67}                             & \cellcolor[HTML]{C6EFCE}{\color[HTML]{006100} 0.664}                            & \cellcolor[HTML]{C6EFCE}{\color[HTML]{006100} 0.49}                             & \cellcolor[HTML]{C6EFCE}{\color[HTML]{006100} 0.724}                               & \cellcolor[HTML]{FFEB9C}{\color[HTML]{9C5700} 0.918}                            & \cellcolor[HTML]{C6EFCE}{\color[HTML]{006100} 0.92}                             & \cellcolor[HTML]{C6EFCE}{\color[HTML]{006100} 0.776}                            & \cellcolor[HTML]{FFEB9C}{\color[HTML]{9C5700} 8.328}                               & \cellcolor[HTML]{C6EFCE}{\color[HTML]{006100} 0.697}                            & \cellcolor[HTML]{FFEB9C}{\color[HTML]{9C5700} 0.698}                            & \cellcolor[HTML]{FFEB9C}{\color[HTML]{9C5700} 0.519}                            & \cellcolor[HTML]{C6EFCE}{\color[HTML]{006100} 0.832}                               & \cellcolor[HTML]{FFEB9C}{\color[HTML]{9C5700} 0.836}                            & \cellcolor[HTML]{FFEB9C}{\color[HTML]{9C5700} 0.859}                            & \cellcolor[HTML]{FFEB9C}{\color[HTML]{9C5700} 0.681}                            & \cellcolor[HTML]{FFEB9C}{\color[HTML]{9C5700} 0.599}                               & \cellcolor[HTML]{C6EFCE}{\color[HTML]{006100} 0.69}                             & \cellcolor[HTML]{C6EFCE}{\color[HTML]{006100} 0.701}                            & \cellcolor[HTML]{C6EFCE}{\color[HTML]{006100} 0.495}                            & \cellcolor[HTML]{C6EFCE}{\color[HTML]{006100} 12.302}                             \\ \hline
\end{tabular}%
}
\end{table*}
\end{landscape}

%% file: tables/benchmark-ptest.tex
\begin{table}[htbp]
\centering
\caption{Significance testing on benchmarking results. \cmark indicate that our metric (\metricname-5 or \metricname-2  has a significantly higher mean PLCC value than the compared metric. Significance testing was done using bootstrapping and one tailed paired t-test (significance level 0.05).}
\label{tab:benchmark-ptest}
\resizebox{\columnwidth}{!}{%
\begin{tabular}{lcc|cc|cc}
\hline
              & \multicolumn{2}{c|}{GamingVideoSET}                                                                 & \multicolumn{2}{c|}{LIVE Livestream}                                                                & \multicolumn{2}{c}{CG-VQD (ours)}                                                                   \\ \hline
              & \metricname-5                                              & \metricname-2                                            & \metricname-5                                              & \metricname-2                                            & \metricname-5                                              & \metricname-2                                            \\ \hline
ABS           & \cellcolor[HTML]{FFC7CE}{\color[HTML]{9C0006} \xmark} & \cellcolor[HTML]{FFC7CE}{\color[HTML]{9C0006} \xmark} & \cellcolor[HTML]{C6EFCE}{\color[HTML]{006100} \cmark} & \cellcolor[HTML]{C6EFCE}{\color[HTML]{006100} \cmark} & \cellcolor[HTML]{C6EFCE}{\color[HTML]{006100} \cmark} & \cellcolor[HTML]{C6EFCE}{\color[HTML]{006100} \cmark} \\
PSNR          & \cellcolor[HTML]{C6EFCE}{\color[HTML]{006100} \cmark} & \cellcolor[HTML]{C6EFCE}{\color[HTML]{006100} \cmark} & \cellcolor[HTML]{C6EFCE}{\color[HTML]{006100} \cmark} & \cellcolor[HTML]{C6EFCE}{\color[HTML]{006100} \cmark} & \cellcolor[HTML]{C6EFCE}{\color[HTML]{006100} \cmark} & \cellcolor[HTML]{C6EFCE}{\color[HTML]{006100} \cmark} \\
SSIM          & \cellcolor[HTML]{C6EFCE}{\color[HTML]{006100} \cmark} & \cellcolor[HTML]{FFC7CE}{\color[HTML]{9C0006} \xmark} & \cellcolor[HTML]{FFC7CE}{\color[HTML]{9C0006} \xmark} & \cellcolor[HTML]{FFC7CE}{\color[HTML]{9C0006} \xmark} & \cellcolor[HTML]{C6EFCE}{\color[HTML]{006100} \cmark} & \cellcolor[HTML]{C6EFCE}{\color[HTML]{006100} \cmark} \\
MS-SSIM       & \cellcolor[HTML]{C6EFCE}{\color[HTML]{006100} \cmark} & \cellcolor[HTML]{FFC7CE}{\color[HTML]{9C0006} \xmark} & \cellcolor[HTML]{C6EFCE}{\color[HTML]{006100} \cmark} & \cellcolor[HTML]{C6EFCE}{\color[HTML]{006100} \cmark} & \cellcolor[HTML]{C6EFCE}{\color[HTML]{006100} \cmark} & \cellcolor[HTML]{C6EFCE}{\color[HTML]{006100} \cmark} \\
IW-SSIM       & \cellcolor[HTML]{C6EFCE}{\color[HTML]{006100} \cmark} & \cellcolor[HTML]{FFC7CE}{\color[HTML]{9C0006} \xmark} & \cellcolor[HTML]{C6EFCE}{\color[HTML]{006100} \cmark} & \cellcolor[HTML]{C6EFCE}{\color[HTML]{006100} \cmark} & \cellcolor[HTML]{C6EFCE}{\color[HTML]{006100} \cmark} & \cellcolor[HTML]{C6EFCE}{\color[HTML]{006100} \cmark} \\
VIFp          & \cellcolor[HTML]{C6EFCE}{\color[HTML]{006100} \cmark} & \cellcolor[HTML]{C6EFCE}{\color[HTML]{006100} \cmark} & \cellcolor[HTML]{C6EFCE}{\color[HTML]{006100} \cmark} & \cellcolor[HTML]{C6EFCE}{\color[HTML]{006100} \cmark} & \cellcolor[HTML]{FFC7CE}{\color[HTML]{9C0006} \xmark} & \cellcolor[HTML]{C6EFCE}{\color[HTML]{006100} \cmark} \\
FSIM          & \cellcolor[HTML]{C6EFCE}{\color[HTML]{006100} \cmark} & \cellcolor[HTML]{FFC7CE}{\color[HTML]{9C0006} \xmark} & \cellcolor[HTML]{C6EFCE}{\color[HTML]{006100} \cmark} & \cellcolor[HTML]{C6EFCE}{\color[HTML]{006100} \cmark} & \cellcolor[HTML]{C6EFCE}{\color[HTML]{006100} \cmark} & \cellcolor[HTML]{C6EFCE}{\color[HTML]{006100} \cmark} \\
SR-SIM        & \cellcolor[HTML]{C6EFCE}{\color[HTML]{006100} \cmark} & \cellcolor[HTML]{FFC7CE}{\color[HTML]{9C0006} \xmark} & \cellcolor[HTML]{C6EFCE}{\color[HTML]{006100} \cmark} & \cellcolor[HTML]{C6EFCE}{\color[HTML]{006100} \cmark} & \cellcolor[HTML]{C6EFCE}{\color[HTML]{006100} \cmark} & \cellcolor[HTML]{C6EFCE}{\color[HTML]{006100} \cmark} \\
GMSD          & \cellcolor[HTML]{FFC7CE}{\color[HTML]{9C0006} \xmark} & \cellcolor[HTML]{FFC7CE}{\color[HTML]{9C0006} \xmark} & \cellcolor[HTML]{C6EFCE}{\color[HTML]{006100} \cmark} & \cellcolor[HTML]{C6EFCE}{\color[HTML]{006100} \cmark} & \cellcolor[HTML]{C6EFCE}{\color[HTML]{006100} \cmark} & \cellcolor[HTML]{C6EFCE}{\color[HTML]{006100} \cmark} \\
MS-GMSD       & \cellcolor[HTML]{FFC7CE}{\color[HTML]{9C0006} \xmark} & \cellcolor[HTML]{FFC7CE}{\color[HTML]{9C0006} \xmark} & \cellcolor[HTML]{C6EFCE}{\color[HTML]{006100} \cmark} & \cellcolor[HTML]{C6EFCE}{\color[HTML]{006100} \cmark} & \cellcolor[HTML]{C6EFCE}{\color[HTML]{006100} \cmark} & \cellcolor[HTML]{C6EFCE}{\color[HTML]{006100} \cmark} \\
VSI           & \cellcolor[HTML]{C6EFCE}{\color[HTML]{006100} \cmark} & \cellcolor[HTML]{FFC7CE}{\color[HTML]{9C0006} \xmark} & \cellcolor[HTML]{C6EFCE}{\color[HTML]{006100} \cmark} & \cellcolor[HTML]{C6EFCE}{\color[HTML]{006100} \cmark} & \cellcolor[HTML]{C6EFCE}{\color[HTML]{006100} \cmark} & \cellcolor[HTML]{C6EFCE}{\color[HTML]{006100} \cmark} \\
DSS           & \cellcolor[HTML]{C6EFCE}{\color[HTML]{006100} \cmark} & \cellcolor[HTML]{FFC7CE}{\color[HTML]{9C0006} \xmark} & \cellcolor[HTML]{FFC7CE}{\color[HTML]{9C0006} \xmark} & \cellcolor[HTML]{FFC7CE}{\color[HTML]{9C0006} \xmark} & \cellcolor[HTML]{C6EFCE}{\color[HTML]{006100} \cmark} & \cellcolor[HTML]{C6EFCE}{\color[HTML]{006100} \cmark} \\
Content-Score & \cellcolor[HTML]{FFC7CE}{\color[HTML]{9C0006} \xmark} & \cellcolor[HTML]{FFC7CE}{\color[HTML]{9C0006} \xmark} & \cellcolor[HTML]{C6EFCE}{\color[HTML]{006100} \cmark} & \cellcolor[HTML]{C6EFCE}{\color[HTML]{006100} \cmark} & \cellcolor[HTML]{C6EFCE}{\color[HTML]{006100} \cmark} & \cellcolor[HTML]{C6EFCE}{\color[HTML]{006100} \cmark} \\
Style-Score   & \cellcolor[HTML]{C6EFCE}{\color[HTML]{006100} \cmark} & \cellcolor[HTML]{C6EFCE}{\color[HTML]{006100} \cmark} & \cellcolor[HTML]{C6EFCE}{\color[HTML]{006100} \cmark} & \cellcolor[HTML]{C6EFCE}{\color[HTML]{006100} \cmark} & \cellcolor[HTML]{C6EFCE}{\color[HTML]{006100} \cmark} & \cellcolor[HTML]{C6EFCE}{\color[HTML]{006100} \cmark} \\
HaarPSI       & \cellcolor[HTML]{C6EFCE}{\color[HTML]{006100} \cmark} & \cellcolor[HTML]{FFC7CE}{\color[HTML]{9C0006} \xmark} & \cellcolor[HTML]{C6EFCE}{\color[HTML]{006100} \cmark} & \cellcolor[HTML]{C6EFCE}{\color[HTML]{006100} \cmark} & \cellcolor[HTML]{C6EFCE}{\color[HTML]{006100} \cmark} & \cellcolor[HTML]{C6EFCE}{\color[HTML]{006100} \cmark} \\
MDSI          & \cellcolor[HTML]{C6EFCE}{\color[HTML]{006100} \cmark} & \cellcolor[HTML]{FFC7CE}{\color[HTML]{9C0006} \xmark} & \cellcolor[HTML]{C6EFCE}{\color[HTML]{006100} \cmark} & \cellcolor[HTML]{C6EFCE}{\color[HTML]{006100} \cmark} & \cellcolor[HTML]{C6EFCE}{\color[HTML]{006100} \cmark} & \cellcolor[HTML]{C6EFCE}{\color[HTML]{006100} \cmark} \\
LPIPS         & \cellcolor[HTML]{C6EFCE}{\color[HTML]{006100} \cmark} & \cellcolor[HTML]{FFC7CE}{\color[HTML]{9C0006} \xmark} & \cellcolor[HTML]{C6EFCE}{\color[HTML]{006100} \cmark} & \cellcolor[HTML]{C6EFCE}{\color[HTML]{006100} \cmark} & \cellcolor[HTML]{C6EFCE}{\color[HTML]{006100} \cmark} & \cellcolor[HTML]{C6EFCE}{\color[HTML]{006100} \cmark} \\
DISTS         & \cellcolor[HTML]{C6EFCE}{\color[HTML]{006100} \cmark} & \cellcolor[HTML]{FFC7CE}{\color[HTML]{9C0006} \xmark} & \cellcolor[HTML]{C6EFCE}{\color[HTML]{006100} \cmark} & \cellcolor[HTML]{C6EFCE}{\color[HTML]{006100} \cmark} & \cellcolor[HTML]{C6EFCE}{\color[HTML]{006100} \cmark} & \cellcolor[HTML]{C6EFCE}{\color[HTML]{006100} \cmark} \\
STRRED        & \cellcolor[HTML]{C6EFCE}{\color[HTML]{006100} \cmark} & \cellcolor[HTML]{C6EFCE}{\color[HTML]{006100} \cmark} & \cellcolor[HTML]{C6EFCE}{\color[HTML]{006100} \cmark} & \cellcolor[HTML]{C6EFCE}{\color[HTML]{006100} \cmark} & \cellcolor[HTML]{C6EFCE}{\color[HTML]{006100} \cmark} & \cellcolor[HTML]{C6EFCE}{\color[HTML]{006100} \cmark} \\
PieAPP        & \cellcolor[HTML]{C6EFCE}{\color[HTML]{006100} \cmark} & \cellcolor[HTML]{FFC7CE}{\color[HTML]{9C0006} \xmark} & \cellcolor[HTML]{C6EFCE}{\color[HTML]{006100} \cmark} & \cellcolor[HTML]{C6EFCE}{\color[HTML]{006100} \cmark} & \cellcolor[HTML]{C6EFCE}{\color[HTML]{006100} \cmark} & \cellcolor[HTML]{C6EFCE}{\color[HTML]{006100} \cmark} \\
FLIP          & \cellcolor[HTML]{FFC7CE}{\color[HTML]{9C0006} \xmark} & \cellcolor[HTML]{FFC7CE}{\color[HTML]{9C0006} \xmark} & \cellcolor[HTML]{C6EFCE}{\color[HTML]{006100} \cmark} & \cellcolor[HTML]{C6EFCE}{\color[HTML]{006100} \cmark} & \cellcolor[HTML]{C6EFCE}{\color[HTML]{006100} \cmark} & \cellcolor[HTML]{C6EFCE}{\color[HTML]{006100} \cmark} \\
FoVVideoVDP   & \cellcolor[HTML]{C6EFCE}{\color[HTML]{006100} \cmark} & \cellcolor[HTML]{FFC7CE}{\color[HTML]{9C0006} \xmark} & \cellcolor[HTML]{FFC7CE}{\color[HTML]{9C0006} \xmark} & \cellcolor[HTML]{FFC7CE}{\color[HTML]{9C0006} \xmark} & \cellcolor[HTML]{C6EFCE}{\color[HTML]{006100} \cmark} & \cellcolor[HTML]{C6EFCE}{\color[HTML]{006100} \cmark} \\
ColorVideoVDP          & \cellcolor[HTML]{FFC7CE}{\color[HTML]{9C0006} \xmark} & \cellcolor[HTML]{FFC7CE}{\color[HTML]{9C0006} \xmark} & \cellcolor[HTML]{FFC7CE}{\color[HTML]{9C0006} \xmark} & \cellcolor[HTML]{FFC7CE}{\color[HTML]{9C0006} \xmark} & \cellcolor[HTML]{C6EFCE}{\color[HTML]{006100} \cmark} & \cellcolor[HTML]{C6EFCE}{\color[HTML]{006100} \cmark} \\
ERQA          & \cellcolor[HTML]{C6EFCE}{\color[HTML]{006100} \cmark} & \cellcolor[HTML]{C6EFCE}{\color[HTML]{006100} \cmark} & \cellcolor[HTML]{C6EFCE}{\color[HTML]{006100} \cmark} & \cellcolor[HTML]{C6EFCE}{\color[HTML]{006100} \cmark} & \cellcolor[HTML]{C6EFCE}{\color[HTML]{006100} \cmark} & \cellcolor[HTML]{C6EFCE}{\color[HTML]{006100} \cmark} \\
VMAF          & \cellcolor[HTML]{C6EFCE}{\color[HTML]{006100} \cmark} & \cellcolor[HTML]{FFC7CE}{\color[HTML]{9C0006} \xmark} & \cellcolor[HTML]{C6EFCE}{\color[HTML]{006100} \cmark} & \cellcolor[HTML]{C6EFCE}{\color[HTML]{006100} \cmark} & \cellcolor[HTML]{C6EFCE}{\color[HTML]{006100} \cmark} & \cellcolor[HTML]{C6EFCE}{\color[HTML]{006100} \cmark} \\ \hline
\end{tabular}%
}
\end{table}

%% file: tables/arch-ptest.tex
\begin{table*}[]
\centering
\caption{Significance testing on experiment results. $t_{ij}=$\cmark indicate that $i^{\textrm{th}}$ row has a significantly higher mean PLCC value than $j^{\textrm{th}}$ column. Significance testing was done using bootstrapping and one tailed paired t-test (significance level 0.05).}
\label{tab:arch-ptest}
\resizebox{\textwidth}{!}{%
\begin{tabular}{lccccccc|ccccccc|ccccccc}
\hline
                                                                       & \multicolumn{7}{c|}{GamingVideoSET}                                                                                                                                                                                                                                                                                                                                                                           & \multicolumn{7}{c|}{LIVE Livestream}                                                                                                                                                                                                                                                                                                                                                                        & \multicolumn{7}{c}{CG-VQD (ours)}                                                                                                                                                                                                                                                                                                                                                                           \\ \hline
                                                                       & LPIPS                                            & C3D                                              & MC3                                              & R(2+1)D                                          & R3D                                              & \begin{tabular}[c]{@{}c@{}}R(2+1)D\\ (No \\ pre-training)\end{tabular} & \begin{tabular}[c]{@{}c@{}}R(2+1)D\\ (No \\ calibration)\end{tabular} & LPIPS                                            & C3D                                              & MC3                                              & R(2+1)D                                          & R3D                                              & \begin{tabular}[c]{@{}c@{}}R(2+1)D\\ (No\\ pre-training)\end{tabular} & \begin{tabular}[c]{@{}c@{}}R(2+1)D\\ (No\\ calibration)\end{tabular} & LPIPS                                            & C3D                                              & MC3                                              & R(2+1)D                                          & R3D                                              & \begin{tabular}[c]{@{}c@{}}R(2+1)D\\ (No\\ pre-training)\end{tabular} & \begin{tabular}[c]{@{}c@{}}R(2+1)D\\ (No\\ calibration)\end{tabular} \\ \hline
LPIPS                                                                  & \cellcolor[HTML]{FFC7CE}{\color[HTML]{9C0006} \xmark} & \cellcolor[HTML]{C6EFCE}{\color[HTML]{006100} \cmark} & \cellcolor[HTML]{FFC7CE}{\color[HTML]{9C0006} \xmark} & \cellcolor[HTML]{FFC7CE}{\color[HTML]{9C0006} \xmark} & \cellcolor[HTML]{FFC7CE}{\color[HTML]{9C0006} \xmark} & \cellcolor[HTML]{C6EFCE}{\color[HTML]{006100} \cmark}                       & \cellcolor[HTML]{FFC7CE}{\color[HTML]{9C0006} \xmark}                      & \cellcolor[HTML]{FFC7CE}{\color[HTML]{9C0006} \xmark} & \cellcolor[HTML]{FFC7CE}{\color[HTML]{9C0006} \xmark} & \cellcolor[HTML]{FFC7CE}{\color[HTML]{9C0006} \xmark} & \cellcolor[HTML]{FFC7CE}{\color[HTML]{9C0006} \xmark} & \cellcolor[HTML]{FFC7CE}{\color[HTML]{9C0006} \xmark} & \cellcolor[HTML]{FFC7CE}{\color[HTML]{9C0006} \xmark}                      & \cellcolor[HTML]{FFC7CE}{\color[HTML]{9C0006} \xmark}                     & \cellcolor[HTML]{FFC7CE}{\color[HTML]{9C0006} \xmark} & \cellcolor[HTML]{FFC7CE}{\color[HTML]{9C0006} \xmark} & \cellcolor[HTML]{FFC7CE}{\color[HTML]{9C0006} \xmark} & \cellcolor[HTML]{FFC7CE}{\color[HTML]{9C0006} \xmark} & \cellcolor[HTML]{FFC7CE}{\color[HTML]{9C0006} \xmark} & \cellcolor[HTML]{FFC7CE}{\color[HTML]{9C0006} \xmark}                      & \cellcolor[HTML]{FFC7CE}{\color[HTML]{9C0006} \xmark}                     \\
C3D                                                                    & \cellcolor[HTML]{FFC7CE}{\color[HTML]{9C0006} \xmark} & \cellcolor[HTML]{FFC7CE}{\color[HTML]{9C0006} \xmark} & \cellcolor[HTML]{FFC7CE}{\color[HTML]{9C0006} \xmark} & \cellcolor[HTML]{FFC7CE}{\color[HTML]{9C0006} \xmark} & \cellcolor[HTML]{FFC7CE}{\color[HTML]{9C0006} \xmark} & \cellcolor[HTML]{FFC7CE}{\color[HTML]{9C0006} \xmark}                       & \cellcolor[HTML]{FFC7CE}{\color[HTML]{9C0006} \xmark}                      & \cellcolor[HTML]{C6EFCE}{\color[HTML]{006100} \cmark} & \cellcolor[HTML]{FFC7CE}{\color[HTML]{9C0006} \xmark} & \cellcolor[HTML]{FFC7CE}{\color[HTML]{9C0006} \xmark} & \cellcolor[HTML]{FFC7CE}{\color[HTML]{9C0006} \xmark} & \cellcolor[HTML]{FFC7CE}{\color[HTML]{9C0006} \xmark} & \cellcolor[HTML]{FFC7CE}{\color[HTML]{9C0006} \xmark}                      & \cellcolor[HTML]{FFC7CE}{\color[HTML]{9C0006} \xmark}                     & \cellcolor[HTML]{FFC7CE}{\color[HTML]{9C0006} \xmark} & \cellcolor[HTML]{FFC7CE}{\color[HTML]{9C0006} \xmark} & \cellcolor[HTML]{FFC7CE}{\color[HTML]{9C0006} \xmark} & \cellcolor[HTML]{FFC7CE}{\color[HTML]{9C0006} \xmark} & \cellcolor[HTML]{FFC7CE}{\color[HTML]{9C0006} \xmark} & \cellcolor[HTML]{FFC7CE}{\color[HTML]{9C0006} \xmark}                      & \cellcolor[HTML]{FFC7CE}{\color[HTML]{9C0006} \xmark}                     \\
MC3                                                                    & \cellcolor[HTML]{C6EFCE}{\color[HTML]{006100} \cmark} & \cellcolor[HTML]{C6EFCE}{\color[HTML]{006100} \cmark} & \cellcolor[HTML]{FFC7CE}{\color[HTML]{9C0006} \xmark} & \cellcolor[HTML]{FFC7CE}{\color[HTML]{9C0006} \xmark} & \cellcolor[HTML]{FFC7CE}{\color[HTML]{9C0006} \xmark} & \cellcolor[HTML]{C6EFCE}{\color[HTML]{006100} \cmark}                       & \cellcolor[HTML]{C6EFCE}{\color[HTML]{006100} \cmark}                      & \cellcolor[HTML]{C6EFCE}{\color[HTML]{006100} \cmark} & \cellcolor[HTML]{C6EFCE}{\color[HTML]{006100} \cmark} & \cellcolor[HTML]{FFC7CE}{\color[HTML]{9C0006} \xmark} & \cellcolor[HTML]{FFC7CE}{\color[HTML]{9C0006} \xmark} & \cellcolor[HTML]{FFC7CE}{\color[HTML]{9C0006} \xmark} & \cellcolor[HTML]{C6EFCE}{\color[HTML]{006100} \cmark}                      & \cellcolor[HTML]{C6EFCE}{\color[HTML]{006100} \cmark}                     & \cellcolor[HTML]{C6EFCE}{\color[HTML]{006100} \cmark} & \cellcolor[HTML]{C6EFCE}{\color[HTML]{006100} \cmark} & \cellcolor[HTML]{FFC7CE}{\color[HTML]{9C0006} \xmark} & \cellcolor[HTML]{FFC7CE}{\color[HTML]{9C0006} \xmark} & \cellcolor[HTML]{FFC7CE}{\color[HTML]{9C0006} \xmark} & \cellcolor[HTML]{C6EFCE}{\color[HTML]{006100} \cmark}                      & \cellcolor[HTML]{C6EFCE}{\color[HTML]{006100} \cmark}                     \\
R(2+1)D                                                                & \cellcolor[HTML]{C6EFCE}{\color[HTML]{006100} \cmark} & \cellcolor[HTML]{C6EFCE}{\color[HTML]{006100} \cmark} & \cellcolor[HTML]{FFC7CE}{\color[HTML]{9C0006} \xmark} & \cellcolor[HTML]{FFC7CE}{\color[HTML]{9C0006} \xmark} & \cellcolor[HTML]{FFC7CE}{\color[HTML]{9C0006} \xmark} & \cellcolor[HTML]{C6EFCE}{\color[HTML]{006100} \cmark}                       & \cellcolor[HTML]{C6EFCE}{\color[HTML]{006100} \cmark}                      & \cellcolor[HTML]{C6EFCE}{\color[HTML]{006100} \cmark} & \cellcolor[HTML]{C6EFCE}{\color[HTML]{006100} \cmark} & \cellcolor[HTML]{FFC7CE}{\color[HTML]{9C0006} \xmark} & \cellcolor[HTML]{FFC7CE}{\color[HTML]{9C0006} \xmark} & \cellcolor[HTML]{FFC7CE}{\color[HTML]{9C0006} \xmark} & \cellcolor[HTML]{C6EFCE}{\color[HTML]{006100} \cmark}                      & \cellcolor[HTML]{C6EFCE}{\color[HTML]{006100} \cmark}                     & \cellcolor[HTML]{C6EFCE}{\color[HTML]{006100} \cmark} & \cellcolor[HTML]{C6EFCE}{\color[HTML]{006100} \cmark} & \cellcolor[HTML]{FFC7CE}{\color[HTML]{9C0006} \xmark} & \cellcolor[HTML]{FFC7CE}{\color[HTML]{9C0006} \xmark} & \cellcolor[HTML]{FFC7CE}{\color[HTML]{9C0006} \xmark} & \cellcolor[HTML]{C6EFCE}{\color[HTML]{006100} \cmark}                      & \cellcolor[HTML]{C6EFCE}{\color[HTML]{006100} \cmark}                     \\
R3D                                                                    & \cellcolor[HTML]{C6EFCE}{\color[HTML]{006100} \cmark} & \cellcolor[HTML]{C6EFCE}{\color[HTML]{006100} \cmark} & \cellcolor[HTML]{C6EFCE}{\color[HTML]{006100} \cmark} & \cellcolor[HTML]{FFC7CE}{\color[HTML]{9C0006} \xmark} & \cellcolor[HTML]{FFC7CE}{\color[HTML]{9C0006} \xmark} & \cellcolor[HTML]{C6EFCE}{\color[HTML]{006100} \cmark}                       & \cellcolor[HTML]{C6EFCE}{\color[HTML]{006100} \cmark}                      & \cellcolor[HTML]{C6EFCE}{\color[HTML]{006100} \cmark} & \cellcolor[HTML]{C6EFCE}{\color[HTML]{006100} \cmark} & \cellcolor[HTML]{FFC7CE}{\color[HTML]{9C0006} \xmark} & \cellcolor[HTML]{FFC7CE}{\color[HTML]{9C0006} \xmark} & \cellcolor[HTML]{FFC7CE}{\color[HTML]{9C0006} \xmark} & \cellcolor[HTML]{C6EFCE}{\color[HTML]{006100} \cmark}                      & \cellcolor[HTML]{C6EFCE}{\color[HTML]{006100} \cmark}                     & \cellcolor[HTML]{C6EFCE}{\color[HTML]{006100} \cmark} & \cellcolor[HTML]{C6EFCE}{\color[HTML]{006100} \cmark} & \cellcolor[HTML]{FFC7CE}{\color[HTML]{9C0006} \xmark} & \cellcolor[HTML]{FFC7CE}{\color[HTML]{9C0006} \xmark} & \cellcolor[HTML]{FFC7CE}{\color[HTML]{9C0006} \xmark} & \cellcolor[HTML]{C6EFCE}{\color[HTML]{006100} \cmark}                      & \cellcolor[HTML]{C6EFCE}{\color[HTML]{006100} \cmark}                     \\
\begin{tabular}[c]{@{}l@{}}R(2+1)D \\ (No\\ pre-training)\end{tabular} & \cellcolor[HTML]{FFC7CE}{\color[HTML]{9C0006} \xmark} & \cellcolor[HTML]{FFC7CE}{\color[HTML]{9C0006} \xmark} & \cellcolor[HTML]{FFC7CE}{\color[HTML]{9C0006} \xmark} & \cellcolor[HTML]{FFC7CE}{\color[HTML]{9C0006} \xmark} & \cellcolor[HTML]{FFC7CE}{\color[HTML]{9C0006} \xmark} & \cellcolor[HTML]{FFC7CE}{\color[HTML]{9C0006} \xmark}                       & \cellcolor[HTML]{FFC7CE}{\color[HTML]{9C0006} \xmark}                      & \cellcolor[HTML]{C6EFCE}{\color[HTML]{006100} \cmark} & \cellcolor[HTML]{C6EFCE}{\color[HTML]{006100} \cmark} & \cellcolor[HTML]{FFC7CE}{\color[HTML]{9C0006} \xmark} & \cellcolor[HTML]{FFC7CE}{\color[HTML]{9C0006} \xmark} & \cellcolor[HTML]{FFC7CE}{\color[HTML]{9C0006} \xmark} & \cellcolor[HTML]{FFC7CE}{\color[HTML]{9C0006} \xmark}                      & \cellcolor[HTML]{FFC7CE}{\color[HTML]{9C0006} \xmark}                     & \cellcolor[HTML]{C6EFCE}{\color[HTML]{006100} \cmark} & \cellcolor[HTML]{C6EFCE}{\color[HTML]{006100} \cmark} & \cellcolor[HTML]{FFC7CE}{\color[HTML]{9C0006} \xmark} & \cellcolor[HTML]{FFC7CE}{\color[HTML]{9C0006} \xmark} & \cellcolor[HTML]{FFC7CE}{\color[HTML]{9C0006} \xmark} & \cellcolor[HTML]{FFC7CE}{\color[HTML]{9C0006} \xmark}                      & \cellcolor[HTML]{FFC7CE}{\color[HTML]{9C0006} \xmark}                     \\
\begin{tabular}[c]{@{}l@{}}R(2+1)D \\ (No \\ calibration)\end{tabular} & \cellcolor[HTML]{C6EFCE}{\color[HTML]{006100} \cmark} & \cellcolor[HTML]{C6EFCE}{\color[HTML]{006100} \cmark} & \cellcolor[HTML]{FFC7CE}{\color[HTML]{9C0006} \xmark} & \cellcolor[HTML]{FFC7CE}{\color[HTML]{9C0006} \xmark} & \cellcolor[HTML]{FFC7CE}{\color[HTML]{9C0006} \xmark} & \cellcolor[HTML]{C6EFCE}{\color[HTML]{006100} \cmark}                       & \cellcolor[HTML]{FFC7CE}{\color[HTML]{9C0006} \xmark}                      & \cellcolor[HTML]{C6EFCE}{\color[HTML]{006100} \cmark} & \cellcolor[HTML]{C6EFCE}{\color[HTML]{006100} \cmark} & \cellcolor[HTML]{FFC7CE}{\color[HTML]{9C0006} \xmark} & \cellcolor[HTML]{FFC7CE}{\color[HTML]{9C0006} \xmark} & \cellcolor[HTML]{FFC7CE}{\color[HTML]{9C0006} \xmark} & \cellcolor[HTML]{C6EFCE}\cmark                                             & \cellcolor[HTML]{FFC7CE}{\color[HTML]{9C0006} \xmark}                     & \cellcolor[HTML]{C6EFCE}{\color[HTML]{006100} \cmark} & \cellcolor[HTML]{C6EFCE}{\color[HTML]{006100} \cmark} & \cellcolor[HTML]{FFC7CE}{\color[HTML]{9C0006} \xmark} & \cellcolor[HTML]{FFC7CE}{\color[HTML]{9C0006} \xmark} & \cellcolor[HTML]{FFC7CE}{\color[HTML]{9C0006} \xmark} & \cellcolor[HTML]{C6EFCE}{\color[HTML]{006100} \cmark}                      & \cellcolor[HTML]{FFC7CE}{\color[HTML]{9C0006} \xmark}                     \\ \hline
\end{tabular}%
}
\end{table*}